\documentclass[journal]{IEEEtran}
\usepackage{amsmath,amsfonts}
\usepackage{algorithmic}
\usepackage[ruled,linesnumbered]{algorithm2e}
\usepackage{array}
\usepackage{textcomp}
\usepackage{stfloats}
\usepackage{url}
\usepackage{verbatim}
\usepackage{graphicx}
\def\BibTeX{{\rm B\kern-.05em{\sc i\kern-.025em b}\kern-.08em
    T\kern-.1667em\lower.7ex\hbox{E}\kern-.125emX}}
\usepackage{balance}
\usepackage{cite}
\usepackage{bm}
\usepackage{subfigure}

\usepackage{multirow,makecell}
\usepackage{booktabs}

\usepackage{clipboard}  
\usepackage{enumerate}
\usepackage[normalem]{ulem}
\usepackage{enumitem}
\usepackage{xcolor}

\ifodd 0
\newcommand{\rev}[1]{{\color{blue}#1}}
  
\newcommand{\revhu}[1]{{\color{blue}#1}}  

\newcommand{\norev}[1]{#1} 
\else
\newcommand{\rev}[1]{#1} 
  
\newcommand{\revhu}[1]{#1}

\newcommand{\norev}[1]{#1} 
\fi

\ifodd 1

\else

\fi 

\newcommand{\mathi}{\mathrm{i}}

\newcommand{\lightc}{\mathrm{c}}

\begin{document}
\newcommand{\name}{WiAnchor\xspace} 
\newcommand{\dataset}{NFS-Fi\xspace}
\title{Cross-Domain Multi-Person Human Activity Recognition via Near-Field Wi-Fi Sensing}
\author{
Xin~Li,~\IEEEmembership{Member,~IEEE},
Jingzhi~Hu,~\IEEEmembership{Member,~IEEE}, 
Yinghui~He,~\IEEEmembership{Member,~IEEE}, \\
Hongbo~Wang,~\IEEEmembership{Graduate Student Member,~IEEE},
Jin~Gan,
and~Jun~Luo,~\IEEEmembership{Fellow,~IEEE}

\thanks{
\IEEEcompsocthanksitem This work has been submitted to the IEEE for possible publication. Copyright may be transferred without notice, after which this version may no longer be accessible.
\IEEEcompsocthanksitem This research is supported by The National Research Foundation Singapore and Infocomm Media Development Authority under its Future Communications Research \& Development Programme, and MOE Tier 1 grant RG16/22.
\IEEEcompsocthanksitem X. Li, J. Hu, Y. He, H. Wang, J. Gan and J. Luo are with the College of Computing and Data Science, Nanyang Technological University, Singapore.
(email: \{l.xin,~yinghui.he,~hongbo001,~jin010,~junluo\}@ntu.edu.sg, jingzhi.hu518@gmail.com).
}
}
\markboth{Submitted Manuscript}%
{\name}

\maketitle
\begin{abstract}
Wi-Fi-based human activity recognition (HAR) provides substantial convenience and has emerged as a thriving research field,
yet the coarse spatial resolution inherent to Wi-Fi significantly hinders its ability to distinguish \emph{multiple subjects}.
By exploiting the near-field domination effect, establishing a dedicated sensing link for each subject through their personal Wi-Fi device offers a promising solution for multi-person HAR under native traffic.
However, due to the subject-specific characteristics and irregular patterns of near-field signals, HAR neural network models require fine-tuning (FT) for cross-domain adaptation, which becomes particularly challenging with certain categories unavailable.
In this paper, we propose \emph{\name}, a novel training framework for efficient cross-domain adaptation in the presence of incomplete activity categories.
This framework processes Wi-Fi signals embedded with irregular time information in three steps:
during pre-training, we enlarge inter-class feature margins to enhance the separability of activities;
in the FT stage, we innovate an \emph{anchor matching} mechanism for cross-domain adaptation, filtering subject-specific interference informed by incomplete activity categories, rather than attempting to extract complete features from them;
finally, the recognition of input samples is further improved based on their feature-level similarity with anchors.
We construct a comprehensive dataset to thoroughly evaluate \name, achieving over 90\% cross-domain accuracy with absent activity categories under multi-person scenarios.
\end{abstract}

\section{Introduction} \label{sec:intro}

\IEEEPARstart{W}{ith} the ubiquitous deployment of its infrastructure, Wi-Fi has become an indispensable part of modern life~\cite{wu2023enabling}.
The ubiquity of Wi-Fi, in turn, sparks significant interest in various research fields, prompting extensive exploration in multiple directions~\cite{Li2020Wireless, ma2019wifi}.
Among these, Integrated Sensing and Communications (ISAC)~\cite{isacot}, which seeks to harness Wi-Fi's sensing capabilities rather than merely treating it as a convenient communication medium, has attracted considerable attention from both academia and industry due to its promising application potential~\cite{Widar2-MobiSys18, mDTrack-MobiCom19, EI-MobiCom18, zhang2021widar3, huang2024wimans, torun2023wi, Wang2017PhaseBeat, hillyard2018experience}.
In particular, Wi-Fi sensing refers to inferring environment conditions or human activities from variations in signal amplitude and phase during propagation~\cite{zeng2019farsense}.
As the pivotal enabler of Wi-Fi sensing, Channel State Information (CSI)~\cite{PicoScenes-IoIJ21} provides an easily accessible signal representation that propels the technology into a wide range of applications,
including localization~\cite{Widar2-MobiSys18, mDTrack-MobiCom19}, human activity recognition (HAR)~\cite{EI-MobiCom18, zhang2021widar3, huang2024wimans, torun2023wi}, and vital sign monitoring~\cite{Wang2017PhaseBeat, hillyard2018experience}.
Among the various applications, HAR stands at the forefront, offering substantial practical values for diverse important scenarios, such as augmented/virtual reality (AR/VR)~\cite{wang2025vr} and health emergency detection~\cite{wang2016rt}.

\begin{figure}[t]
	\centering 
	\setlength{\abovecaptionskip}{6pt}
		\includegraphics[width=0.92\linewidth]{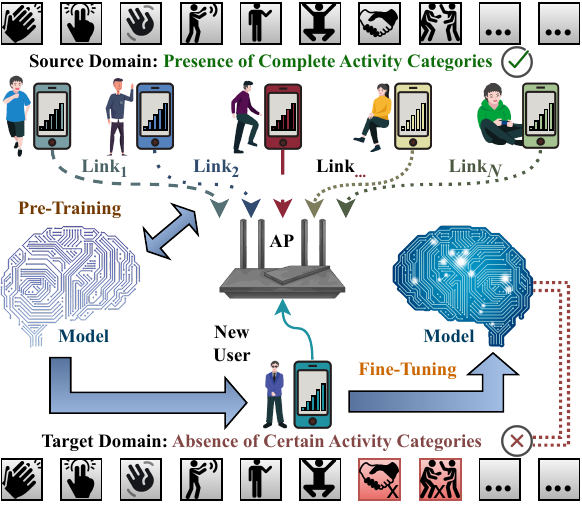}  
	\caption{\rev{Constructing dedicated links via smart devices holds promise for multi-person HAR, but the subject-specific characteristics necessitate model fine-tuning for cross-domain adaptation, which is hindered by the absence of certain activity categories.}}
	\label{fig:teaser}
\vspace{-1.em}
\end{figure}
%

However, \revhu{existing deployable Wi-Fi sensing methods (as exemplified by Widar3.0~\cite{zhang2021widar3}) 
are often designed for single-person rather than \textit{multi-person}%
\footnote{\revhu{Here, multi-person HAR falls into three progressively more challenging sensing targets: a specific subject under multi-person interference, the independent activities of each subject, and the interactions among multiple subjects.}} HAR}, 
and thus cannot address the increasingly complex requirements of real-world scenarios.
This is because multi-person HAR demands sufficient spatial resolution to distinguish different subjects, which is inherently constrained by the limited channel bandwidth of Wi-Fi systems.
Since the primary task of Wi-Fi systems remains communication, excessive expansion of channel bandwidth is particularly restricted to reduce co-channel interference~\cite{chintalapudi2012wifi,luo2021energy}.
Consequently, makeshift solutions, including decomposing CSI into multiple source components~\cite{MultiSense-UbiComp20} or
employing deep neural networks to overfit CSI~\cite{huang2024wimans}, have been explored for multi-person sensing;
yet these approaches fail to scale up 
due to the lack of \textit{physical-layer diversity}.
Another line of work collects signals across different antenna arrays~\cite{mDTrack-MobiCom19, karanam2019tracking, song2024siwis} or channels~\cite{xie2015precise, Chronos-NSDI16, li2024uwb} to compensate the limited bandwidth with spatial or temporal diversity, but it often requires complex system modifications and may disrupt normal communications.
Consequently, developing a realistic multi-person Wi-Fi sensing method is an essential and urgent step toward realizing the ISAC ambition.

Fortunately, the ubiquity of Wi-Fi-connected personal smart devices, such as smartphones, enables a promising framework for multi-person sensing through multi-link utilization, as shown in Fig.~\ref{fig:teaser} (upper panel).
While early studies~\cite{tan2019multitrack, ren2022gopose} have demonstrated that constructing multiple links using several fixed devices can marginally improve spatial resolution, 
they overlook a critical \textit{near-field domination effect}: given the close proximity of each subject to its Wi-Fi-connected user equipment~(UE), the activity-induced CSI impact on the Wi-Fi link is sufficiently strong to make interference from other subjects negligible~\cite{hu2023password,wang2024muki,cong2024near}.
This effect implies a unique correspondence between a subject and the link that its UE established with the access point (AP), making multi-person HAR feasible with commercial off-the-shelf (COTS) devices under prevalent communication configurations~\cite{hu2023musefi}.

\norev{Nevertheless,}
the HAR models that rely on the near-field domination effect still face several inherent challenges, as shown in Fig.~\ref{fig:teaser} (lower panel).
First, unlike existing Wi-Fi sensing systems that benefit from a high and regular CSI sampling rate (up to 1000 frames/s~\cite{Widar2-MobiSys18,WiPose-MobiCom20}), the frame arrival rate per link in multi-user default communication scenarios is significantly lower and highly irregular due to the contention-based multi-access nature of Wi-Fi, which severely degrades sensing performance.
Second, the strong subject-specific characteristics of near-field channel samples necessitate the calibration, i.e., fine-tuning (FT), of pre-trained models with the CSI samples for all potential activity classes to 
achieve cross-domain adaptation~\cite{Hu25TWC_Cross}.
\rev{While the source domain contains data covering all activity classes, in real-world scenarios it is impractical to require target users to perform all activity classes for extensive data collection due to usability and safety concerns~\cite{hu2024poison}.
However, FT with only a limited number of samples from an incomplete set of activity classes in the target domain leads to unsatisfactory adaptation performance.}
Last but not least, currently available datasets~\cite{huang2024wimans, yousefi2017survey, palipana2018falldefi, ma2018signfi,guo2019wiar,brinke2019dataset,zhang2021widar3, baha2020dataset,hu2021wigr,bocus2022operanet, yang2022efficientfi, yang2023mm, zhao2024finding, wang2024xrf55,lan2025xrf}  
have no support for the near-field multi-person HAR under default communication configurations, and are built with NICs that implement outdated Wi-Fi standards.

\rev{In this paper, we propose a practical multi-person Wi-Fi sensing system leveraging the near-field domination effect.}
We then develop a novel framework, \emph{\name}, which facilitates efficient cross-domain adaptation using only a small number of samples, even in the complete absence of samples for certain activity categories, thereby enabling real-world deployment of Wi-Fi-based multi-person HAR.
Specifically, we design a time information embedding algorithm that encodes the highly non-uniform frame arrival time into temporal features.
During the pre-training (PT) stage, we introduce an inter-class margin enlarging strategy to encourage the HAR neural model to extract discriminative activity features.
During the FT stage, features from sub-sampled portion of the PT dataset are used as anchors, and target domain features are encouraged to align with them to filter out subject-specific interference.
In the inference stage, we adopt a composite strategy that combines model logits with feature similarity to the anchors to yield accurate activity recognition.
Finally, we construct a comprehensive dataset with approximately 65,000 samples and conduct a thorough evaluation of the proposed framework using it.
In summary, our main contributions are:
\begin{itemize}
    \item We present a practical \rev{system} for multi-person sensing based on near-field domination effect, leveraging COTS Wi-Fi devices without requiring hardware modifications.
    \item We design a time information embedding algorithm to effectively capture and represent the highly non-uniform CSI sampling patterns \rev{of the sensing system}.
    \item We propose \name framework \rev{for multi-person HAR}, which facilitates efficient cross-domain adaptation in the absence of certain categories via a two-stage training strategy and a composite decision mechanism.
    \item We construct the first multi-person near-field Wi-Fi sensing dataset, containing approximately 65,000 samples collected under default communication configurations, with a diverse set of subjects and environments.
    \item We conduct a comprehensive evaluation of \name framework, showing a 56.8\% improvement in recognition accuracy for categories without FT samples and an overall accuracy exceeding 90\%.
\end{itemize}

The rest of our paper is structured as follows:
Section~\ref{sec:motivation} presents theoretical and practical evidence for the multi-person HAR system based on the near-field domination effect,
along with associated challenges.
Section~\ref{sec:method} formulates our \name framework.
Section~\ref{sec:dataset} details the experiment setup and dataset construction.
Section~\ref{sec:evaluation} presents the evaluation results.
The conclusion and discussion are presented in Section~\ref{sec:conclusion}.

\vspace{-0.5em}
\section{Wi-Fi Sensing under Near-Field Domination} \label{sec:motivation}

In this section, we first introduce the fundamentals of Wi-Fi sensing and analyze existing studies.
We then demonstrate the feasibility of multi-person sensing under the near-field domination effect.
Finally, we present experiments that illustrate the challenges and potential solutions for fine-tuning HAR models in the absence of certain activity categories.

\vspace{-0.7em}
\subsection{Wi-Fi Sensing Basics} \label{ssec:basic}

We begin with a general Wi-Fi sensing system, comprising an AP-UE pair and multiple sensed subjects within the wireless network.
The $k$-th path in this system at time $t$ can be described by the tuple $(\tau_{k,t}, \theta_{k,t})$, where $\tau$ and $\theta$ are the \textit{time of flight} (ToF) and \textit{angle of arrival} (AoA), respectively.
The CSI $[\bm H]_{n,m,t} = h_{n,m,t}$ received at AP can be modeled as:
\begin{equation}
\begin{aligned}
  \!\!\!\!\!\!\!&h_{n,m,t}
  =
   \sum_{k=1}^{K}
   \alpha_{n,m,k,t} \cdot
   h_{m,k,t}^\mathrm{ToF} \cdot
   h_{n,k,t}^\mathrm{AoA} + \zeta_t \\
  \!\!\!\!\!\!&=
   \sum_{k=1}^{K}
   \alpha_{n,m,k,t}
   \mathrm{e}^{-\mathi2 \pi(f_{\mathrm{c}}  \pm  (m-1)f_{\mathrm{b}} )\tau_{k,t}}
   \mathrm{e}^{-\mathi2 \pi (n-1)d\cos(\theta_{k,t})\frac{f_{\mathrm{c}}}{\mathrm{c}}}+ \zeta_t, \!\!\!\!\!\!\!\!\!\!\!\!\!
   \label{eqn:CSI_simple}
\end{aligned}
\end{equation}
where AP antennas are linearly arranged with a spacing of $d$,
$n$ and $m$ respectively index the antenna and subcarrier,
$f_{\mathrm{c}}$ and $f_{\mathrm{b}}$ respectively denote channel centre frequency and subcarrier bandwidth,
$\alpha$ represents channel gain,
$\mathrm{c}$ is the speed of light,
and $\zeta$ indicates noise introduced by the environment and hardware.
For multi-person sensing, the multipath components in the Wi-Fi system need to be distinguished to extract subject-specific information, which requires sufficient spatial resolution.
According to Eqn.~\eqref{eqn:CSI_simple}, spatial resolution can be improved by enhancing the \textit{range} (ToF) and/or \textit{bearing} (AoA) resolutions.
Based on~\cite{adib20143d}, range resolution, $\Delta L = \frac{\lightc}{W}$, increases linearly with the effective sensing bandwidth $W$.
Given the impracticality of excessively expanding a single channel’s bandwidth, previous works~\cite{xie2015precise, Chronos-NSDI16, li2024uwb} fuse multiple channels to attain a larger effective sensing bandwidth, thereby enhancing $\Delta L$.
Additionally, as shown in~\cite{johnson1992array}, bearing resolution, $\Delta \theta = \frac{\lambda}{(N-1)d}$, increases linearly with the number of antennas $N$, a fact leveraged by previous works~\cite{mDTrack-MobiCom19, karanam2019tracking, song2024siwis} to facilitate multi-person sensing.

As forward-looking prototypes, these methods
require modifications to COTS devices; hence, current Wi-Fi-based HAR research~\cite{yousefi2017survey, ma2018signfi,guo2019wiar,brinke2019dataset,zhang2021widar3, baha2020dataset,hu2021wigr,bocus2022operanet, yang2022efficientfi, yang2023mm, zhao2024finding, wang2024xrf55,lan2025xrf} has primarily concentrated on single-person scenarios and corresponding dataset development.
Even though multi-person HAR approaches, such as FallDeFi~\cite{palipana2018falldefi}, which can detect the fall of one subject in a two-person environment by applying time-frequency analysis to CSIs, their scalability to more complex multi-person scenarios remains unvalidated.
Besides, WiMANS~\cite{huang2024wimans} claims to support HAR for up to five subjects using CSI from a system with 20~\!MHz bandwidth and three antennas;
however, its performance is heavily dependent on the neural network’s fitting capacity due to the lack of additional physical-layer information to compensate for limited frequency diversity,
thereby undermining generalization.
Therefore, developing a practical Wi-Fi multi-person HAR system using COTS devices is critical for advancing its deployment in real-world scenarios.

\vspace{-.3em}
\subsection{Feasibility of Near-Field Sensing} \label{ssec:near_field}

Fortunately, the widespread availability of personal smart devices facilitates the construction of multi-link systems for multi-person HAR.
In contrast to systems composed of multiple fixed devices that function as a distributed multi-antenna array, our approach establishes a dedicated link for each subject, thereby enabling scalable multi-person sensing based on the near-field domination effect.
To demonstrate the feasibility of near-field sensing,
we consider a scenario in which each subject is equipped with a UE connected to an AP.
The multipath signal of a given link can then be decomposed into four components:
target reflections $h_i(t)$ for the $i$-th ($i \in [1, \mathcal{Q}]$) subject, non-target reflections ${\sum_{j \neq q}^Q} h_{j}(t)$ from other subjects, static components $h^{\mathrm{S}}(t)$ due to the environment and the line-of-sight (LoS) path, and dynamic components $h^{\mathrm{D}}(t)$ resulting from surrounding movements and hardware fluctuations.
Thus, Eqn.~\eqref{eqn:CSI_simple} can be reformulated as:
\begin{equation}
\vspace{-0.2em}
    h(t) = h_i(t) + {\sum_{j \neq i}^Q} h_{j}(t) + h^{\mathrm{S}}(t) + h^{\mathrm{D}}(t),
    \label{eqn:CSI_comp}
\end{equation}
where indices $n$ and $m$ are omitted for brevity.
Considering that both the channel gain $\alpha$ and phase depend on the propagation distance, we denote the distances from the subject to the UE and AP as $L^{\mathcal{U},\mathcal{S}_i}$ and $L^{\mathcal{S}_i,\mathcal{A}}$, respectively.
The component $h_i(t)$ is modeled as:
\begin{equation}
    h_i(t) = \frac{\lambda^2 \sqrt{G_i} \mathrm{exp}\left(-\mathi 2 \pi (L^{\mathcal{U},\mathcal{S}_i}(t) + L^{\mathcal{S}_i,\mathcal{A}}(t))/\lambda \right)}{(4\pi)^2 \left(L^{\mathcal{U},\mathcal{S}_i}(t) L^{\mathcal{S}_i,\mathcal{A}}(t)\right)^{\sigma/2}},
    \label{eqn:subject}
\end{equation}
where wavelength $\lambda = \lightc/f_{\mathrm{c}}$, $G$ denotes a coefficient determined by the antenna gain and the subject’s reflection properties, and $\sigma \approx 4$ according to~\cite{rappaport2024wireless}.
For the $i$-th subject located near or within the near-field region of its associated UE (approximately 0.2~\!m~\cite{hu2023musefi}), the variation in $h(t)$ is primarily determined by $h_i(t)$.
This phenomenon, termed the \textit{near-field domination effect}, facilitates practical multi-person HAR.

We begin by providing a theoretical demonstration to support the feasibility of sensing based on the near-field domination effect, i.e., near-field sensing.
The variation in $h_i(t)$ is quantified using the power of channel variation $\mathcal{P}_i$, defined as the squared magnitude of its partial derivative w.r.t. time $t$:
\begin{equation}
\begin{aligned}
    &\mathcal{P}_i = |\frac{\partial h_i(t)}{\partial t}|^2 \\
    &\approx \frac{G_i \lambda^4 v_i^2}{(4\pi)^4(L^{\mathcal{U},\mathcal{S}_i} L^{\mathcal{S}_i,\mathcal{A}})^{\sigma}}
    \left[\frac{\sigma^2}{4}\left(\frac{L^{\mathcal{U},\mathcal{S}_i} + L^{\mathcal{S}_i,\mathcal{A}}}{L^{\mathcal{U},\mathcal{S}_i} L^{\mathcal{S}_i,\mathcal{A}}}\right)^2 + \frac{16\pi^2}{\lambda^2} \right],
    \label{eqn:power_var}
\end{aligned}
\end{equation}
where $t$ is omitted for brevity, $v_i$, representing the velocity of the $i$-th subject's motion, is simplified as $v_i = \partial L^{\mathcal{U},\mathcal{S}_i} / \partial t \approx \partial L^{\mathcal{S}_i,\mathcal{A}} / \partial t$.
The first and second terms in the bracket correspond to amplitude and phase variations, respectively.
In typical 5~\!GHz Wi-Fi near-field sensing systems, phase variations induced by the subject dominate, rendering the amplitude-related term negligible.
As an illustrative example, consider $ L^{\mathcal{U},\mathcal{S}_i} = 0.2$~\!m, $L^{\mathcal{S}_i,\mathcal{A}}= 5$~\!m, and $\lambda = 0.06$~\!m;
in this case, the second term is over 400 times larger than the first, further justifying its omission.
Thus, Eqn.~\eqref{eqn:power_var} can be simplified as:
\begin{equation}
    \mathcal{P}_i \approx \frac{G_i \lambda^4 v_i^2}{(4\pi)^4(L^{\mathcal{U},\mathcal{S}_i} L^{\mathcal{S}_i,\mathcal{A}})^{\sigma}}
    \frac{16\pi^2}{\lambda^2} 
    = \tilde{G}_i v_i^2  (L^{\mathcal{U},\mathcal{S}_i} L^{\mathcal{S}_i,\mathcal{A}})^{-\sigma},
    \label{eqn:power_final}
\end{equation}
where $\tilde{G}_i = G_i (\lambda/4\pi)^2$ is considered a constant.
Similarly, the power of channel variation for the $j$-th subject can be in the same form as $\mathcal{P}_j = \tilde{G}_j v_j^2  (L^{\mathcal{U},\mathcal{S}_j} L^{\mathcal{S}_{j},\mathcal{A}})^{-\sigma}$.
Since all subjects are generally far from the AP and move at similar speeds (i.e., $L^{\mathcal{S}_i,\mathcal{A}} \approx L^{\mathcal{S}_{j},\mathcal{A}}$,  $ v_i \approx v_{j}$), and the $i$-th subject is in the near-field of its own UE (i.e., $L^{\mathcal{U},\mathcal{S}_i} < L^{\mathcal{U},\mathcal{S}_{j}}$), the near-field domination effect ($\propto (L^{\mathcal{U},\mathcal{S}_i})^{-\sigma}$) leads to $\mathcal{P}_i \gg \mathcal{P}_j$.
\rev{This indicates that the UE–AP link is dominated by the motion of the nearby $i$-th subject when $L^{\mathcal{U},\mathcal{S}_i}$ is small.}
Thus, by sniffing CSI from different links and associating each link with a subject via its MAC address, we can effectively distinguish multiple subjects for HAR, with the further advantage of mitigating the impact of environment factors.

\begin{figure}[t]
	\centering 
	\setlength{\abovecaptionskip}{2pt}  
	\subfigure[Experiment setting.] 
		{
		 \centering
		 \includegraphics[height=0.57\linewidth]{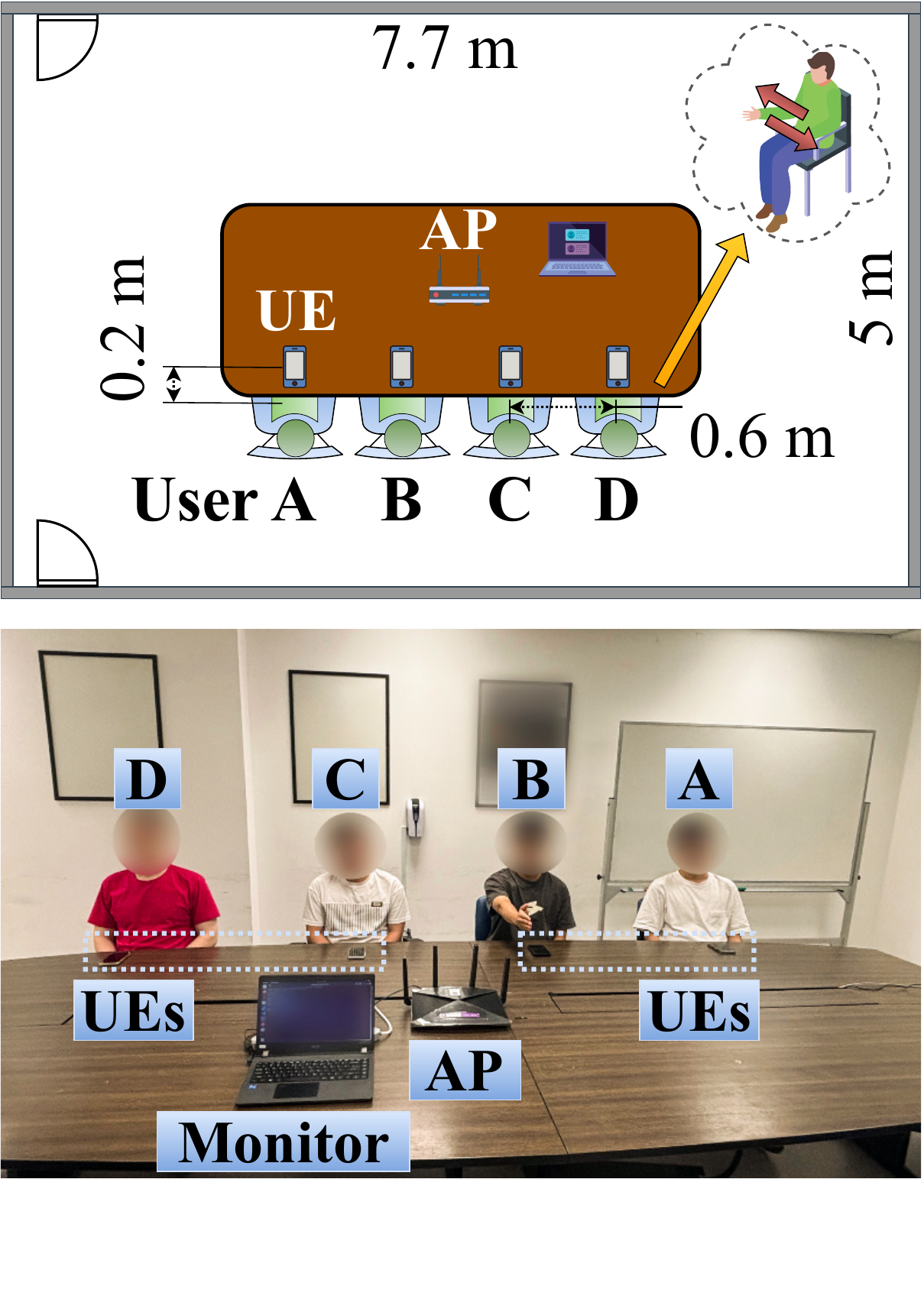}
		 \label{sfig:pre_layout}    
		}
        \hfill
	\subfigure[CSI phase variations.]
		{
			\centering         
			\includegraphics[height=0.57\linewidth]{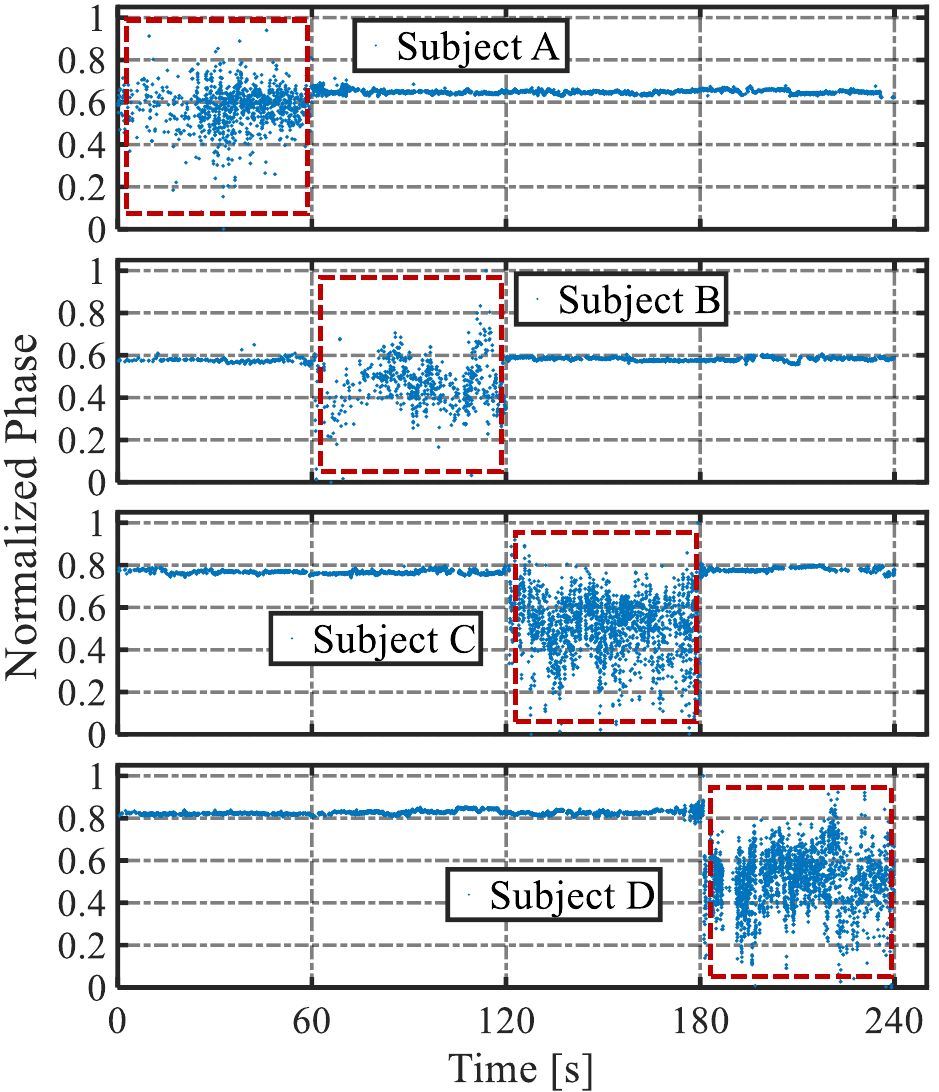}
			\label{sfig:pre_var}  
		}
	\caption{Experiments on near-field sensing. The results indicate that the subject in the near-field of its corresponding UE has a dominant influence on the CSI.}
     \label{fig:pre_exp_near}
 \vspace{-1em}
\end{figure}

\begin{figure}[b]
\vspace{-1.em}
	\centering 
	\setlength{\abovecaptionskip}{2pt}  
		 \includegraphics[width=0.92\linewidth]{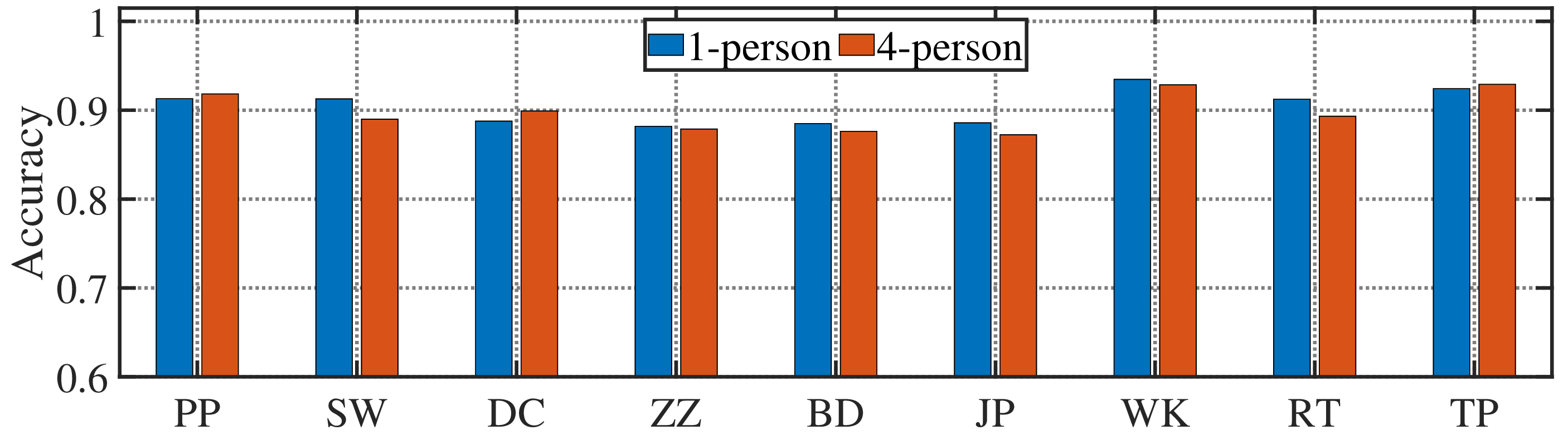}
	\caption{\rev{Comparison of 1-person and 4-person scenarios.}}
     \label{fig:recog_1vs4}
\end{figure}

To provide an intuitive insight, we conduct an experiment to validate the near-field domination effect.
As shown in Fig.~\ref{sfig:pre_layout}, four subjects are seated in a meeting room, each with a UE placed 20~\!cm in front of them; with a 60~\!cm spacing between body centers, the subjects are in close proximity, corresponding to typical adult body sizes.
Each subject performs a sweeping motion in turn, while their respective UEs maintain communication with the AP by streaming video.
We sniff CSIs from all four links and show their phase variations in Fig.~\ref{sfig:pre_var}.
The results illustrate that only the link corresponding to the active subject exhibits significant phase fluctuations, with minimal interference observed on the other links,
demonstrating the practicality of near-field sensing for enabling multi-person HAR, as it effectively mitigates interference from other subjects and the environment.
\rev{Meanwhile, non-uniformity of samples across links is also observed.}
\rev{We further compare the HAR accuracy between single-person and four-person scenarios. A simple GRU model (see Section~\ref{ssec:sys_pt}) is used to recognize nine activities that can be performed by a single person (see Section~\ref{ssec:exp_set}, with handshaking removed).
As shown in Fig.~\ref{fig:recog_1vs4}, no significant difference in accuracy exists, suggesting that multi-person sensing achieves performance comparable to the single-person case within our near-field sensing system.
Therefore, we focus on multi-person scenarios in the remainder of this work.}

\vspace{-0.7em}
\subsection{Fine-Tuning for Cross-Domain Adaptation} \label{ssec:fine_tuning}

\subsubsection{Multi-person HAR via Near-field Sensing} \label{sssec:pre_train}

Due to its twofold nature, the near-field domination effect enhances CSI responsiveness to the intended subject and simplifies surrounding interference into a single-subject abstraction, while also increasing subject (domain) specificity in HAR.
Conventional CSI-based HAR approaches~\cite{EI-MobiCom18, zhang2021widar3, gao2022towards, niu2021understanding, liu2023wisr, wang2022airfi},
which are widely adopted, are initially employed in attempts to address cross-domain adaptation.
Among these approaches, the first strategy~\cite{zhang2021widar3, gao2022towards, niu2021understanding} focuses on applying time-frequency transformations to CSIs in order to extract subject motion features such as speed and direction, which remain invariant across domains;
for example, Widar3.0~\cite{zhang2021widar3} extracts a body-coordinate velocity profile (BVP) to serve this purpose.
The second strategy~\cite{EI-MobiCom18, liu2023wisr, wang2022airfi} adopts adversarial learning to extract domain-invariant representations, as exemplified by the EI framework proposed by~\cite{EI-MobiCom18}.
\rev{In addition, sophisticated neural network architectures~\cite{lu2026spiking, zhang2026wi} have also been studied in recent years, such as the latest Wi-CBR~\cite{zhang2026wi}.}
To evaluate their cross-domain adaptation in the context of the task considered in this work, we conduct further analyses.

\begin{figure}[t]
	\centering 
	\setlength{\abovecaptionskip}{2pt}  
	\subfigure[\rev{Accuracy on the source domain across different leave-one-out users.}] 
		{
		 \centering
		 \includegraphics[width=0.91\linewidth]{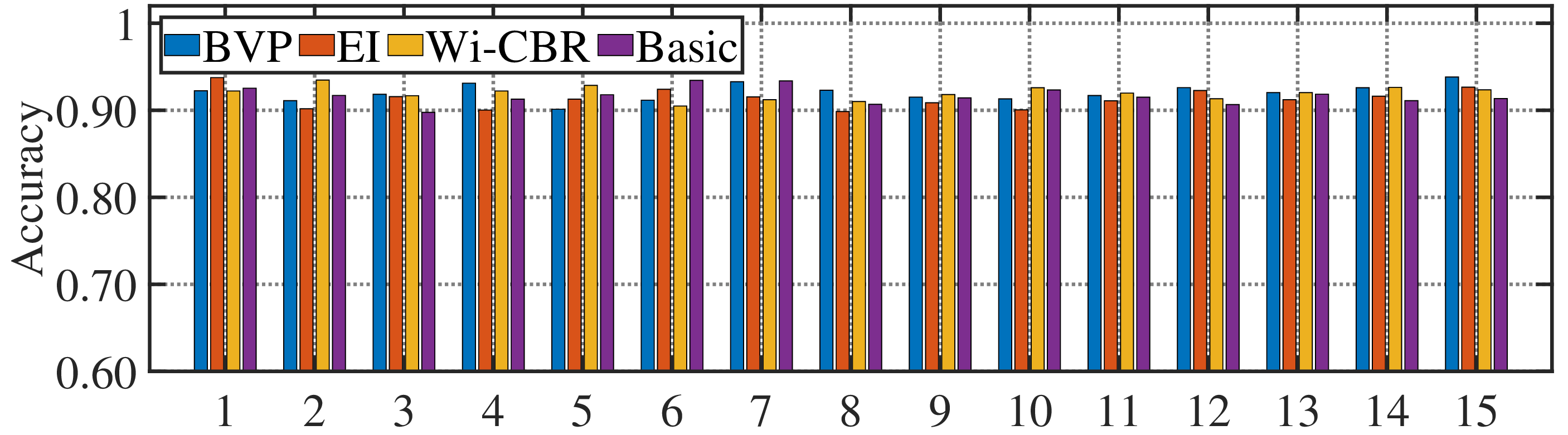}
		 \label{sfig:pre_source}    
		}
        \\
	\subfigure[\rev{Accuracy on the target domain across different leave-one-out users.}]
		{
			\centering         
			\includegraphics[width=0.91\linewidth]{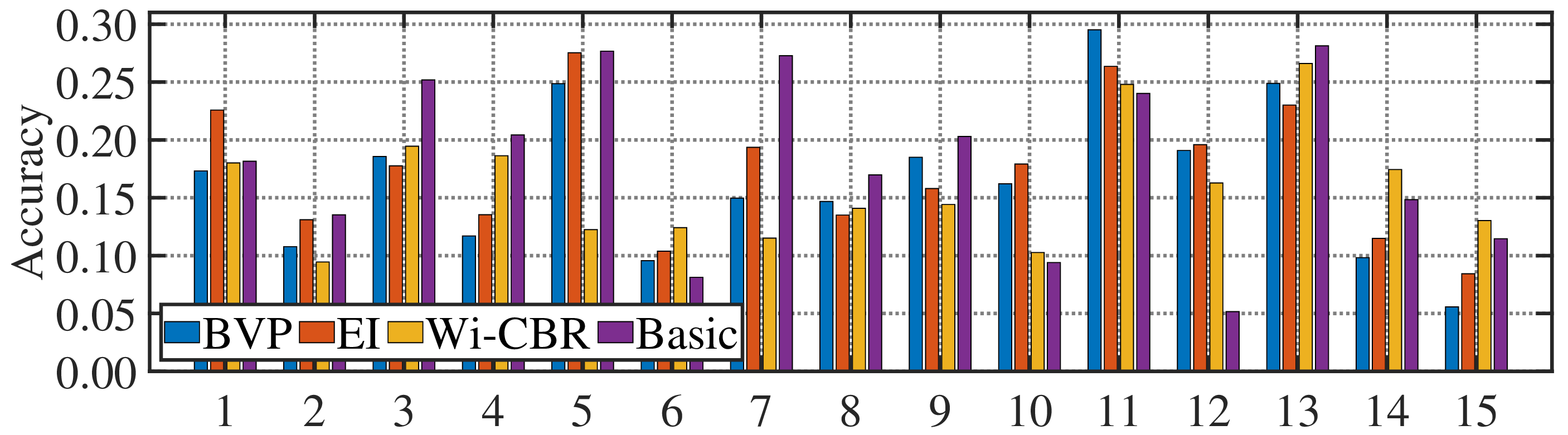}
			\label{sfig:pre_target}  
		}
	\caption{Source domain accuracy vs. target domain accuracy.}
     \label{fig:pre_training}
 \vspace{-1em}
\end{figure}

For preliminary analysis, we extract data involving 2–4 concurrently active users from 15 subjects performing 10 types of activities, including gestures and body movements (see Section~\ref{ssec:exp_set} for details).
We evaluate the models’ cross-domain performance using the leave-one-out method~\cite{wong2015performance}: data from one subject is used as the test set (target domain), while data from the remaining 14 subjects (source domain) is split into training and validation sets at a $9\!:\!1$ ratio.
In addition to the BVP, EI, \rev{and Wi-CBR} approaches, we also analyze the CSI using a simple GRU model (see Section~\ref{ssec:sys_pt} for details) as the basic approach.
For all methods, the irregular CSI sequences are interpolated to obtain uniformly structured data for processing.
Fig.~\ref{sfig:pre_source} illustrates the accuracy achieved in the source domain, showing that all \rev{four} methods reliably exceed 90\% recognition across the leave-one-user-out scenarios.
However, as shown in Fig.~\ref{sfig:pre_target}, the accuracy in target domain drops sharply to below 30\%, indicating that these approaches do not generalize well to near-field channel samples.
Further analysis reveals that two factors contribute to the degradation:
first, the near-field domination effect imparts subject-specific characteristics to the CSIs, causing signals from different domains to exhibit substantial physical variability; second, CSIs driven by native traffic are highly irregular and deviate significantly from the uniform traffic assumed in prior studies.

\subsubsection{Fine-Tuning with Categories Absence} \label{sssec:fine_tuning}

\begin{figure}[b]
\vspace{-1.em}
	\centering 
	\setlength{\abovecaptionskip}{2pt}  
	\subfigure[\rev{Impact of sample quantity.}] 
		{
		 \centering
		 \includegraphics[width=0.365\linewidth]{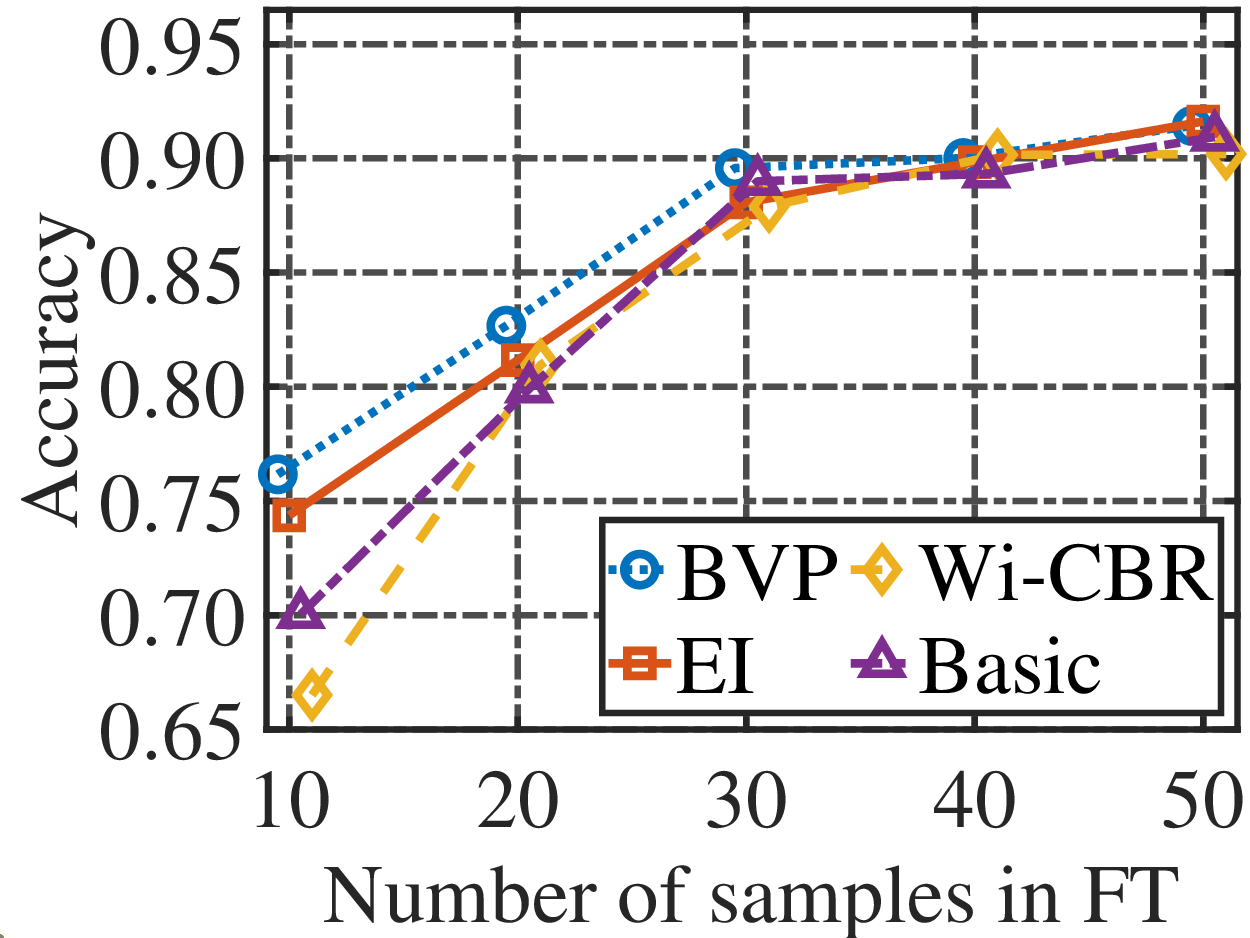}
		 \label{sfig:FT_size}    
		}
        \hfill
	\subfigure[\rev{FT performance with categories absence.}]
		{
			\centering         
			\includegraphics[width=0.557\linewidth]{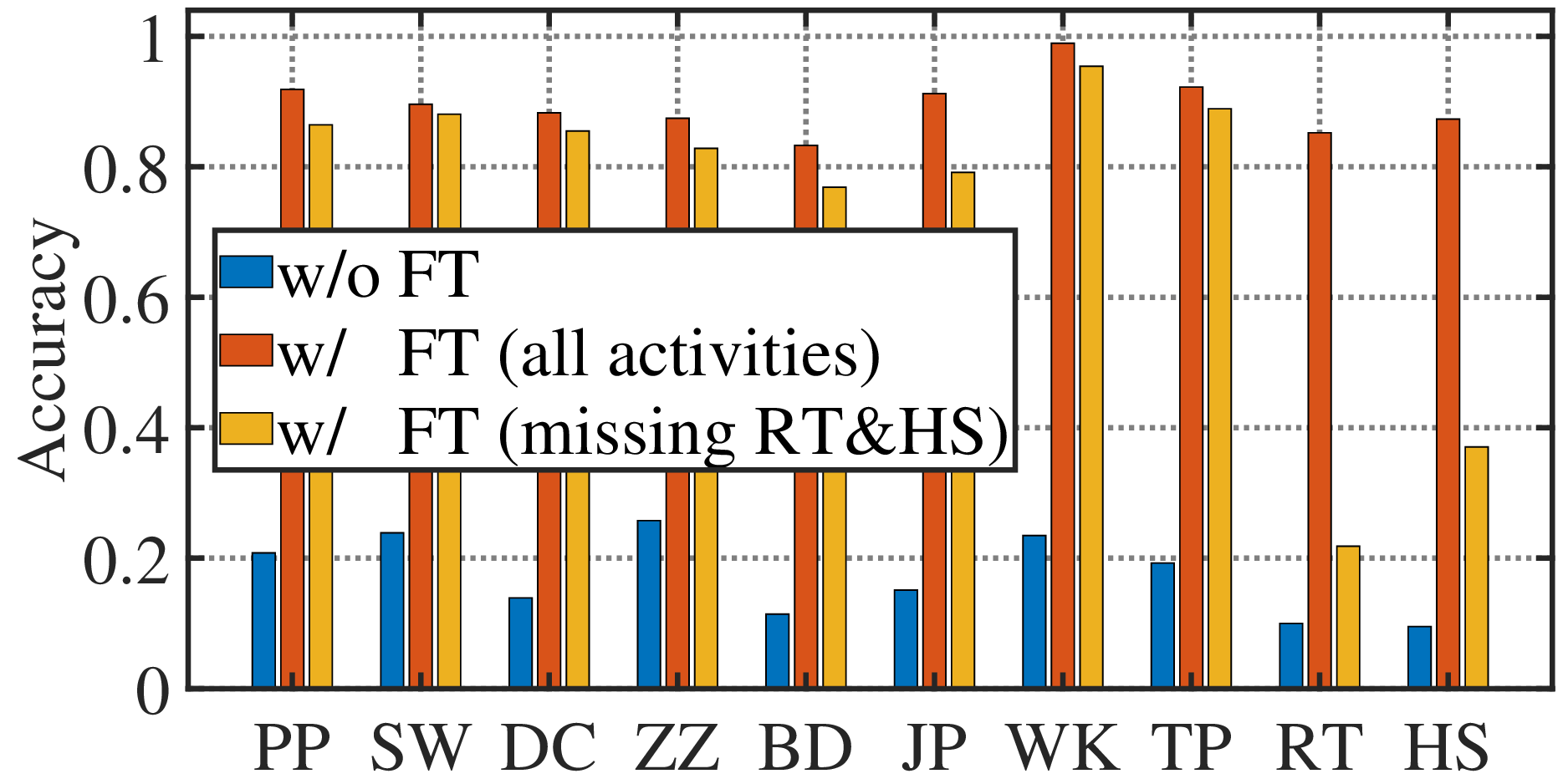}
			\label{sfig:FT_miss}  
		}
	\caption{FT for domain adaptation. The (a) limited number and (b) absence of samples from specific categories significantly degrade accuracy.}
     \label{fig:pre_FT}
\end{figure}

Since CSIs from various domains exhibit significant differences, fine-tuning a pre-trained model on a small subset of target domain data is an effective approach for cross-domain adaptation.
As shown in Fig.~\ref{sfig:FT_size}, the recognition accuracy in the target domain improves steadily as the number of FT samples increases, and it saturates at around 30 samples per category, indicating that the model has acquired sufficient information.
However, in practical scenarios, it is often unrealistic to assume the availability of data from every category, as the user experience burden of repeatedly performing a number of activities and other constraints may render the collection of FT data difficult or even infeasible.
For example, the handshaking (HS) gesture is difficult to perform with only one person present, and the rotating (RT) action, which is often used to detect hazardous events for elderly people such as falls or medical emergencies, is not feasible or safe to collect for FT.
To investigate the impact of missing category-specific data, we remove HS and RT samples from the FT process of the Basic model, while maintaining 30 samples per category for all other activities.
As illustrated in Fig.~\ref{sfig:FT_miss},
although HS and RT achieved an average recognition accuracy of 29.5\%, reflecting a slight improvement over the results without FT, the performance remains substantially lower than when data from all categories are available.
Nevertheless, these absent categories often correspond to activities that a HAR model must reliably recognize.
\rev{Notably, in the absence of FT samples, HS achieves higher accuracy than RT, which may be attributed to the fact that HS, as an interactive activity, exhibits more distinctive features.}

\begin{figure}[t]
	\centering 
	\setlength{\abovecaptionskip}{2pt}  
	\subfigure[Target domain (all).]  
		{
		 \centering
		 \includegraphics[width=0.29\linewidth]{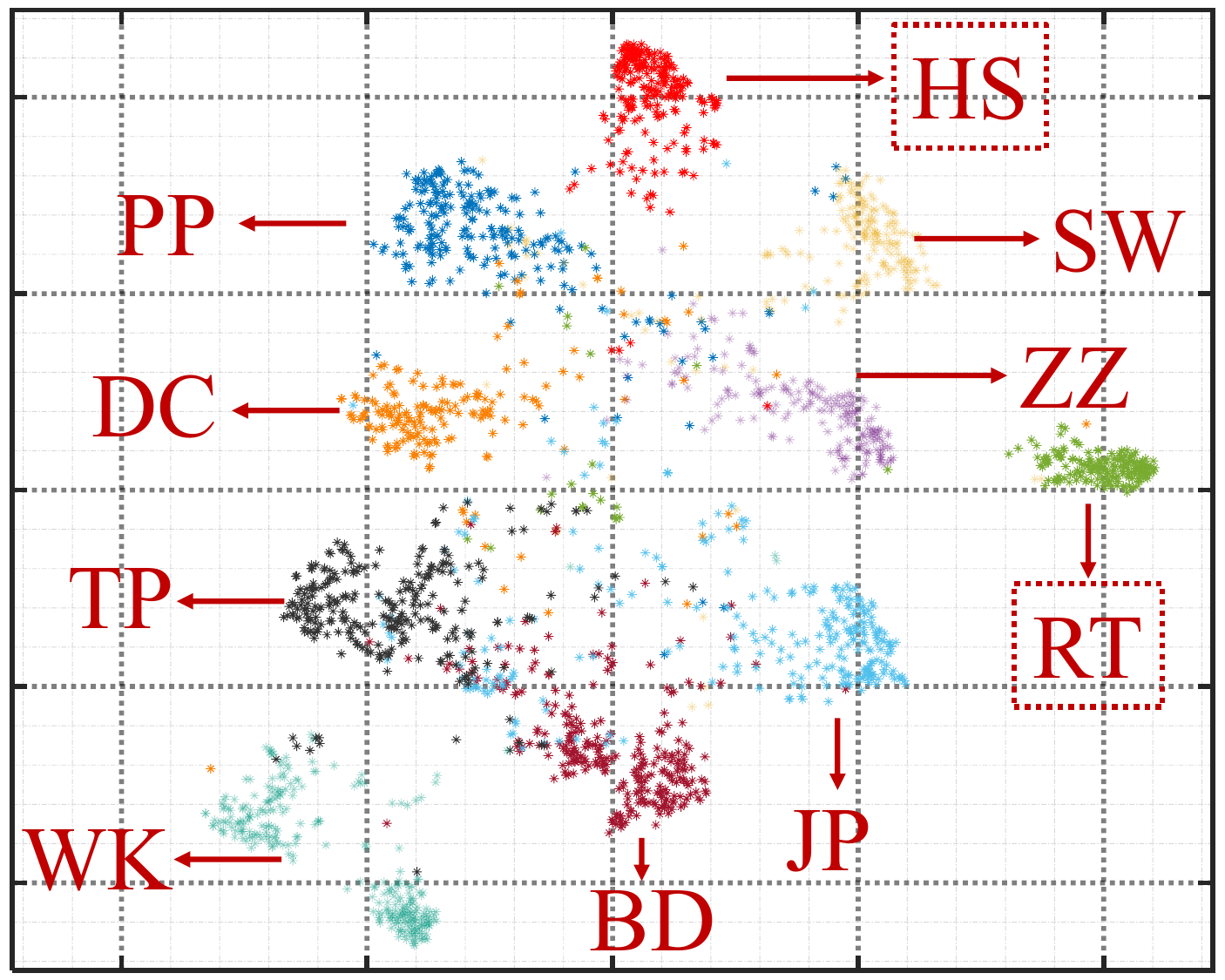}
		 \label{sfig:tsne_target_all}    
		}
        \hfill
	\subfigure[Target domain.]
		{
			\centering         
			\includegraphics[width=0.29\linewidth]{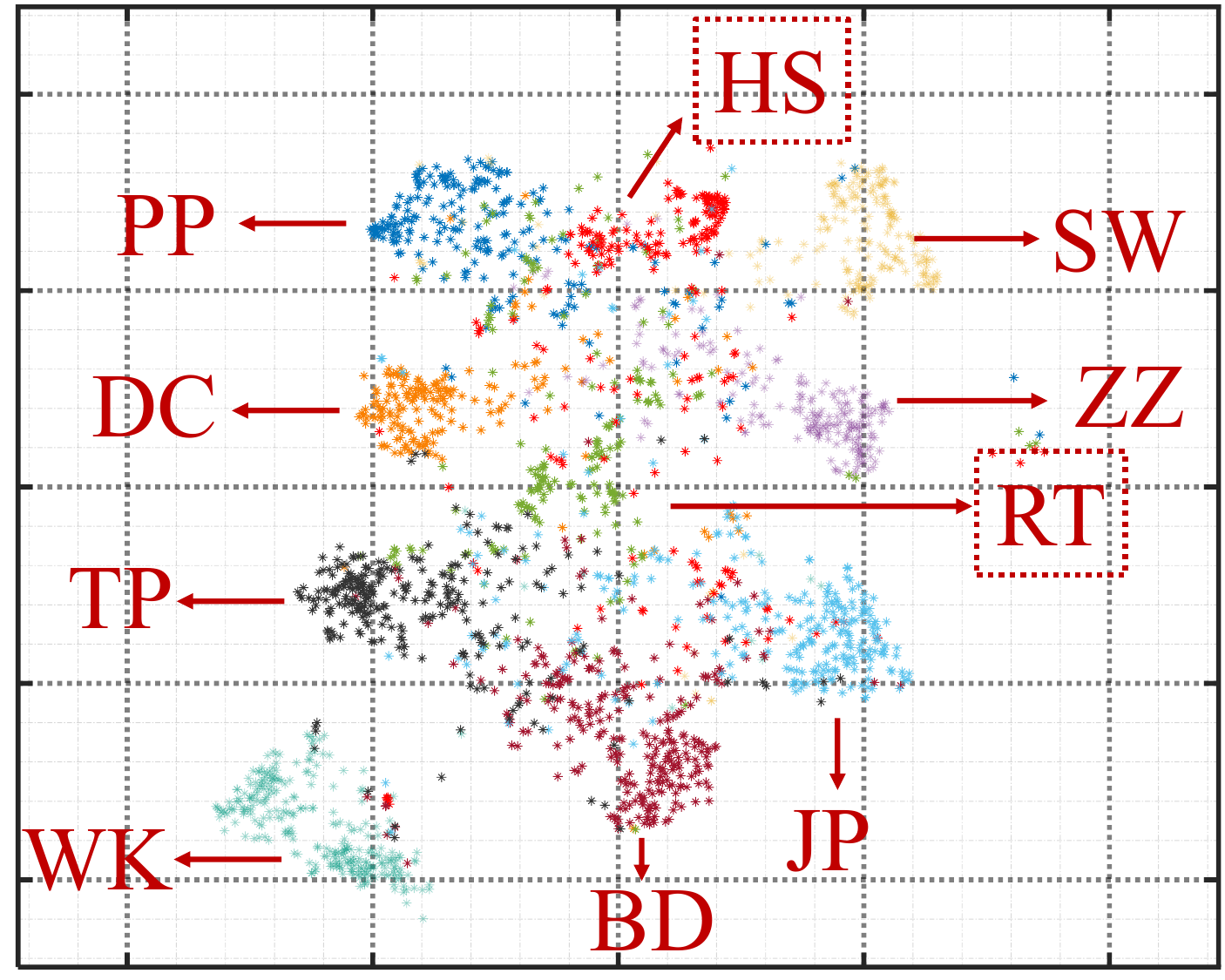}
			\label{sfig:tsne_target}  
		}
         \hfill
	\subfigure[Source domain.]
		{
			\centering         
			\includegraphics[width=0.29\linewidth]{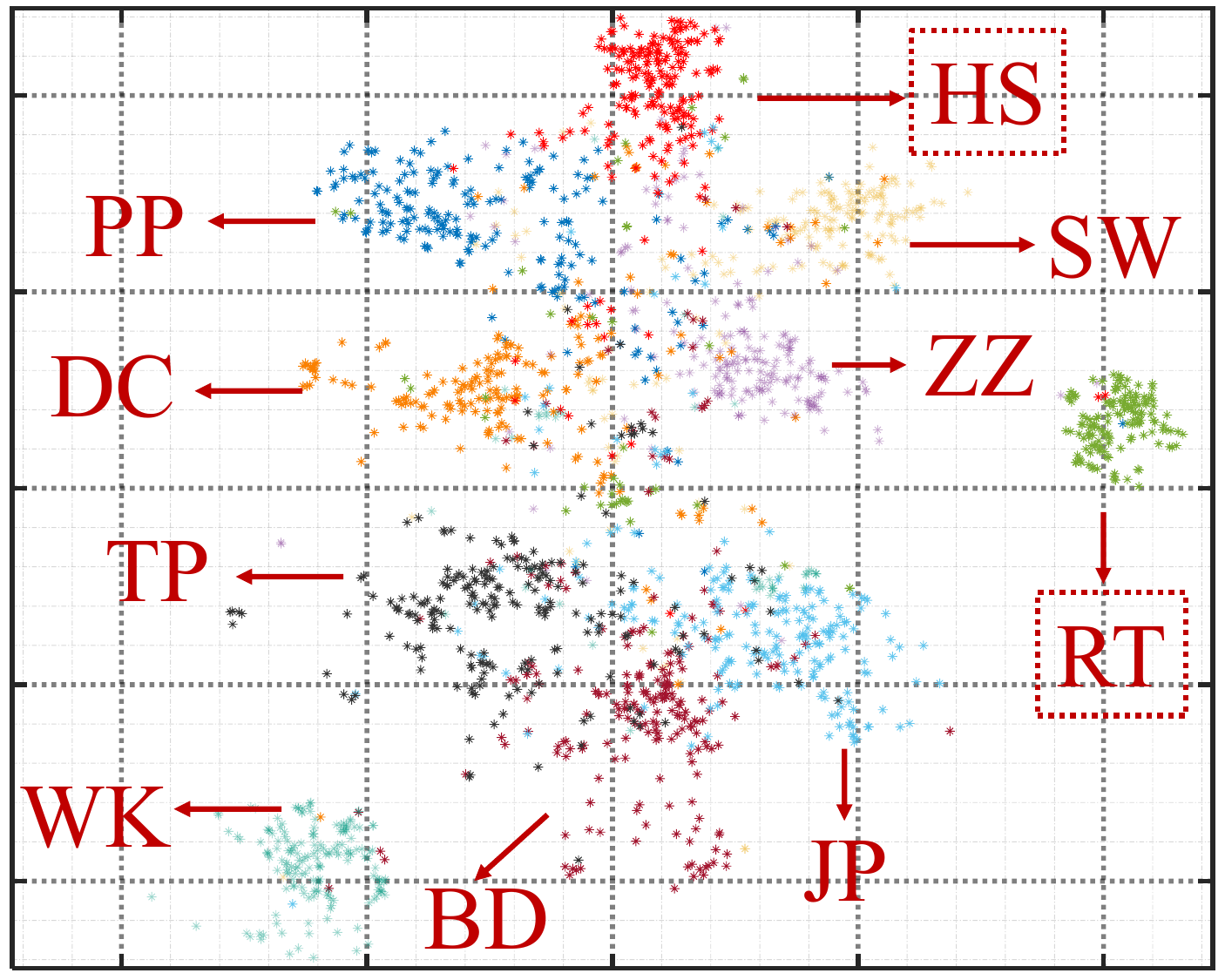}
			\label{sfig:tsne_source}  
		}
	\caption{Visualization with t-SNE. The absence of category-specific data negatively affects feature extraction for all categories.}
     \label{fig:vis_tsne}
 \vspace{-1.em}
\end{figure}
\begin{figure*}[b]
\vspace{-.5em}
	\centering 
	\setlength{\abovecaptionskip}{6pt}
		\includegraphics[width=0.95\linewidth]{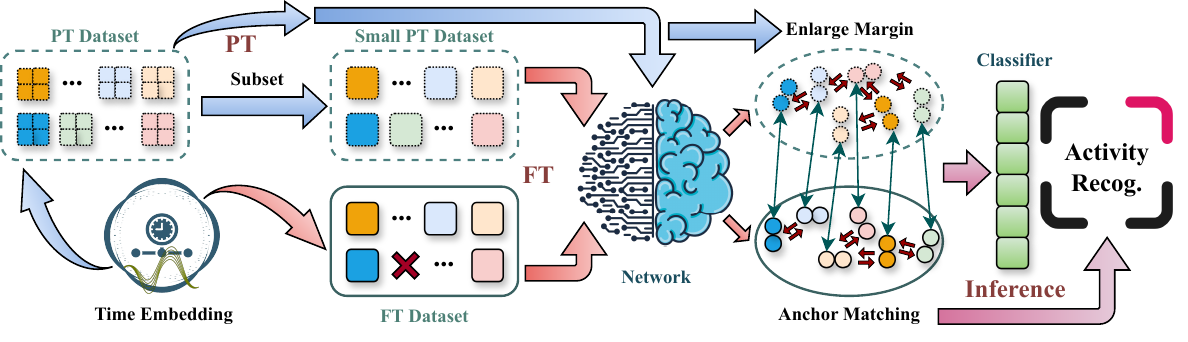} 
	\caption{\name framework overview.}
	\label{fig:compressive sensing spectrum}
\end{figure*}

To gain deeper insights, we visualize the features using t-distributed Stochastic Neighbor Embedding (t-SNE).
As shown in Fig.~\ref{sfig:tsne_target_all}, the target domain features from the model fine-tuned with complete category data form well-defined clusters with distinct decision boundaries.
In contrast, when RT and HS samples are excluded from FT, as shown in Fig.~\ref{sfig:tsne_target}, these two categories become less distinguishable, and the inter-class separation of the remaining categories also diminishes, aligning with the slight drop in recognition accuracy observed in Fig.~\ref{sfig:FT_miss}.
A further examination of the source domain features under the same FT setting, as shown in Fig.~\ref{sfig:tsne_source}, reveals that while these categories are identifiable, they remain densely packed (even compared to Fig.~\ref{sfig:tsne_target_all}), indicating that FT with incomplete categories reduces the inter-class margins.
In addition, the feature distributions in Fig.~\ref{sfig:tsne_target} and Fig.~\ref{sfig:tsne_source} are not entirely consistent, suggesting that the target domain features may not be accurately extracted.
Accordingly, two principal strategies for efficient FT with incomplete categories can be identified: enlarging the inter-class margins of features and shifting toward filtering subject-specific interference, rather than merely extracting incomplete features.
In the following sections, we will design a training framework based on these two guiding principles.

\section{Methodology \rev{for Cross-Domain Adaptation}} \label{sec:method}

\revhu{
In this section, we define the \name framework for cross-domain adaptation in Wi-Fi-based multi-person HAR under missing FT samples for certain categories.
As illustrated in Fig.~\ref{fig:compressive sensing spectrum}, the framework consists of a time embedding algorithm to capture the irregular patterns of near-field CSIs and a three-stage pipeline comprising PT, FT, and inference:
}

\begin{itemize}
    \item In the PT stage, a strategy is proposed to reward the extraction of features with large inter-class margins, thereby enhancing category separability.
    \item In the FT stage, a small subset of source domain samples is used as anchors, and the target domain data are guided to learn subject-specific denoising characteristics driven by matched filtering.
    \item In the inference \revhu{stage}, a composite strategy combining the model logits with the similarity to anchors is introduced to further improve recognition accuracy.
\end{itemize}

\revhu{
In Section~\ref{ssec:sys_time}, we introduce the time embedding design for modeling the irregular patterns of near-field CSIs.
In Section~\ref{ssec:sys_pt}, we present the PT-stage design for enhancing category separability in the source domain.
In Section~\ref{ssec:sys_ft}, we describe the FT-stage design that uses source-domain anchors to guide subject-specific denoising adaptation.
In Section~\ref{ssec:sys_descion}, we detail the composite decision strategy used during inference.
}

\vspace{-.5em}
\subsection{Time Information Embedding} \label{ssec:sys_time}

To address the temporal irregularity of near-field CSIs collected under native traffic, we first propose a time embedding module.
\revhu{Its role in \name is to construct a temporally informative representation of CSI input sequences, establishing a solid foundation for accurate HAR.}

Specifically, to effectively capture the temporal irregular patterns of sequences, the embedding process \revhu{consists of} two components: time vector embedding and CSI data preprocessing.
We design an adaptive embedding scheme based on time differences, following this insight: in sparse regions of the sequence, long-term trends should be emphasized, while in dense regions, short-term fluctuations should be captured, rather than uniformly encoding all temporal information~\cite{Vaswani2017attention, gehring2017convolutional}.
Assuming $[\Delta t_1,\cdots,\Delta t_i,\cdots]$ is the time-difference vector obtained by differentiating the raw time vector of the received packets,
the time embedding $\bm{e}^{\mathrm{t}}$ is then defined \rev{to capture its multi-scale patterns} as: 
\begin{equation}
    \left\{
    \begin{aligned}
    \bm{e}^{\mathrm{t}}(i,2j-1) &=  \sin \left(\frac{\Delta t_i}{\mathcal{T}^{2j/D} \Delta t^{\mathrm{Ref}}}\right) \\
    \bm{e}^{\mathrm{t}}(i,2j) &=  \cos \left(\frac{\Delta t_i}{\mathcal{T}^{2j/D} \Delta t^{\mathrm{Ref}}}\right),
    \end{aligned}
    \right.
    \label{eqn:time_encode}
\end{equation}
where $\Delta t^{\mathrm{Ref}}$ denotes the reference interval, which can be obtained through statistical analysis of the data, $\mathcal{T}$ represents the duration of the activity, and $D$ is the embedding dimension.
\rev{In contrast to existing methods that map data onto a regular temporal grid~\cite{yang2025wirelessgpt, zhao2025csi}, our design explicitly transforms temporal irregularity into informative features, thereby enabling the model to better interpret irregular packet arrivals.}

Given the analysis in Section~\ref{ssec:near_field} indicating that subject motions in the near field primarily induce variations in CSI phase, \revhu{the CSI preprocessing design focuses on enhancing phase-related cues while preserving complementary signal information.}
\revhu{To suppress shared interference and emphasize motion-sensitive phase variation,} the signal received by the first antenna is \revhu{selected} as a reference for conjugate multiplication, \revhu{yielding} $\hat{h}=\bar{h}_{1,m,t} h_{n\neq 1,m,t}$.
\revhu{To preserve} phase \revhu{continuity despite the rapid value hopping between 0 and $2\pi$, the resulting phase is represented on the continuous unit circle as} $\varphi = [\sin(\angle \hat{h}),\cos(\angle \hat{h})]$.
Finally, \revhu{the} processed phase $\varphi$, the normalized CSI amplitude $\mathsf{norm}(|\hat{h}|)$, and the normalized Received Signal Strength Indicator (RSSI) $\mathsf{norm}(\mathrm{RSSI})$ are \revhu{integrated into} a unified input representation for the neural network model:
\begin{equation}
    z = (\bm{e}^{\mathrm{t}},\mathsf{norm}(\mathrm{RSSI}),\mathsf{norm}(|\hat{h}|),\varphi).
    \label{eqn:dataset}
\end{equation}
%


\revhu{After padding, a sample can be represented as $x\in \mathbb R^{T\times S}$, where $T$ is the maximum number of packets, and $S$ indicates the dimension of $z$. 
Then, a CSI dataset is defined as $\mathcal{D}=(\mathcal{X},\mathcal{Y})$, where $\mathcal{X}=\{x_i\}_i$ and $\mathcal{Y}=\{y_i\}_i$ are the sets of sample representations and ground-truth labels, respectively.}

\vspace{-0.5em}
\subsection{Pre-training \revhu{Stage}} 
\label{ssec:sys_pt}

\revhu{To obtain discriminative activity representations that remain robust after subsequent adaptation, the PT stage aims to learn a compact feature space in which different activity categories are more distinguishable from one another.
Our main idea is to train a lightweight feature extractor with an inter-class margin enlargement objective, enabling it to capture transferable activity cues.
Within \name, this stage provides the source-domain feature extractor and activity classifier, establishing the foundation for the anchor-guided FT.}

\begin{figure}[t]
	\centering 
	\setlength{\abovecaptionskip}{6pt}
		\includegraphics[width=0.98\linewidth]{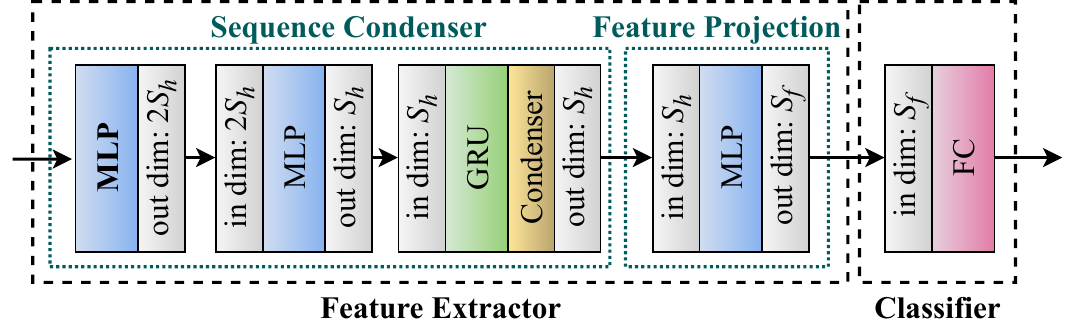}  
	\caption{Basic neural network architecture.}
	\label{fig:network}
\vspace{-1.em}
\end{figure}

We begin with a basic network model composed of three simple components:
sequence condenser, feature projection, and a classifier\footnote{This basic model is empirically designed to extract a compact feature representation, and Section~\ref{sssec:factor_arc} further demonstrates that our training framework remains effective across diverse network architectures.}, as shown in Fig.~\ref{fig:network}.
Specifically, the sequence condenser consists of two Multi-Layer Perceptron (MLP) modules, a Gated Recurrent Unit (GRU) module, and a condenser operator.
The two MLPs form a lightweight encoder-decoder (ED) structure that reconstructs input information and adjusts the feature dimension, denoted as $x \rightarrow x^{\mathrm{MLP}}$.
\revhu{The GRU is introduced to capture} the contextual features of CSIs across continuous motions, yielding $x^{\mathrm{MLP}} \rightarrow x^{\mathrm{GRU}}$.
\revhu{To handle the irregular lengths of CSI data under native traffic, the condenser retains the final non-padded element from each GRU output, i.e., $x^{\mathrm{SC}} = x^{\mathrm{GRU}}[\check{t},:]$, enabling temporal aggregation with reduced redundancy to mitigate overfitting.}
The feature projection module, \revhu{implemented with} a simple MLP, further compresses $x^{\mathrm{SC}}$ into a low-dimensional space.
These modules jointly form the feature extractor, effectively defining the mapping $\aleph = \phi^{\mathrm{FP}}(\phi^{\mathrm{SC}}(x))$ \revhu{for compact representation learning.}
Finally, a fully connected (FC) layer serves as \revhu{the} classifier, producing the output $y = \phi^{\mathrm{CLS}}(\aleph)$.

\revhu{During the PT stage, we enlarge inter-class margins to ensure that the learned features remain distinguishable, even if these margins contract during the subsequent FT stage.}
Cross-entropy (CE) loss between the one-hot encoded HAR prediction $\hat{y}$ and ground-truth label $y$ \revhu{provides the basic supervision, i.e.,}
%
\begin{equation}
    \mathcal{L}^{\mathrm{CE}} = -\sum^{C}_{i=1} p_{i}\log(\hat{p}_i),
    \label{eqn:CE_loss}
\end{equation}
where $p_i=\frac{\exp(y_i)}{\sum_{j=1}^{C}\exp(y_j)}$ is the softmax output representing the predicted probability distribution and $C$ denotes the number of activity categories.

\revhu{To enlarge the inter-class margins, we project activity features into a space with stronger class separability, while ensuring the features remain informative and stable.
We penalize features associated with misclassification to avoid overconfidence in narrow cues, thereby achieving the objective:}
$\sum_{i=1}^{C}(p_{i}-\hat{p}_{i})^2 \cdot \|\aleph\|^2_2$.
\revhu{Meanwhile, we encourage cluster centers of features $\aleph_{i}^{\mathsf{C}}$ for different classes to stay apart by maximizing their average Euclidean distance, given by}
$\frac{1}{S_h}\cdot \frac{1}{C(C-1)}\sum_{i \neq j}\|\aleph_{i}^{\mathsf{C}}-\aleph_{j}^{\mathsf{C}}\|_2$,
where $S_h$ is the feature dimension.
Accordingly, the loss $\mathcal{L}^{\mathrm{FE}}$ aimed at enlarging the inter-class margins is formulated as:
\begin{equation}
    \mathcal{L}^{\mathrm{FE}}
    = 
    \lambda_{11} \sum_{i=1}^{C}(p_{i}-\hat{p}_{i})^2 \cdot \|\aleph\|^2_2
    -
    \frac{\lambda_{12}}{S_h C(C-1)}\sum_{i \neq j}\|\aleph_{i}^{\mathsf{C}}-\aleph_{j}^{\mathsf{C}}\|_2,
    \label{eqn:FE_loss}
\end{equation}
where $\lambda_{11}$ and $\lambda_{12}$ are weighting parameters. 
The final loss $\mathcal{L}^{\mathrm{PT}}$ in the PT stage is defined as:
\begin{equation}
    \mathcal{L}^{\mathrm{PT}}\left(\phi^{\mathrm{FE}}, \phi^{\mathrm{CLS}}\right) =
    \mathcal{L}^{\mathrm{CE}}\left(\phi^{\mathrm{FE}}, \phi^{\mathrm{CLS}}\right) +
    \mathcal{L}^{\mathrm{FE}}\left(\phi^{\mathrm{FE}}\right),
    \label{eqn:PT_opt}
\end{equation}
where $\phi^{\mathrm{FE}} = \phi^{\mathrm{FP}}(\phi^{\mathrm{SC}}(\cdot))$.
With the dataset $\mathcal{D}^{\mathrm{PT}}$ in this stage, the optimization problem is formulated as ${\min_{\phi^{\mathrm{FE}}, \phi^{\mathrm{CLS}}}}\mathbb{E}_{(x,y)\sim \mathcal{D}^{\mathrm{PT}}}[\mathcal{L}^{\mathrm{PT}}(\phi^{\mathrm{FE}}, \phi^{\mathrm{CLS}})]$.

\vspace{-0.5em}
\subsection{Fine-tuning \revhu{Stage}} 
\label{ssec:sys_ft}

\revhu{
The FT stage tackles the challenge of cross-domain adaptation in the absence of samples from certain activity categories in the target domain.
Since no target-domain samples are available for the absent categories, standard supervised adaptation cannot directly improve target-domain feature extraction for these categories.
Our main idea is to preserve the activity knowledge learned in PT while utilizing the available target-domain samples to learn how to filter out subject-specific interference.
Within \name, this stage yields a target-adapted feature extractor that preserves the discriminative structure for activities learned in PT. 
The resulting representations are aligned with the source-domain anchors and are subsequently used by the composite decision strategy during inference.
}

Catastrophic forgetting~\cite{goodfellow2013empirical} is \revhu{a central challenge in cross-domain adaptation, as target-domain updates can distort the activity structure learned from the source domain.}
\revhu{To mitigate this issue while maintaining training and memory efficiency, we retain a small subset of source-domain data, denoted as $\tilde{\mathcal{D}}^{\mathrm{PT}}$, during FT, ensuring that adaptation to the target domain preserves the activity knowledge learned in PT.}
\rev{Given that the loss to be designed involves structured terms related to cluster center separation ($\mathcal{L}^{\mathrm{FE}}$ in Eqn.~\eqref{eqn:FE_loss}) and similarity ($\mathcal{L}^{\mathrm{AC}}$ in Eqn.~\eqref{eqn:loss_match}),
mini-batches generated by randomly sampling the mixed dataset of $\tilde{\mathcal{D}}^{\mathrm{PT}}$ and $\mathcal{D}^{\mathrm{FT}}$ may lead to instability in optimization and convergence due to variations in the composition pattern.}
\rev{Therefore,}
we compute their losses $\tilde{\mathcal{L}}^{\mathrm{PT}}$ and $\tilde{\mathcal{L}}^{\mathrm{FT}}$ separately and then combine them into the final loss for the FT stage:
\begin{equation}
    \mathcal{L}^{\mathrm{FT}} = \lambda_{21}\tilde{\mathcal{L}}^{\mathrm{PT}} + \lambda_{22}\tilde{\mathcal{L}}^{\mathrm{FT}},
    \label{eqn:FT_opt}
\end{equation}
where $\lambda_{21}$ and $\lambda_{22}$ are weighting parameters.

We now introduce the design of each sub-loss function.
\revhu{To prevent the feature extractor from forgetting the activity knowledge learned in PT, the PT objective $\mathcal{L}^{\mathrm{PT}}$ is retained in FT and is performed over the source-domain subset $\tilde{\mathcal{D}}^{\mathrm{PT}}$.
In addition, with limited target-domain data, updating the entire model can easily distort the discriminative boundaries of activities established in PT, leading to severe overfitting. 
To avoid this issue, our FT preserves the classifier and adapts only the feature extractor, ensuring that subject-specific interference is filtered while the original category boundaries are maintained.}
Accordingly, the loss $\tilde{\mathcal{L}}^{\mathrm{PT}}$ \revhu{can be expressed} as:
\begin{equation}
    \tilde{\mathcal{L}}^{\mathrm{PT}}(\phi^{\mathrm{FE}})
    =
    \mathcal{L}^{\mathrm{CE}}\left(\phi^{\mathrm{FE}}\right) +
    \mathcal{L}^{\mathrm{FE}}\left(\phi^{\mathrm{FE}}\right).
    \label{eqn:basic_loss}
\end{equation}
Therefore, the sub-optimization problem is formulated as ${\min_{\phi^{\mathrm{FE}}}}\mathbb{E}_{(\tilde{x},\tilde{y})\sim \tilde{\mathcal{D}}^{\mathrm{PT}}}[\tilde{\mathcal{L}}^{\mathrm{PT}}(\phi^{\mathrm{FE}})]$.
This design enables the neural network to preserve its original decision boundary and activity features, while adapting its filtering characteristics to the target domain distribution under controlled FT. 

\begin{algorithm}[t]
    \caption{{\name framework in the FT stage for cross-domain adaptation}}
    \label{alg:train_FT}
    
    \KwIn{Pre-trained model $\phi_0$, datasets $\tilde{\mathcal{D}}^{\mathrm{PT}}$ and $\mathcal{D}^{\mathrm{FT}}$, learning rate $\eta(\varphi)$, number of available activity categories $C^{\mathrm{FT}}$, and training epochs $\mathcal{E}$.}
    
    \KwOut{Fine-tuned model $\phi_{\mathcal{E}}$.}

    \vspace{.3em}
    \For{$\epsilon=1,\dots,\mathcal{E}$}
    {
    Sample batches $(\tilde{x},\tilde{y}) \sim \tilde{\mathcal{D}}^{\mathrm{PT}}$ and
    $(x,y) \sim {\mathcal{D}}^{\mathrm{FT}}$;\\
    Compute activity features: $\tilde{\aleph} = \phi^{FE}(\tilde{x})$ and $\aleph = \phi^{FE}(x)$;\\
    Compute category-wise cluster centers: $\tilde{\aleph}^{\mathsf{C}}_i \leftarrow \tilde{\aleph}$ and $\aleph^{\mathsf{C}}_i \leftarrow \aleph$, $\forall i \le C^{\mathrm{FT}}$ with valid category $i$;\\
    $\mathcal{L}^{\mathrm{AC}} \leftarrow \textbf{Loss}(\aleph^{\mathsf{C}}_i, \tilde{\aleph}^{\mathsf{C}}_i)$ based on Eqn.~\eqref{eqn:loss_match};\\
    $\tilde{\mathcal{L}}^{FT} \leftarrow \textbf{Loss}(\phi_{\epsilon}(x), y) + \mathcal{L}^{\mathrm{AC}}$ based on Eqn.~\eqref{eqn:loss_FTdata};\\
    $\tilde{\mathcal{L}}^{\mathrm{PT}} \leftarrow \textbf{Loss}(\phi_{\epsilon}(\tilde{x}), \tilde{y})$ based on Eqn.~\eqref{eqn:basic_loss};\\
    Update $\phi_{\epsilon}$ based on Eqn.~\eqref{eqn:train_proc}.
   }
\end{algorithm}
\setlength{\textfloatsep}{8pt}

To facilitate cross-domain adaptation, 
\revhu{the FT design learns target-domain matched filtering characteristics, which can be interpreted as suppressing subject-specific interference in the extracted feature $\aleph$, ensuring that it approximates an ideal activity representation $\tilde{\aleph}$.}
In conventional neural network training strategies, the filtering behavior can only be shaped indirectly through label supervision, limiting explicit control and ultimately hindering cross-domain adaptation.
Fortunately, to mitigate catastrophic forgetting, we have intentionally introduced $\tilde{\mathcal{D}}^{\mathrm{PT}}$, from which ideal features can be extracted to serve as anchors for learning well-behaved filtering characteristics.
Nevertheless, directly matching a large number of target domain features with those from the source domain may lead to overfitting and misalignment of structural patterns across domains;
therefore, we use the cluster centers of features as anchors to improve generalization.
Let the activity features from the source domain be denoted as $\tilde{\aleph} = \phi^{FE}(\tilde{x})$ ($\tilde{x} \in \tilde{\mathcal{D}}^{\mathrm{PT}}$), which are clustered by category to obtain the cluster centers $\tilde{\aleph}^{\mathsf{C}}_i$ ($i \in \{1,\cdots, C\}$).
Similarly, the target domain activity features $\aleph = \phi^{FE}(x)$ ($x \in \mathcal{D}^{\mathrm{FT}}$) yield cluster centers $\aleph^{\mathsf{C}}_i$ ($i \in \{1,\cdots, C^{\mathrm{FT}}\}$), where $\{1,\cdots,C^{\mathrm{FT}}\}$ and $\{C^{\mathrm{FT}}+1,\cdots,C\}$ correspond to present and absent categories, respectively.
Cosine similarity is employed to measure the discrepancy between them, leading to the anchor matching loss function $\mathcal{L}^{\mathrm{AC}}$ defined as:
\begin{equation}
    \mathcal{L}^{\mathrm{AC}} = \lambda_{23}\sum_{i=1}^{C^{\mathrm{FT}}}\left(1-\cos(\aleph^{\mathsf{C}}_i, \tilde{\aleph}^{\mathsf{C}}_i)\right),
    \label{eqn:loss_match}
\end{equation}
where $\lambda_{23}$ is a weighting parameter.
In addition, to ensure accurate recognition of activities in the target domain, we further \revhu{integrate} the $\mathcal{L}^{\mathrm{CE}}$ and $\mathcal{L}^{\mathrm{FE}}$ losses on the dataset $\mathcal{D}^{\mathrm{FT}}$.
The overall loss $\tilde{\mathcal{L}}^{FT}$ is then defined as:
\begin{equation}
    \tilde{\mathcal{L}}^{FT}(\phi^{\mathrm{FE}}) = 
    \mathcal{L}^{\mathrm{AC}}\left(\phi^{\mathrm{FE}}\right) +
    \mathcal{L}^{\mathrm{CE}}\left(\phi^{\mathrm{FE}}\right) +
    \mathcal{L}^{\mathrm{FE}}\left(\phi^{\mathrm{FE}}\right).
    \label{eqn:loss_FTdata}
\end{equation}
%
\revhu{Following the same classifier-preserving principle, the corresponding target-adaptation objective is}
${\min_{\phi^{\mathrm{FE}}}}\mathbb{E}_{(x,y)\sim \mathcal{D}^{\mathrm{FT}}}[\tilde{\mathcal{L}}^{\mathrm{FT}}(\phi^{\mathrm{FE}})]$.

Substituting Eqns.~\eqref{eqn:basic_loss} and~\eqref{eqn:loss_FTdata} into Eqn.~\eqref{eqn:FT_opt} \revhu{gives the complete FT objective.}
\revhu{Furthermore,} to finely control the training dynamics, module-specific learning rates $\eta(\varphi)$ ($\varphi \in \{\phi^{\mathrm{SC}}, \phi^{\mathrm{FP}}\}$) are introduced.
The parameter update is expressed as:
\begin{equation*}
    \mathcal{G} = \nabla_{\phi}\left( \lambda_{21}\mathbb{E}_{(\tilde{x},\tilde{y})\sim \tilde{\mathcal{D}}^{\mathrm{PT}}}\tilde{\mathcal{L}}^{\mathrm{PT}}+\lambda_{22}\mathbb{E}_{(x,y)\sim \mathcal{D}^{\mathrm{FT}}}\tilde{\mathcal{L}}^{\mathrm{FT}} \right)
\end{equation*}
\begin{equation}
    \phi_{(\epsilon+1)} \leftarrow \phi_{\epsilon} - \eta(\varphi) \circ \mathcal{G},
    \label{eqn:train_proc}
\end{equation}
where $\circ$ denotes Hadamard product and $\epsilon$ indicates the iteration index.
\revhu{In summary}, the training algorithm for the FT stage is detailed in Algorithm~\ref{alg:train_FT}.

\vspace{-0.5em}
\subsection{Inference \revhu{Stage}} \label{ssec:sys_descion}

\revhu{During inference, relying solely on the predicted probability distribution may not fully leverage the feature geometry established by the PT and FT stages.}
\revhu{This is because the softmax prediction essentially reflects the similarity between activity features and classifier weights~\cite{liu2017sphereface}, but it may still overlook the clustering structure and geometric organization of features in the embedding space.}
%
\revhu{Our main idea is to combine probability-based prediction with explicit similarity to class-wise feature centers, ensuring that classification is guided by both decision-boundary information and embedding-space structure.}
\revhu{Within \name, this stage produces the final activity prediction by fusing these complementary cues, thereby translating the learned representations into more robust target-domain recognition.}

Specifically, the fine-tuned model $\phi_{\mathcal{E}}$ first processes the deliberately constructed dataset $\tilde{\mathcal{D}}^{\mathrm{PT}}$ and performs category-wise clustering to obtain the cluster centers $\breve{\aleph}^{\mathsf{C}}_i$ ($i \in \{1,\cdots,C\}$). 
For each test sample $\breve{x}\in \mathcal{X}^{\mathrm{Test}}$, both the predicted probability distribution 
$\breve{p}_i$ and the normalized similarity $\breve{q}_i = \cos(\phi_{\mathcal{E}}^{\mathrm{FE}}(\breve{x}),\breve{\aleph}^{\mathsf{C}}_i)$ between its feature and the cluster centers from $\tilde{\mathcal{D}}^{\mathrm{PT}}$ are computed.
The final decision result is then denoted as:
\begin{equation}
    \breve{y} = \arg \max_{i \in \{1,\cdots,C\}} (\breve{p}_{i} + \lambda_{3}\breve{q}_i).
    \label{eqn:decision}
\end{equation}
where $\lambda_3$ is a weighting parameter.
This inference strategy captures both discriminative decision boundaries and the semantic consistency of features,
thereby further improving the accuracy of HAR in the target domain.

\section{\dataset Dataset} \label{sec:dataset}

In this section, we first construct a Wi-Fi Near-Field Sensing (\dataset) dataset\footnote{The dataset is available via \url{https://github.com/DeepWiSe888/NFS-Fi}.}.
and provide a brief statistical analysis.

\vspace{-0.5em}
\subsection{Dataset Collection}  \label{ssec:exp_set}

To advance Wi-Fi sensing towards practical multi-person sensing and ISAC development, we build a multi-person HAR dataset,
\dataset,
consisting of near-field channel samples generated under native traffic, leveraging up-to-date NICs.
We begin by setting up the data collection system, followed by a detailed description of the experiment setup.

\begin{figure}[t]
	\centering 
	\setlength{\abovecaptionskip}{2pt}  
	\subfigure[Hardware components.] 
		{
		 \centering
		 \includegraphics[width=0.46\linewidth]{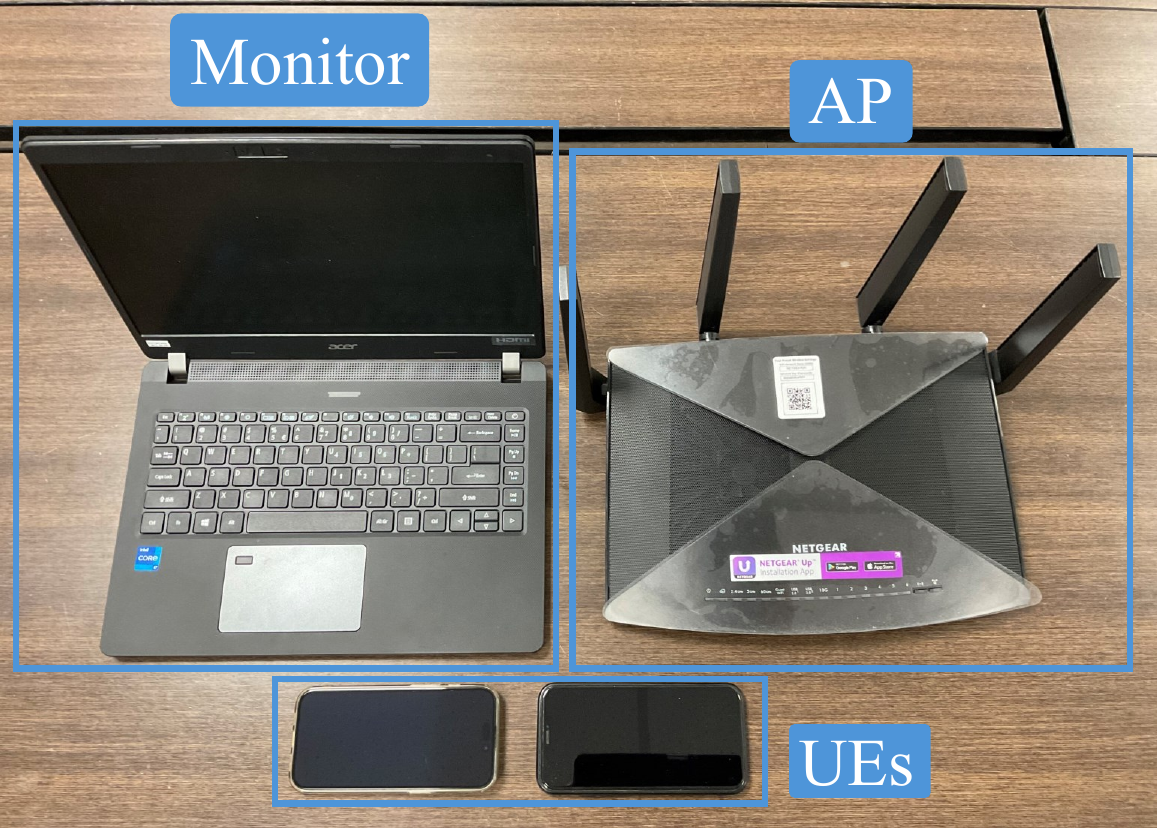}
		 \label{sfig:hardware}    
		}
        \hfill
	\subfigure[Experiment setup.]
		{
			\centering         
			\includegraphics[width=0.46\linewidth]{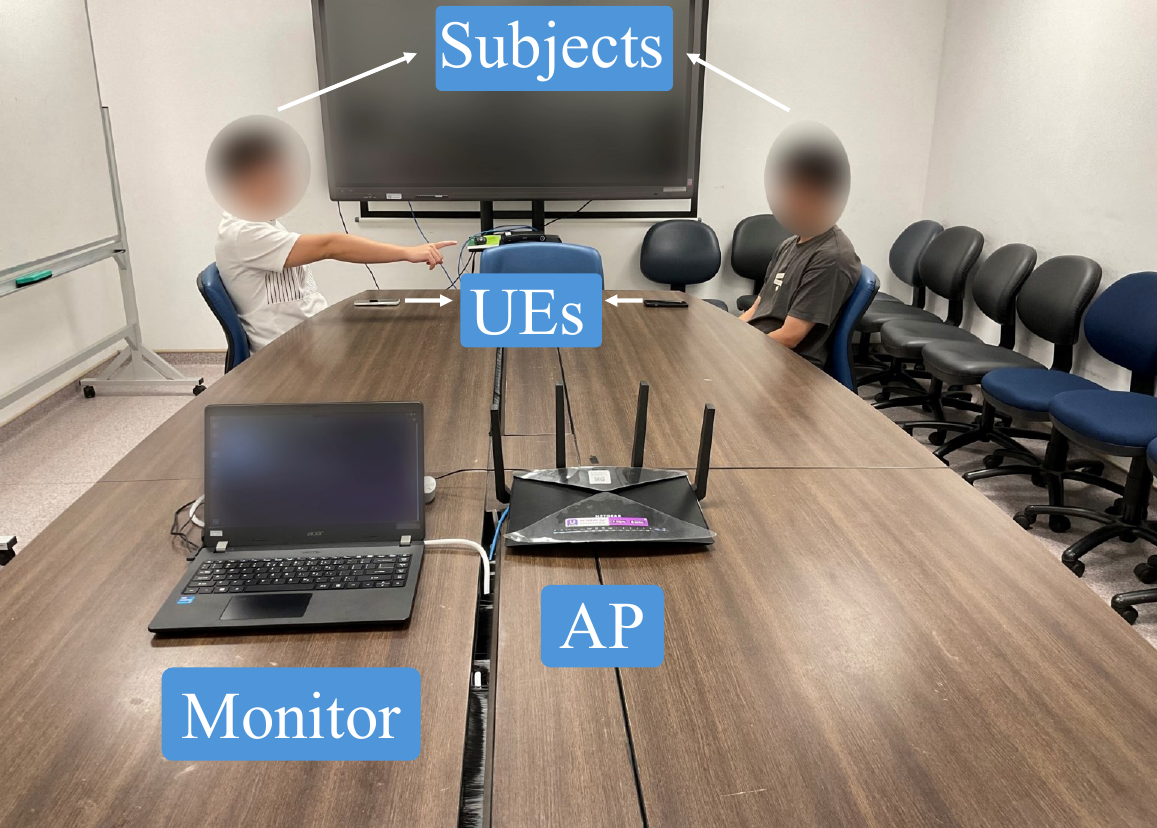}
			\label{sfig:setup}  
		}
	\caption{Data collection system.}
     \label{fig:experiment}
\end{figure}

Our data collection system consists of an AP and several UEs.
The AP is a Netgear Nighthawk X10 router compliant with the IEEE 802.11ac standard, operating on a 5260~\!MHz carrier frequency with a 40~\!MHz channel bandwidth.
The UEs are smartphones running Android or iOS, equipped with NICs compliant with the IEEE 802.11ax standard, and placed approximately 20~\!cm in front of the subjects to induce the near-field domination effect.
During the experiment, the UEs connect to the AP and generate uplink traffic through video meetings, while the subjects engage in various activities,
as shown in Fig.~\ref{fig:experiment}.
A laptop equipped with an Intel AX210 NIC, which also adheres to the IEEE 802.11ax standard, serves as the monitor, and the PicoScenes tool~\cite{PicoScenes-IoIJ21} is employed to capture the Wi-Fi signals.
Among these signals, the QoS Data packets are extracted and parsed to obtain the required sensing information, including timestamp, RSSI, and CSI data.
The raw CSI structure is a $2\times 117$ complex matrix, representing the number of receiving antennas and subcarriers, respectively.
Owing to its versatility, this data collection system finds applicability across diverse near-field sensing applications.

\begin{figure}[b]
\vspace{-1.em}
	\centering 
	\setlength{\abovecaptionskip}{2pt}
	\subfigure[Meeting room.] 
		{
		 \centering
		 \raisebox{3pt}{\includegraphics[width=0.3\linewidth]{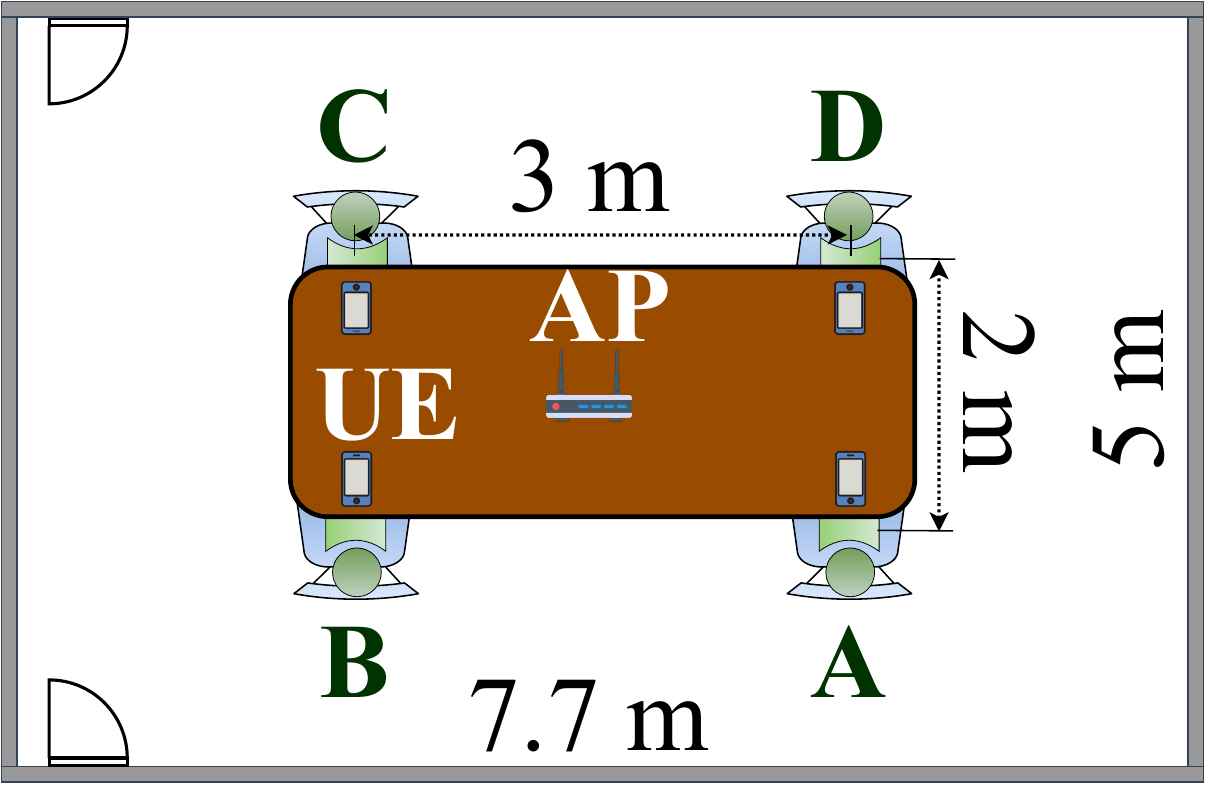}
		 \label{sfig:MR}    
		}}
        \hfill
        \subfigure[Lecture room.] 
		{
		 \centering
		 \includegraphics[width=0.3\linewidth]{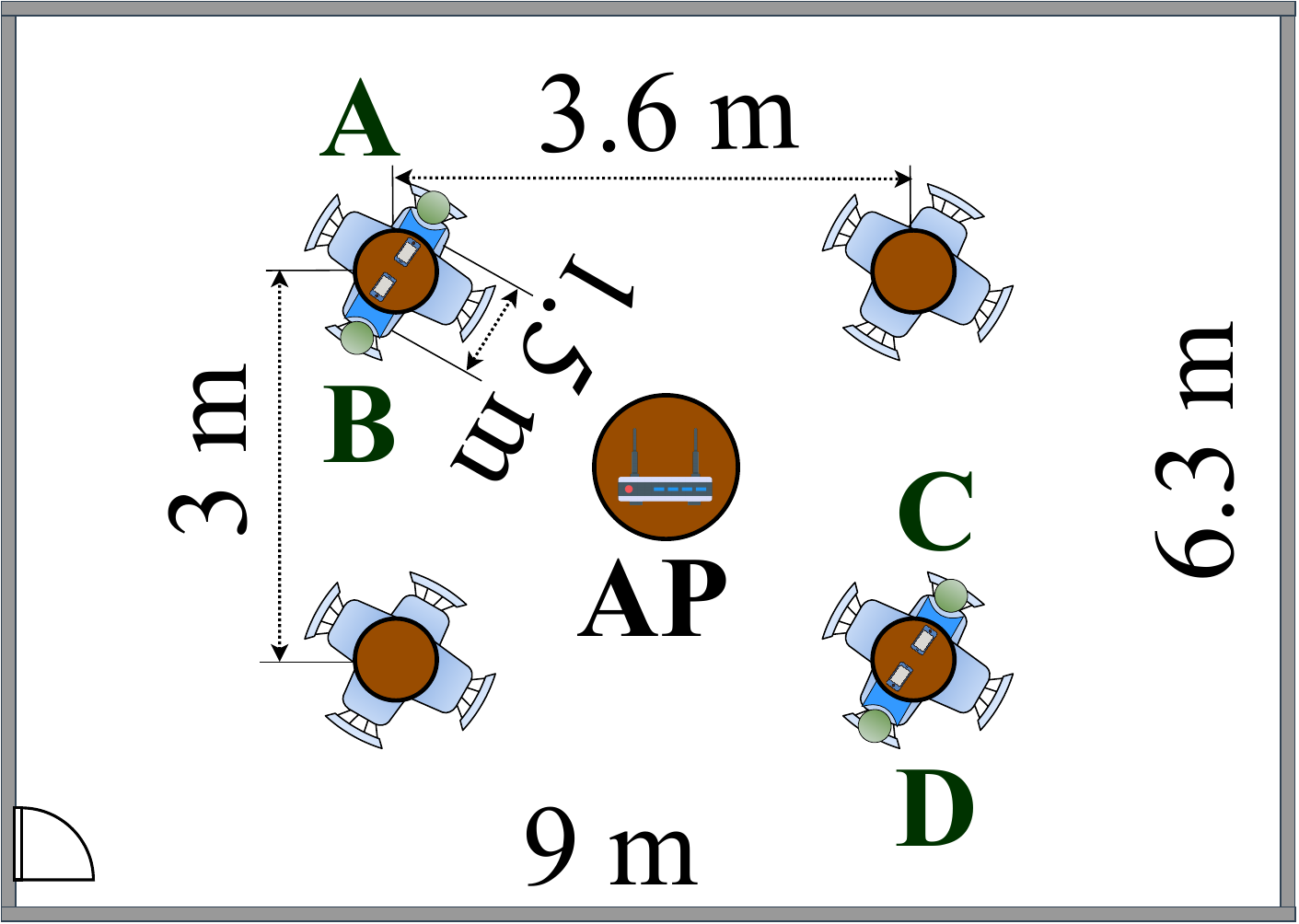}
		 \label{sfig:LR}    
		}
        \hfill
        \subfigure[Discu. room.]
		{
			\centering         
			\raisebox{3pt}{\includegraphics[width=0.2\linewidth]{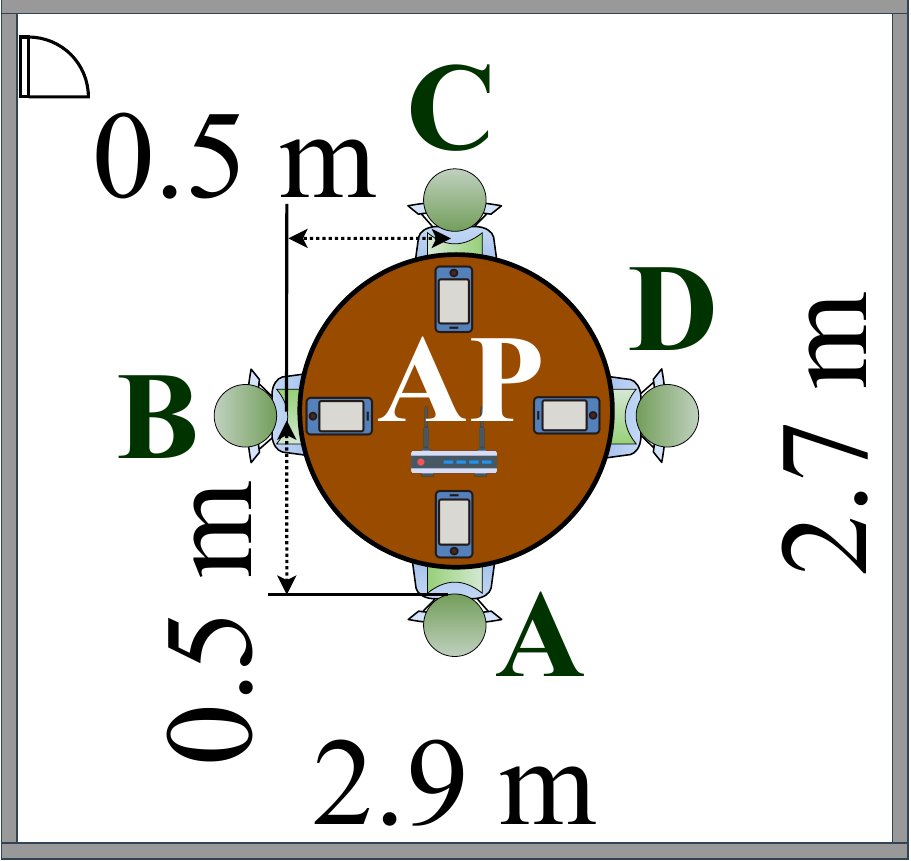}
			\label{sfig:SR}  
		}}
        %
        \\
	\subfigure[Classroom.]
		{
			\centering         
			\includegraphics[width=0.34\linewidth]{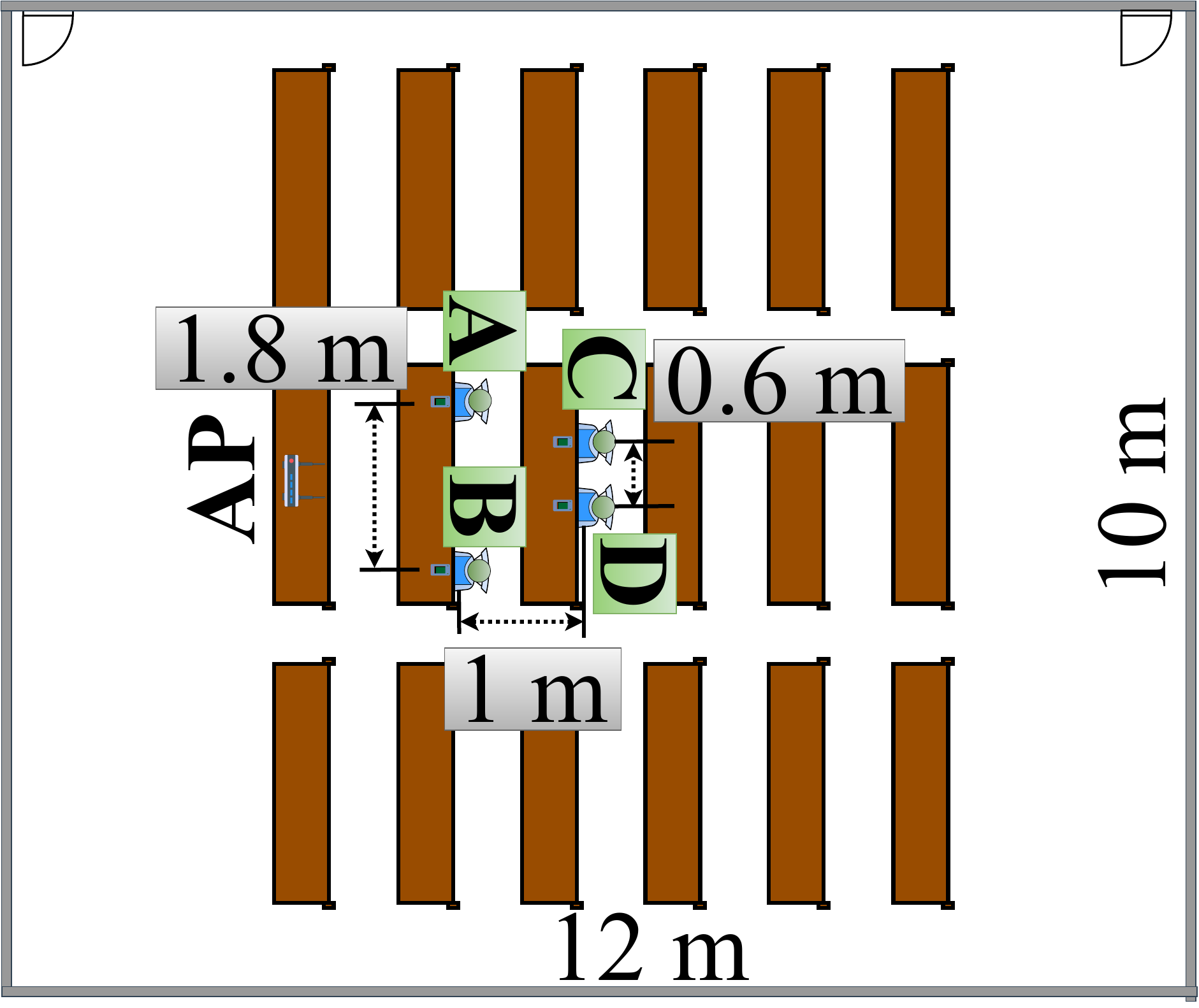}
			\label{sfig:CR}  
		}
        \hfill
        \subfigure[Office room.]
		{
			\centering         
			\raisebox{9pt}{\includegraphics[width=0.21\linewidth]{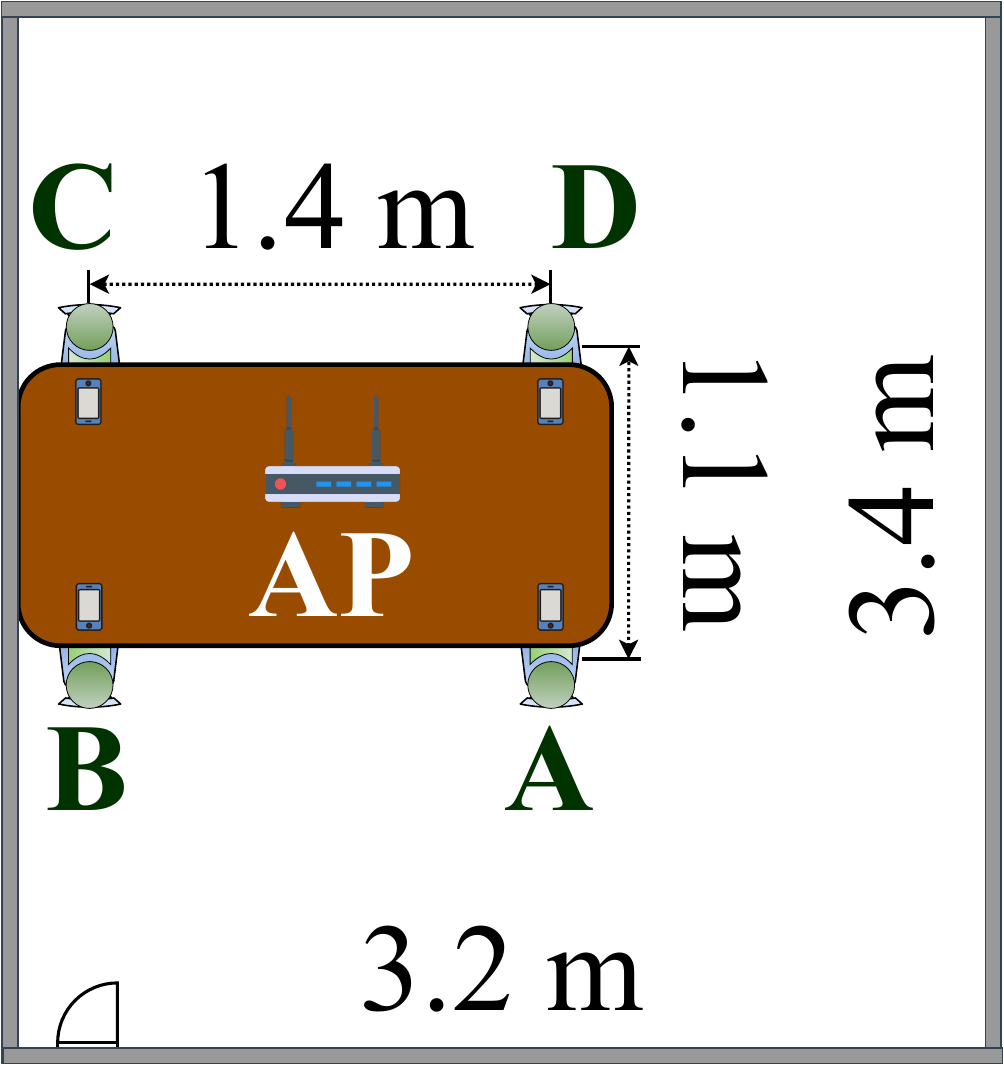}
			\label{sfig:OR}  
		}}
        \hfill
        \subfigure[Self-study room.] 
		{
		 \centering
		 \raisebox{5pt}{\includegraphics[width=0.296\linewidth]{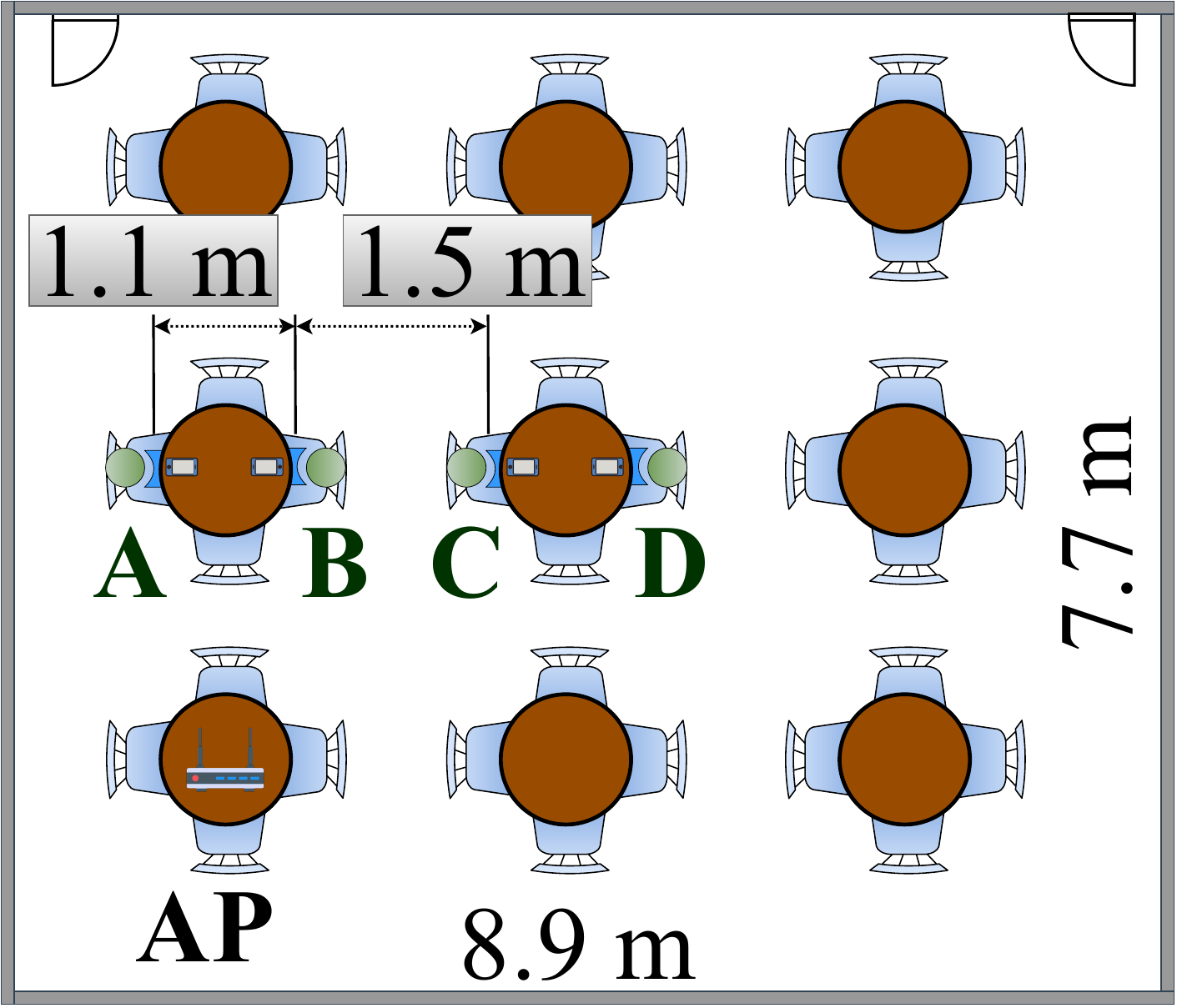}
		 \label{sfig:DR}    
		}}
	\caption{Environment layouts.}
     \label{fig:layout}
 \vspace{-0.5em}
\end{figure}

To build our dataset, we recruit 56 participants, including 36 males and 20 females, aged between 20 and 55 years, with heights ranging from 155~\!cm to 185~\!cm.
Our experiment involves six different environments: meeting room~(MR), lecture room~(LR), discussion room~(DR), classroom~(CR), office room~(OR), and self-study room~(SR). as shown in Fig.~\ref{fig:layout}.
Each subject performs activities in two distinct environments: Subjects 1–16 in MR and LR, Subjects 17–36 in DR and CR, and Subjects 37–56 in OR and SR.
They execute 10 activities in total, including 4 hand gestures, 2 interactive gestures, and 4 body activities,
\norev{designed to cover the three levels defined in Section~\ref{sec:intro}.}
Specifically, these are push\&pull~(PP), sweeping~(SW), drawing circle~(DC), zig\&zag~(ZZ), typing on a phone~(TP), handshaking~(HS), bending~(BD), jumping~(JP), rotating~(RT), and walking~(WK).
Each experiment involves 2 to 4 concurrent participants who perform the activities at their own pace, while being instructed to complete each activity within 2 seconds, followed by a short 1-second pause before starting the next round to facilitate data segmentation.
In evaluations, we extract only the first 2 seconds of data for HAR.
These experiments have strictly followed the IRB of our institute.
Informed consent was obtained from all participants.

\vspace{-1em}
\subsection{Analysis of the Dataset} \label{ssec:data_anal}

\begin{table*}[!b]
    \centering
    \caption{Comparison with public Wi-Fi sensing datasets for human activity recognition}
    \begin{tabular}{>{\rule{0pt}{4ex}}p{2.5cm}ccccccccc}
        \toprule
        \textbf{Dataset} & \textbf{\makecell[cc]{Dataset\\Size}} & \textbf{\makecell[cc]{Concurrent\\Users}} & \textbf{\makecell[cc]{No.\\Act.}} & \textbf{\makecell[cc]{No.\\Participants}} & \textbf{\makecell[cc]{No.\\Env.}} & \textbf{Sampling} & \textbf{\makecell[cc]{Bandwidth\\(MHz)}} & \textbf{\makecell[cc]{Wi-Fi Band\\(GHz)}} & \textbf{Standard} \\
        \midrule
        UT-HAR~\cite{yousefi2017survey} & 5173 & 1 & 7 & 6 & 1 & 1000~\!Hz & 20 & 5 & 802.11n \\
        
        FallDeFi~\cite{palipana2018falldefi} & 1070 & 1--2 & 28 & 3 & 5 & 1000~\!Hz & 20 & 5 & 802.11n \\ 
        
        SignFi~\cite{ma2018signfi} & 14280 & 1 & 276 & 5 & 2 & 200~\!Hz & 20 & 5 & 802.11n \\ 
        
        WiAR~\cite{guo2019wiar} & 4800 & 1 & 16 & 10 & 3 & 30~\!Hz & 20 & 5 & 802.11n \\ 
        
        Brinke et al.~\cite{brinke2019dataset} & 4199 & 1 & 6 & 9 & 1 & 20~\!Hz & 20 & 2.4 & 802.11n \\ 
        
        Widar3.0~\cite{zhang2021widar3} & 258575 & 1 & 16 & 16 & 3 & 1000~\!Hz & 20 & 5 & 802.11n \\
        
        Baha et al.~\cite{baha2020dataset} & 9000 & 1 & 12 & 30 & 3 & 320~\!Hz & 20 & 2.4 & 802.11n \\
        
        CSIDA~\cite{hu2021wigr} & 3000 & 1 & 6 & 5 & 2 & 1000~\!Hz & 40 & 5 & 802.11n \\
        
        OPERAnet~\cite{bocus2022operanet} & 6235 & 1 & 6 & 6 & 2 & 1600~\!Hz & 20 & 5 & 802.11n \\
        
        NTU-HAR~\cite{yang2022efficientfi} & 2400 & 1 & 6 & 20 & 1 & 500~\!Hz & 40 & 5 & 802.11n \\
        
        MM-Fi~\cite{yang2023mm} & 1080 & 1 & 27 & 40 & 4 & 1000~\!Hz & 40 & 5 & 802.11n \\
        
        CSI-BERT~\cite{zhao2024finding} & 3360 & 1 & 7 & 8 & 1 & 100~\!Hz & 20 & 2.4 & 802.11n\\
        
        XRF55~\cite{wang2024xrf55} & 429000 & 1 & 55 & 39 & 4 & 200~\!Hz & 20 & 5 & 802.11n \\
        
        WiMANS~\cite{huang2024wimans} & 11286 & 0--5 & 9 & 6 & 3 & 1000~\!Hz & 20 & 2.4/5 & 802.11n \\
        
        XRF V2~\cite{lan2025xrf} & 853 & 1 & 45 & 16 & 3 & 200~\!Hz & 20 & 5 & 802.11n \\
        
        \textbf{\textit{\dataset}} & \textbf{\textit{64823}} & \textbf{\textit{2-4}} & \textbf{\textit{10}} & \textbf{\textit{56}} & \textbf{\textit{6}} & {\makecell[cc]{\textbf{\textit{Native}}\\\textbf{\textit{Traffic}}}} & \textbf{\textit{40}} & \textbf{\textit{5}} & \textbf{\textit{802.11ac/ax}} \\
        \bottomrule
    \end{tabular}
    \label{tab:dataset_survey}
\end{table*}

Our \dataset dataset is the first practical Wi-Fi multi-person sensing dataset, offering three key advantages over existing datasets~\cite{huang2024wimans, yousefi2017survey, palipana2018falldefi, ma2018signfi,guo2019wiar,brinke2019dataset,zhang2021widar3, baha2020dataset,hu2021wigr,bocus2022operanet, yang2022efficientfi, yang2023mm, zhao2024finding, wang2024xrf55,lan2025xrf}, as summarized in Table~\ref{tab:dataset_survey}.
First, leveraging diverse physical information, the multi-link near-field sensing strategy enables practical multi-person sensing beyond simple scenarios~\cite{palipana2018falldefi} or mere reliance on neural network fitting~\cite{huang2024wimans}.
Second, the dataset contains native traffic from the normal operation of smart devices, without injecting evenly spaced sensing packets, thus avoiding interference with default communication and reflecting realistic conditions.
Third, the dataset is built using up-to-date NICs compliant with IEEE 802.11ac/ax standards, keeping Wi-Fi sensing aligned with the latest technological developments.

\begin{figure}[t]
	\centering 
	\setlength{\abovecaptionskip}{2pt}  
		 \includegraphics[width=0.99\linewidth]{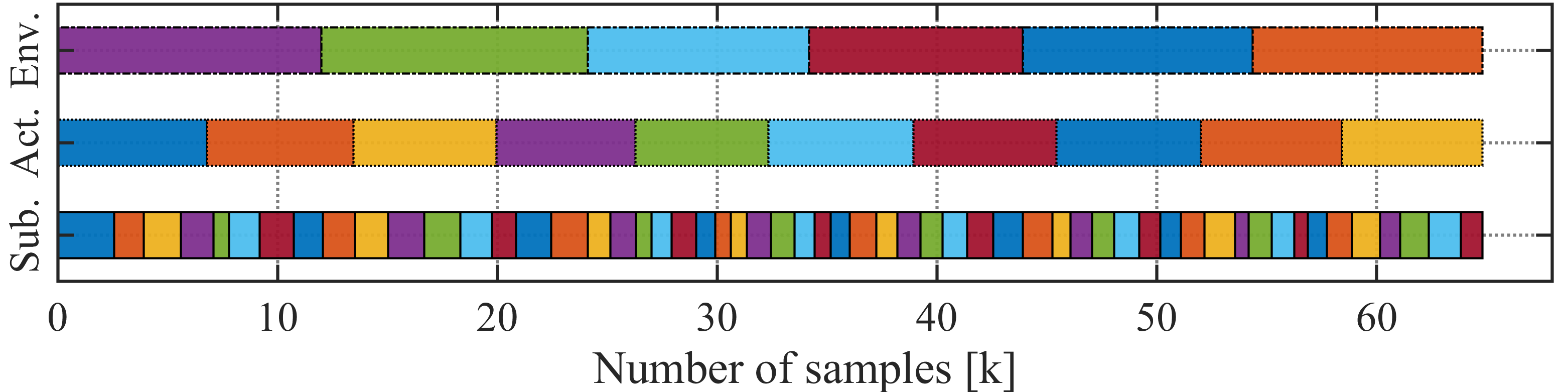}
	\caption{Statistics of samples across subjects, activities, and environments.}
     \label{fig:sta_sample}
 \vspace{-0.5em}
\end{figure}
\begin{figure}[t]
	\centering 
	\setlength{\abovecaptionskip}{2pt}  
	\subfigure[Number of CSIs.]  
		{
		 \centering
		 \includegraphics[width=0.29\linewidth]{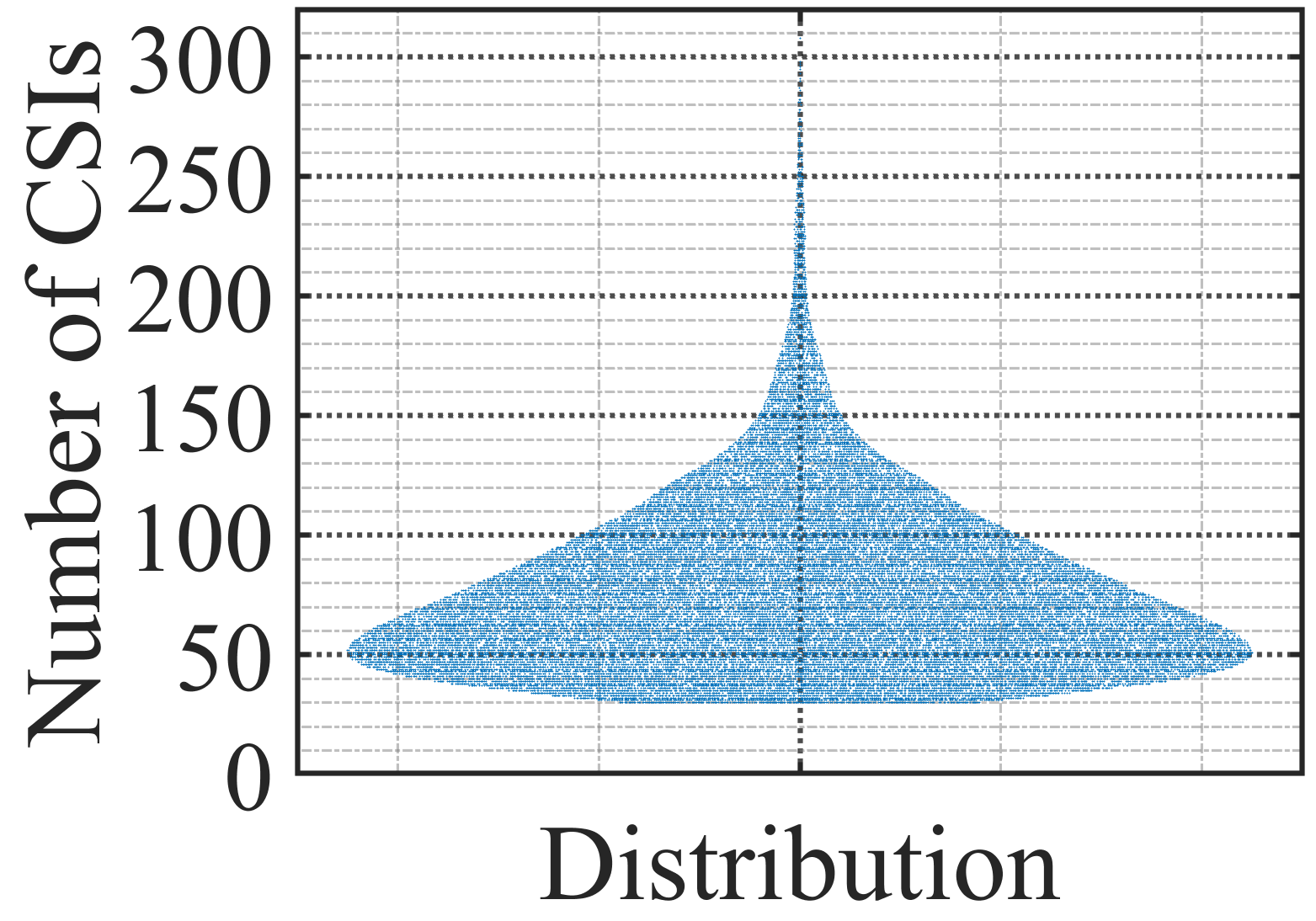}
		 \label{sfig:stat_number}    
		}
        \hfill
	\subfigure[Max. TI.]
		{
			\centering         
			\includegraphics[width=0.29\linewidth]{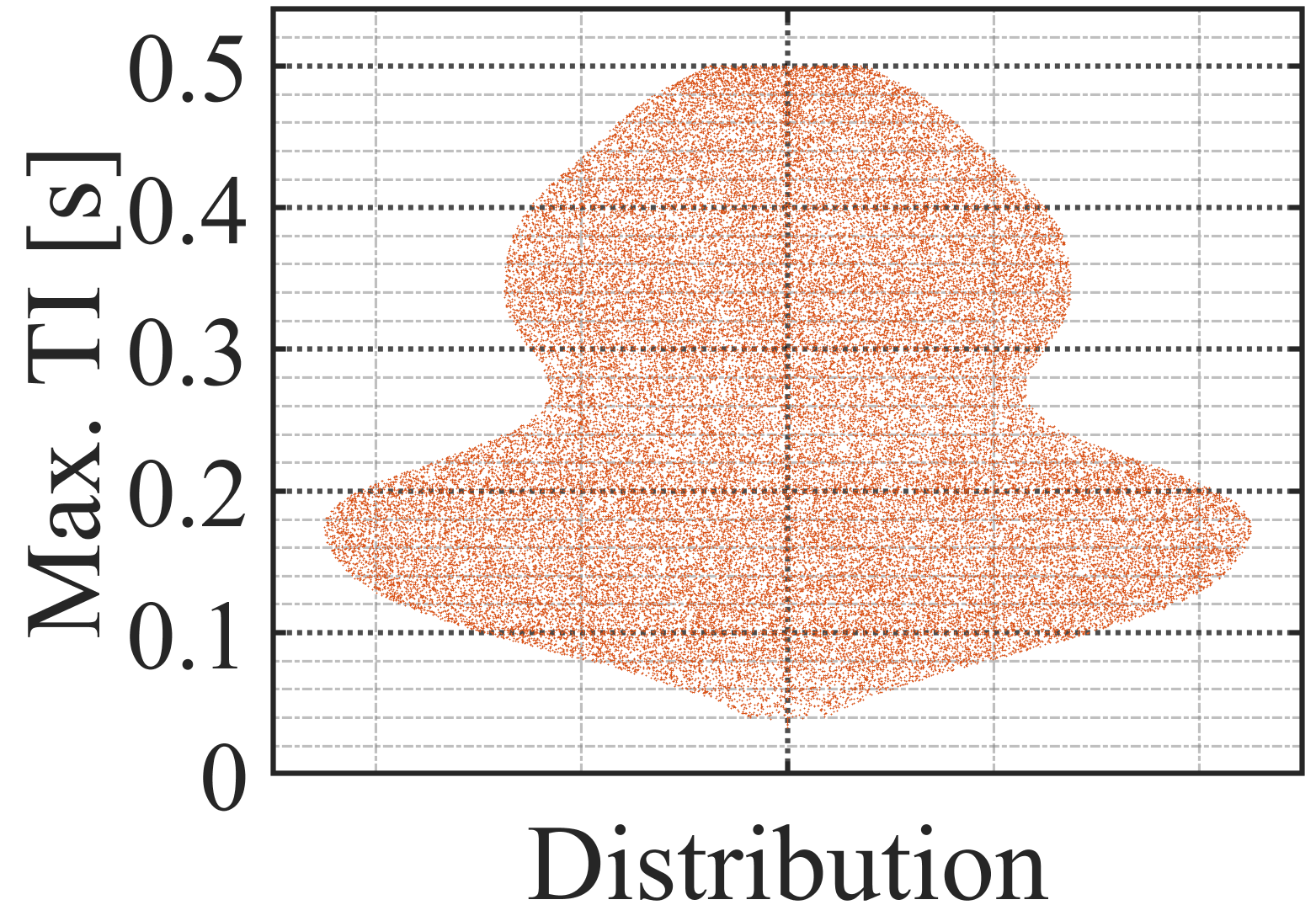}
			\label{sfig:stat_inter}  
		}
         \hfill
	\subfigure[CSI Duration.]
		{
			\centering         
			\includegraphics[width=0.29\linewidth]{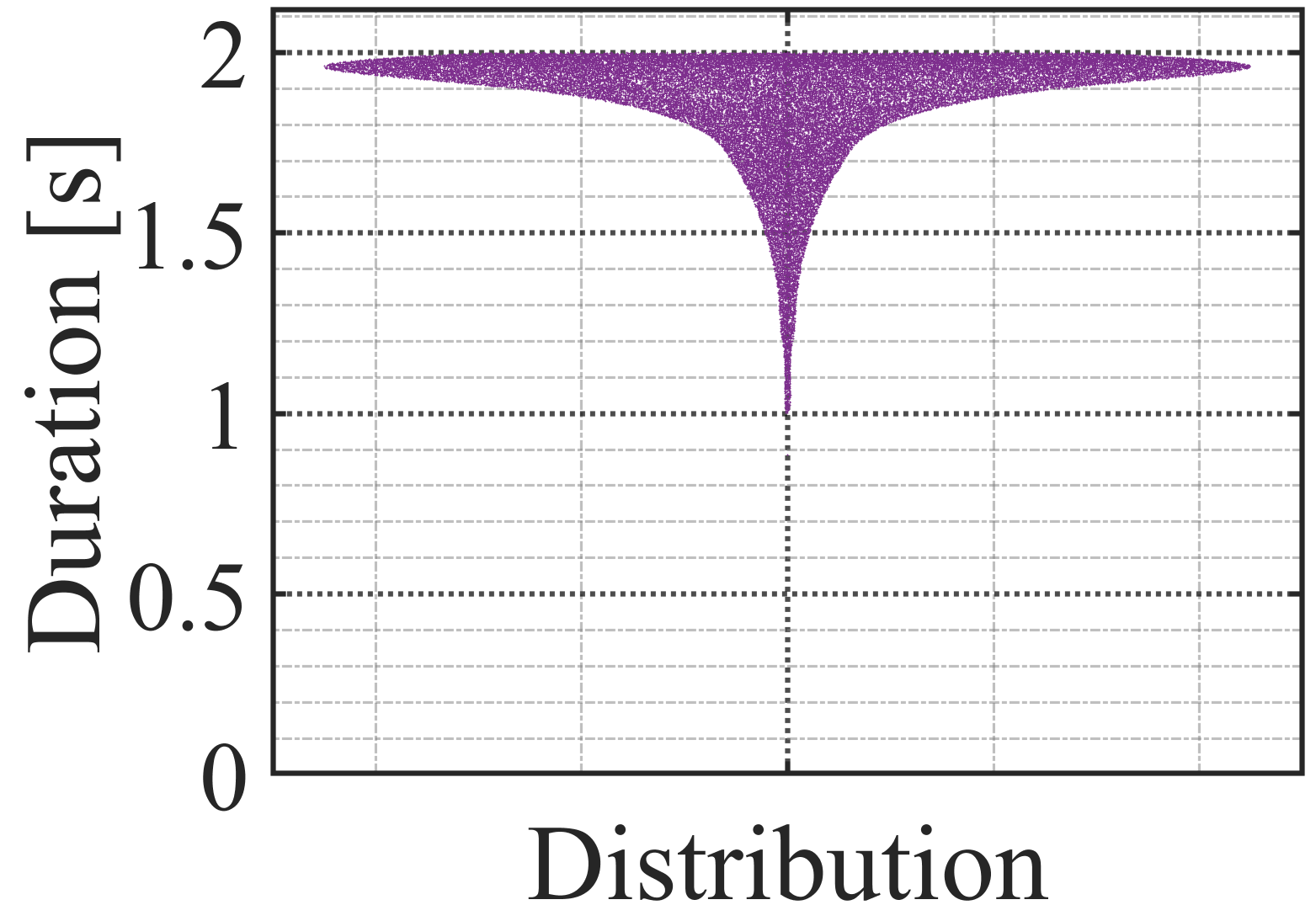}
			\label{sfig:sata_max}  
		}
	\caption{Statistics of CSI entries across samples.}
     \label{fig:stat_CSI}
 \vspace{-0.5em}
\end{figure}

Beyond those advantages, our \dataset dataset provides sufficient diversity to capture real-world scenarios.
It comprises 64,823 samples, with Subject 1 contributing the most valid activities (2,563) and Subject 49 the fewest (624).
Among all activities, PP has the most samples (6,780), while TP has the fewest (6,075).
Across the six environments, LR contains the most samples (12,121) and CR the fewest (9,735).
The detailed distribution is shown in Fig.~\ref{fig:sta_sample}.
Furthermore, on average, each sample contains 77 CSI entries, with a maximum time interval (Max. TI) of approximately 0.25~\!s and an average data collection duration of 1.86~\!s.
The CSI entry statistics for each sample are presented in Fig.~\ref{fig:stat_CSI}.
These results confirm that the dataset collected under native traffic conditions can effectively capture activity cycles, ensuring its usability.

\section{Evaluations} \label{sec:evaluation}

In this section, we conduct a comprehensive evaluation of \name framework using \dataset dataset, beginning with the evaluation setup, followed by a micro-benchmark study, comparison to baselines, and analysis of impact factors.

\subsection{Evaluation Setup} \label{ssec:setup}

The architecture of our basic GRU model is illustrated in Fig.~\ref{fig:network}, with a hidden size of $S_h=64$ and a single layer.
In each evaluation round, one subject’s data is used as the target domain, and data from 6 randomly selected subjects, excluding the target subject and the environment where that subject is recorded, serve as the source domain.
A batch size of 64 is used throughout the entire training process.
During the PT stage, the learning rate is set to $10^{-3}$ for 50 epochs.
In the FT stage, the FT dataset $\mathcal{D}^{\mathrm{FT}}$ contains 10 samples for each available activity, while absent categories contain 0 samples;
the anchor dataset $\tilde{\mathcal{D}}^{\mathrm{PT}}$ includes all classes with 30 samples per category.
The SC and FP modules use learning rates of $7\times10^{-4}$ and $5\times10^{-4}$, respectively, while the other modules are frozen to prevent overfitting, and the model is fine-tuned for 200 epochs.
\rev{Weight $\lambda_3$ is set to 2 in the inference stage.}

\subsection{Micro-benchmark Study} \label{ssec:benckmark}

\begin{figure}[b]
	\centering 
	\setlength{\abovecaptionskip}{2pt}  
        \subfigure[Source domain.]
		{
			\centering         
			\includegraphics[width=0.43\linewidth]{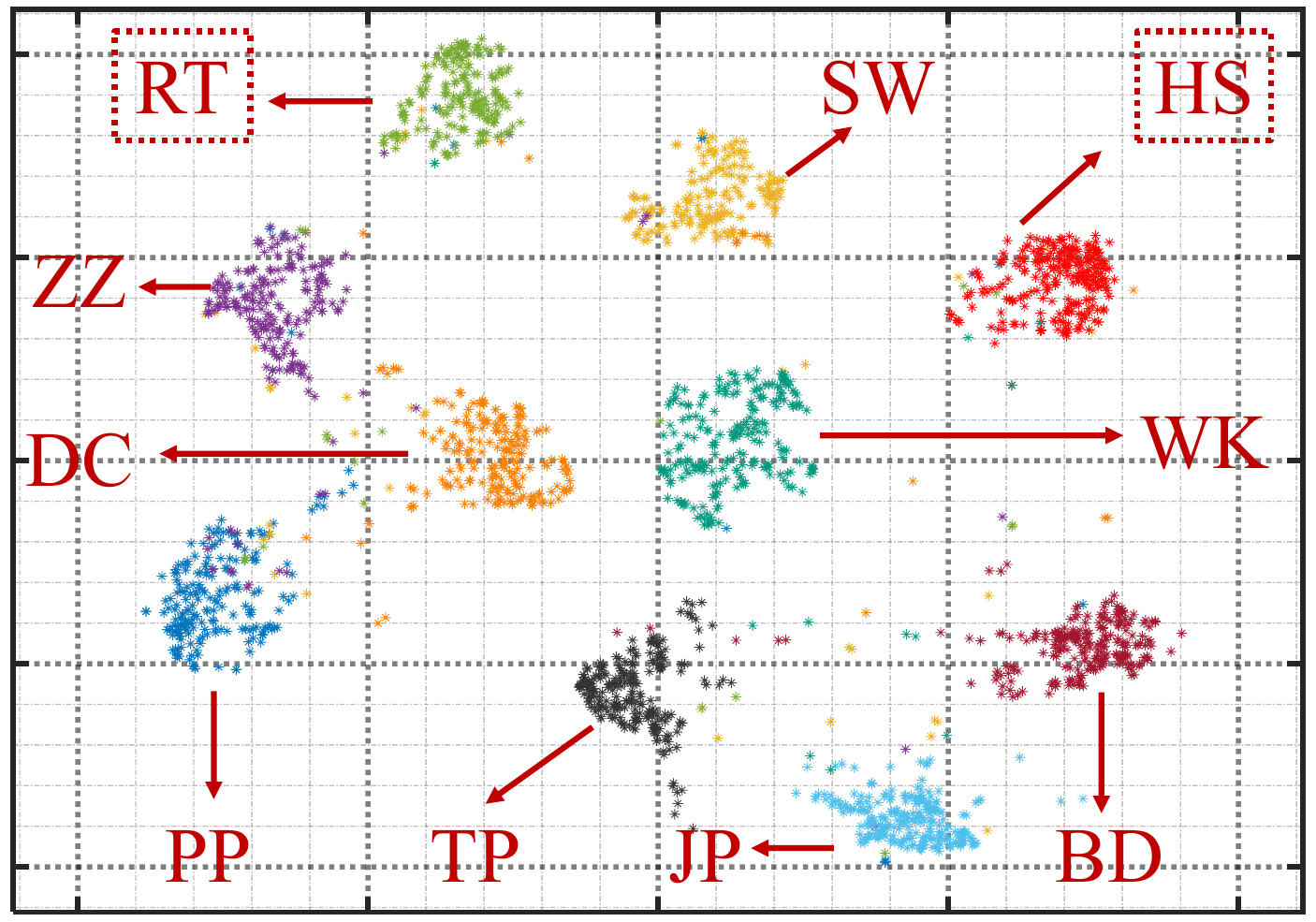}
			\label{sfig:exp_tsne_source}  
		}
        \hspace{8pt}
	\subfigure[Target domain.] 
		{
		 \centering
		 \includegraphics[width=0.43\linewidth]{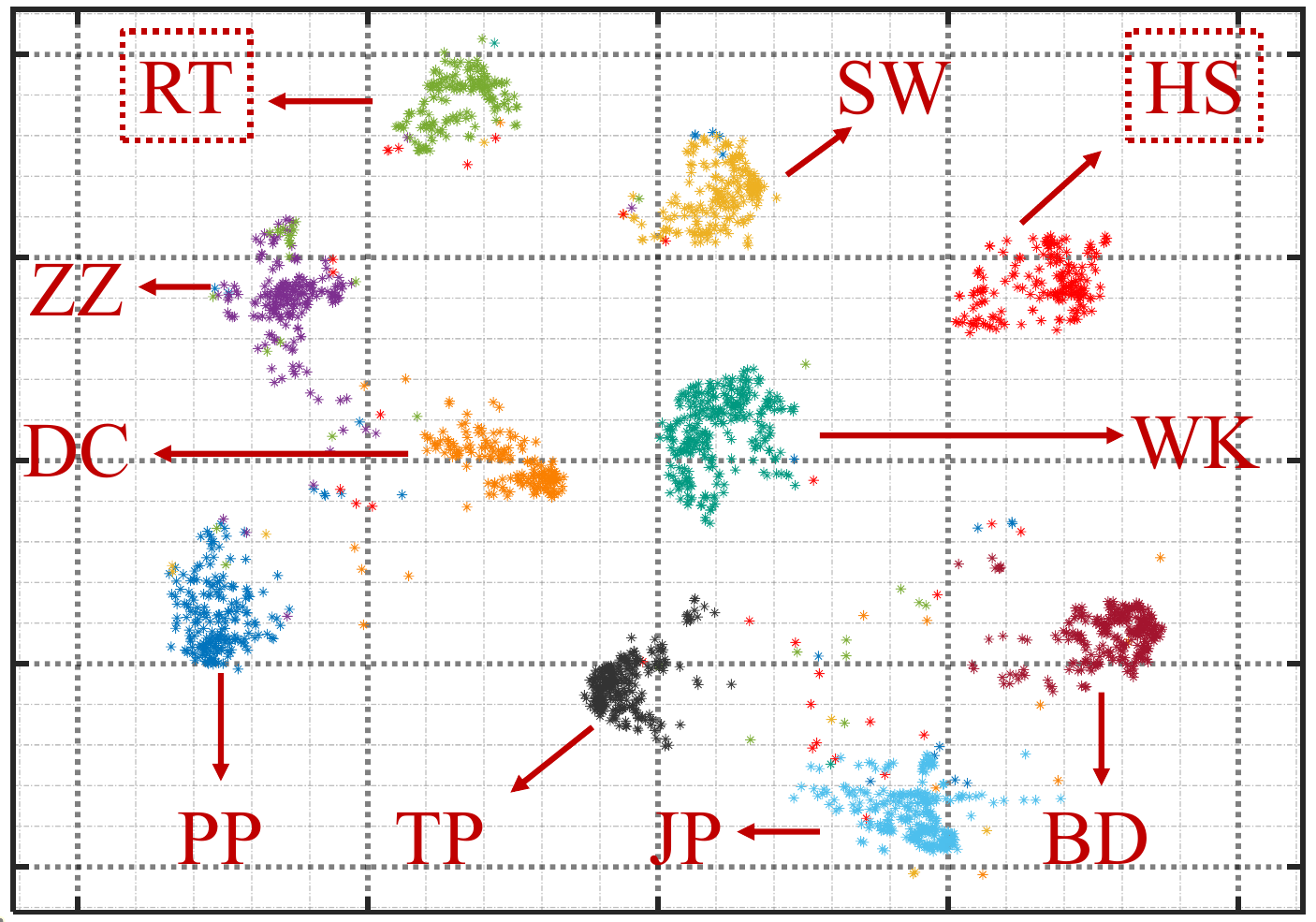}
		 \label{sfig:exp_tsne_target}    
		}
	\caption{t-SNE visualization. Large inter-class margin and feature alignment between the target and source domains demonstrate \name's effectiveness.}
     \label{fig:exp_tsne}
\end{figure}

To analyze the effectiveness of our \name framework, we use t-SNE to visualize the extracted features.
As shown in Fig.~\ref{sfig:exp_tsne_source}, benefiting from the inter-class margin enlarging strategy, the features in the source domain present well-defined clustering patterns, where samples from the same category are tightly grouped and different categories are clearly separated.
In Fig.~\ref{sfig:exp_tsne_target}, leveraging a small subset of data from source domain as anchors during the FT stage results in target domain features closely aligning with the source domain features, exhibiting a consistent distribution without noticeable shift.
Thanks to these strategies and the matched filter-driven mechanism, subject-specific interference in the RT and HS categories of the target domain is effectively eliminated during the FT stage, even without samples, achieving feature separation nearly comparable to that of categories with sufficient samples in the source domain.
These results demonstrate that our \name framework, proposed in Section~\ref{sec:method}, is consistent with the insights discussed in Section~\ref{sssec:fine_tuning}, achieving satisfactory recognition accuracy with certain categories absent.

\subsection{Overall Performance} \label{ssec:overall}

\begin{figure}[b]
\vspace{-1.em}
	\centering 
	\setlength{\abovecaptionskip}{2pt}  
	\subfigure[\name.] 
		{
		 \centering
		 \includegraphics[width=0.46\linewidth]{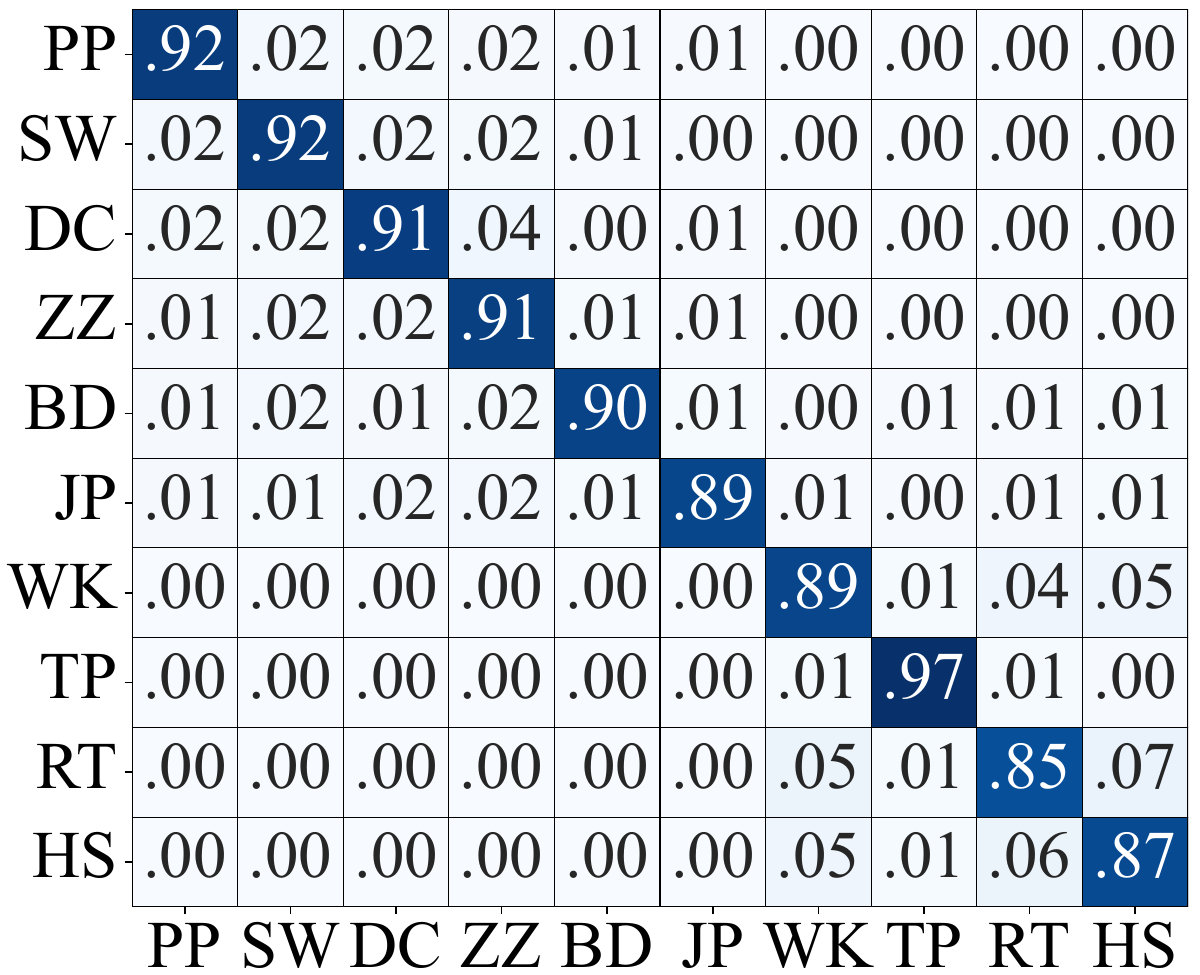}
		 \label{sfig:overall_our}    
		}
        \hfill
	\subfigure[Class-sensitive learning.]
		{
			\centering         
			\includegraphics[width=0.46\linewidth]{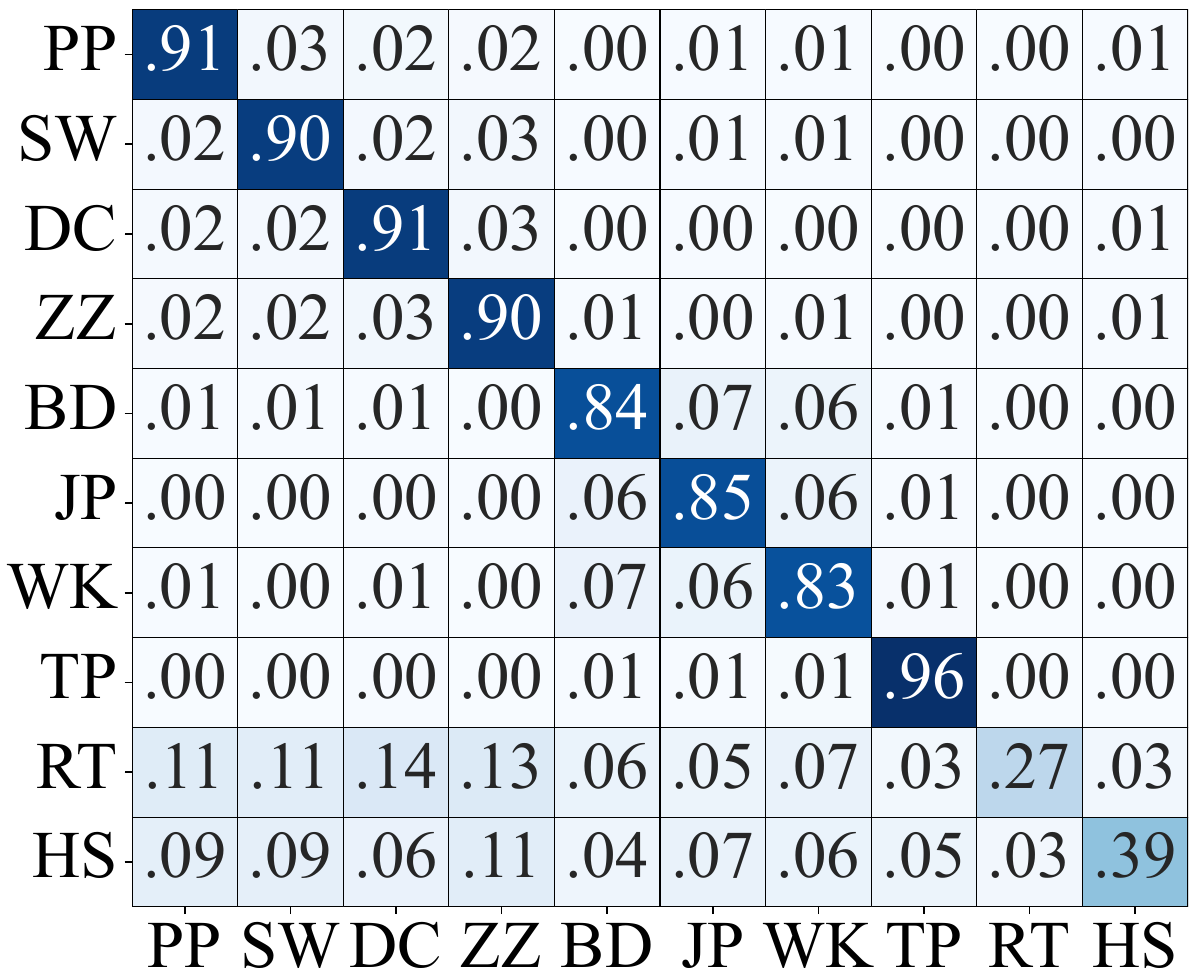}
			\label{sfig:overall_csl}  
		}
        \\
        \subfigure[Data augmentation.] 
		{
		 \centering
		 \includegraphics[width=0.46\linewidth]{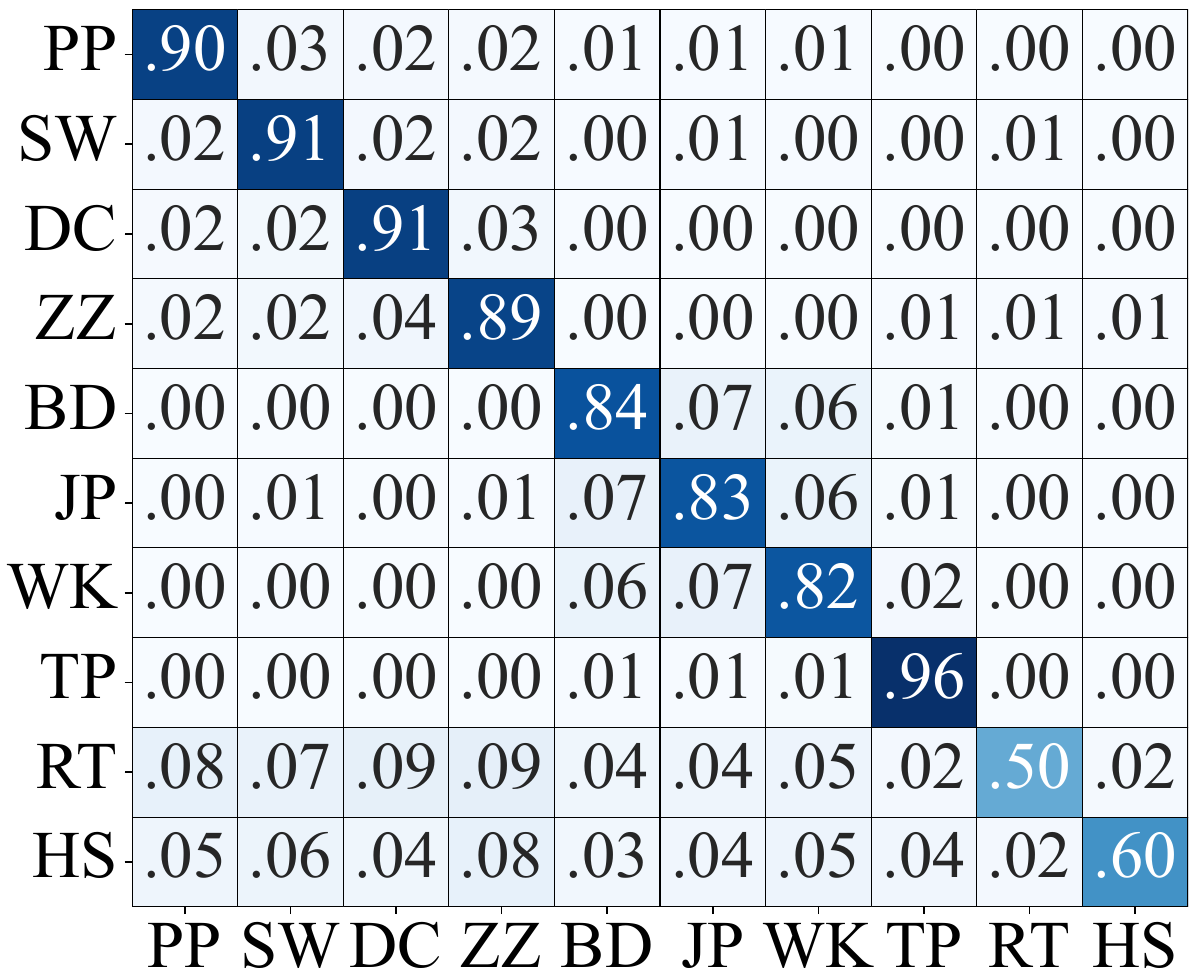}
		 \label{sfig:overall_da}    
		}
        \hfill
	\subfigure[Module optimization.]
		{
			\centering         
			\includegraphics[width=0.46\linewidth]{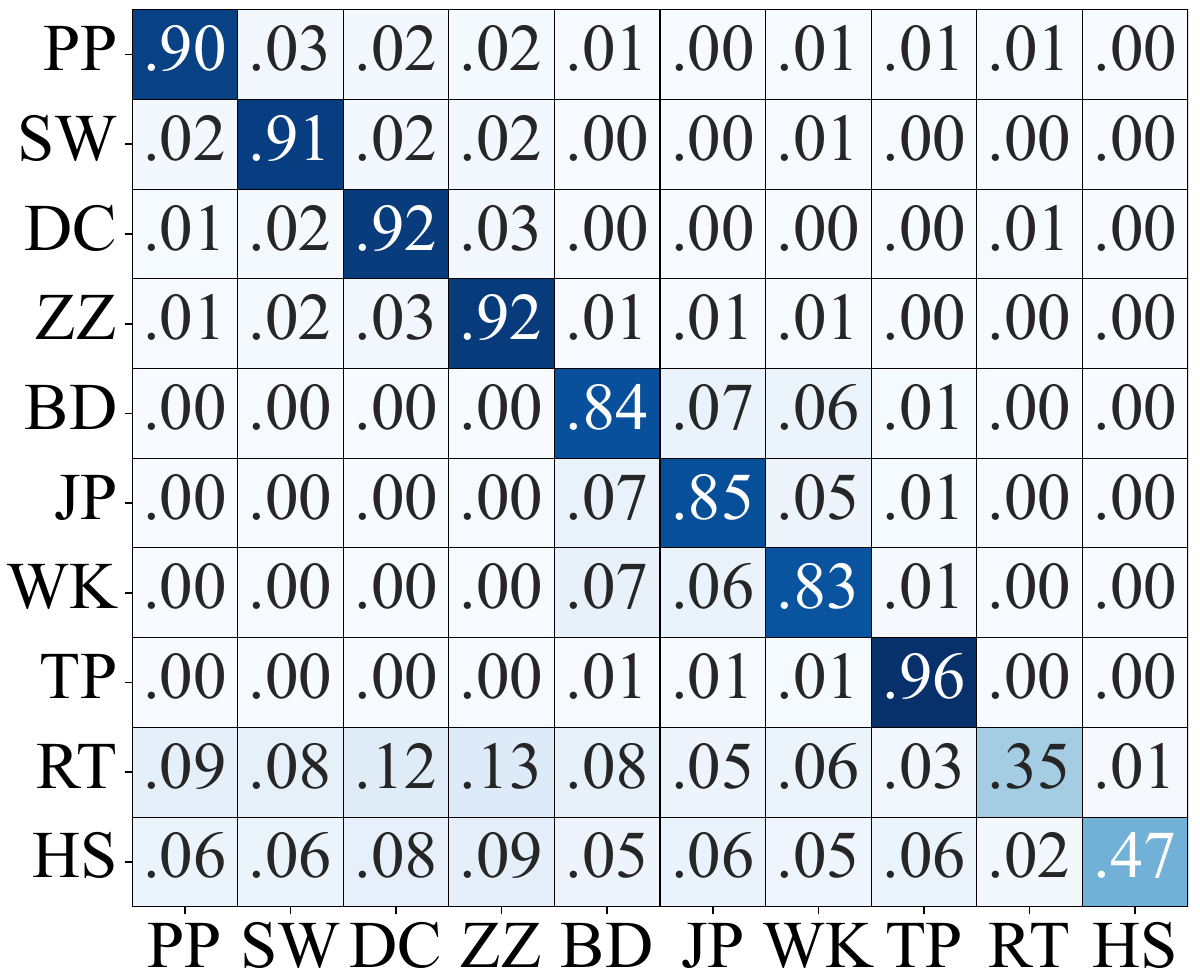}
			\label{sfig:overall_mo}  
		}
	\caption{Overall HAR performance of (a) \name, (b) class-sensitive learning, (c) data augmentation, and (d) module optimization frameworks.}
     \label{fig:overall}
\end{figure}

To evaluate the overall performance of our \name framework, we analyze activity recognition accuracy and compare it with representative baselines.
\rev{Since no existing approaches, to the best of our knowledge, can be directly applied to this unique task in our novel multi-person Wi-Fi sensing system,}
we adopt widely used methods from three aspects~\cite{zhang2023deep}, namely class-sensitive learning, data augmentation, and module optimization, as baselines.
Specifically:
\begin{itemize}
    \item Class-sensitive Learning: During both the PT and FT stages, the softmax loss is reweighted across categories to balance uneven gradients~\cite{park2021influence}, while label smoothing is applied to mitigate overconfident predictions, thereby improving recognition of activities without FT samples.
    \item Data Augmentation: A generative model~\cite{liu2025efficient} is trained on cluster centers derived from abundant source domain data during the PT stage, and subsequently generates absent-category samples from limited target domain data in the FT stage to enhance HAR performance.
    \item Module Optimization: A scale-invariant cosine classifier~\cite{wu2021adversarial} is employed in both PT and FT stages to eliminate the effect of feature and weight scales by constraining vectors on a hypersphere.
    During the FT stage, $\mathcal{D}^{\mathrm{FT}}$ and $\tilde{\mathcal{D}}^{\mathrm{PT}}$ are jointly sampled to promote intra-class similarity and inter-class dissimilarity within each batch, thereby enhancing the feature extractor.
\end{itemize}
To ensure a fair comparison, all baselines are trained on the data processed according to Section~\ref{ssec:sys_time}, using the basic neural network model presented in Section~\ref{ssec:sys_pt}.
RT and HS, two activities inherently difficult to collect, are treated as categories without available samples during the FT stage.

We sequentially designate the data from 56 different subjects as the target domain and compute their overall HAR performance, as shown in Fig.~\ref{fig:overall}. 
It can be observed from Fig.~\ref{sfig:overall_our} that after FT with our \name framework, the overall recognition accuracy reaches approximately 90.4\%. The categories with only a few FT samples achieve an average accuracy of about 91.4\%, while the categories without FT samples, namely RT and HS, attain an average accuracy of approximately 86.3\%.
The RT and HS exhibit an improvement of about 56.8\% over the approximately 29.5\% accuracy shown in Fig.~\ref{sfig:FT_miss}, demonstrating the feasibility of our \name.

In contrast, Fig.~\ref{sfig:overall_csl} shows that the class-sensitive learning framework yields an overall accuracy of 77.7\%, while the average accuracy of RT and HS is only about 33\%.
Since there are no RT and HS samples from the target domain for FT, the framework can only adjust the loss of source domain data to emphasize certain categories; consequently, this strategy still fails to capture the subject-specific features of the target domain effectively.
Fig.~\ref{sfig:overall_da} shows that the data augmentation framework achieves an overall accuracy of 81.6\%.
Although RT and HS show noticeable improvement, their average accuracy remains limited to about 55\%.
This limitation arises from the inherent complexity and ambiguity of Wi-Fi signals, which inevitably introduce discrepancies between real and generated data, thereby restricting recognition to partially similar samples.
Moreover, since the absent categories are not fixed, such methods require maintaining multiple additional generative models, which further increases the overall system complexity.
In Fig.~\ref{sfig:overall_mo}, the module optimization framework yields an overall accuracy of 79.4\%, with RT and HS achieving an average accuracy of approximately 40\%, which falls between the results of the previous two baselines.
This is primarily because, although such methods can promote target domain feature extraction to some extent,
they capture only the local sample distributions of $\mathcal{D}^{\mathrm{FT}}$ and $\tilde{\mathcal{D}}^{\mathrm{PT}}$ within small batches; these limitations, compounded by the reliance on complex classifier, often lead to gradient conflicts and ultimately result in unstable optimization.
These results fully demonstrate the superiority of our \name.

\vspace{-0.5em}
\subsection{Impact Factors} \label{ssec:factor}

In this section, we first evaluate the potential impact factors to demonstrate the generalization capability of our \name framework.
For conciseness, the metrics are defined as the average accuracies in the target domain for categories with FT samples and for those without FT samples.
Finally, we conduct an ablation study to assess the contribution of each algorithm module within the framework.

\subsubsection{Environment and Subject} \label{sssec:factor_env_subj}


\begin{figure}[t]
	\centering 
	\setlength{\abovecaptionskip}{2pt}  
	\subfigure[Impact of environment.] 
		{
		 \centering
		 \includegraphics[width=0.46\linewidth]{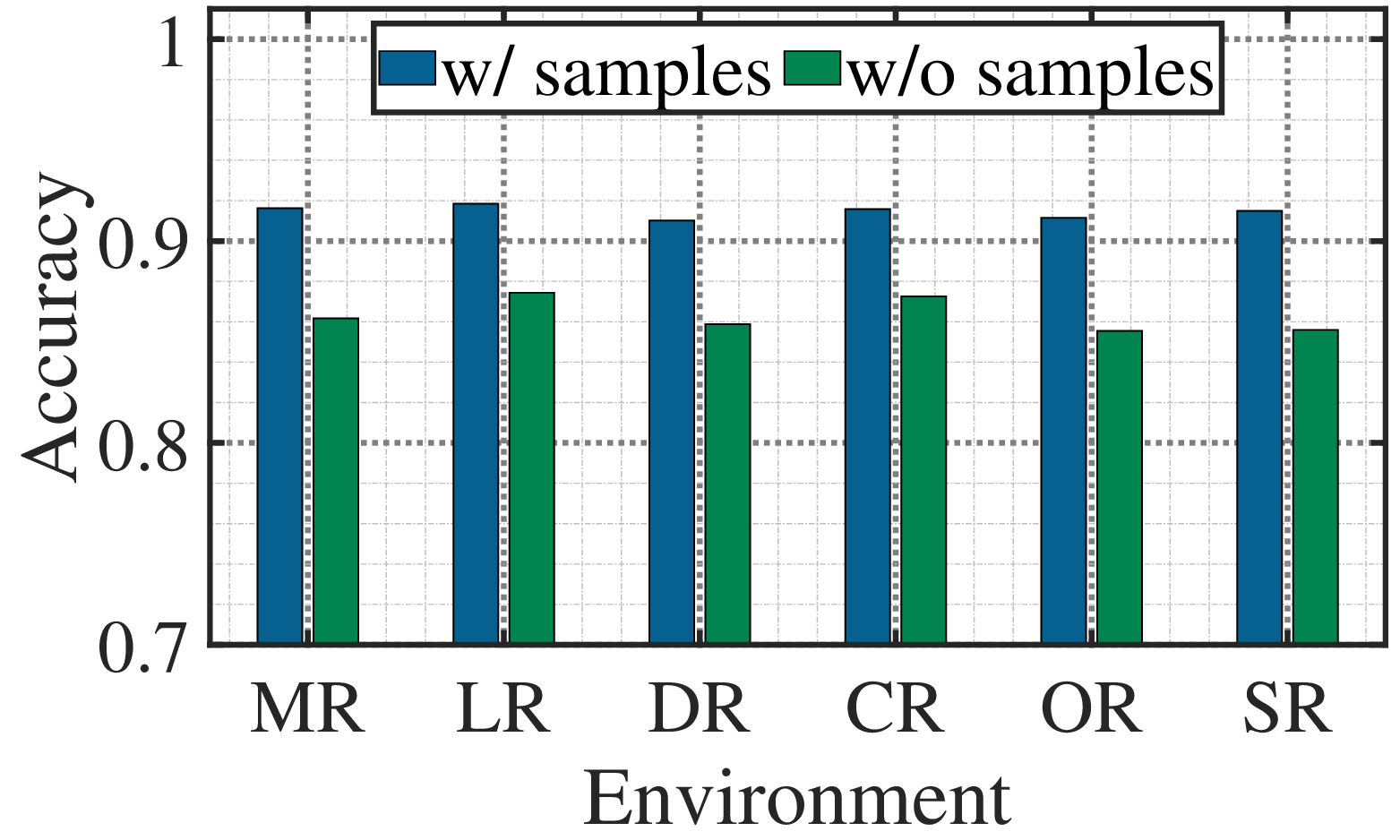}
		 \label{sfig:imp_env}    
		}
        \hfill
        \subfigure[\rev{Impact of concurrent subjects.}] 
		{
		 \centering
		 \includegraphics[width=0.46\linewidth]{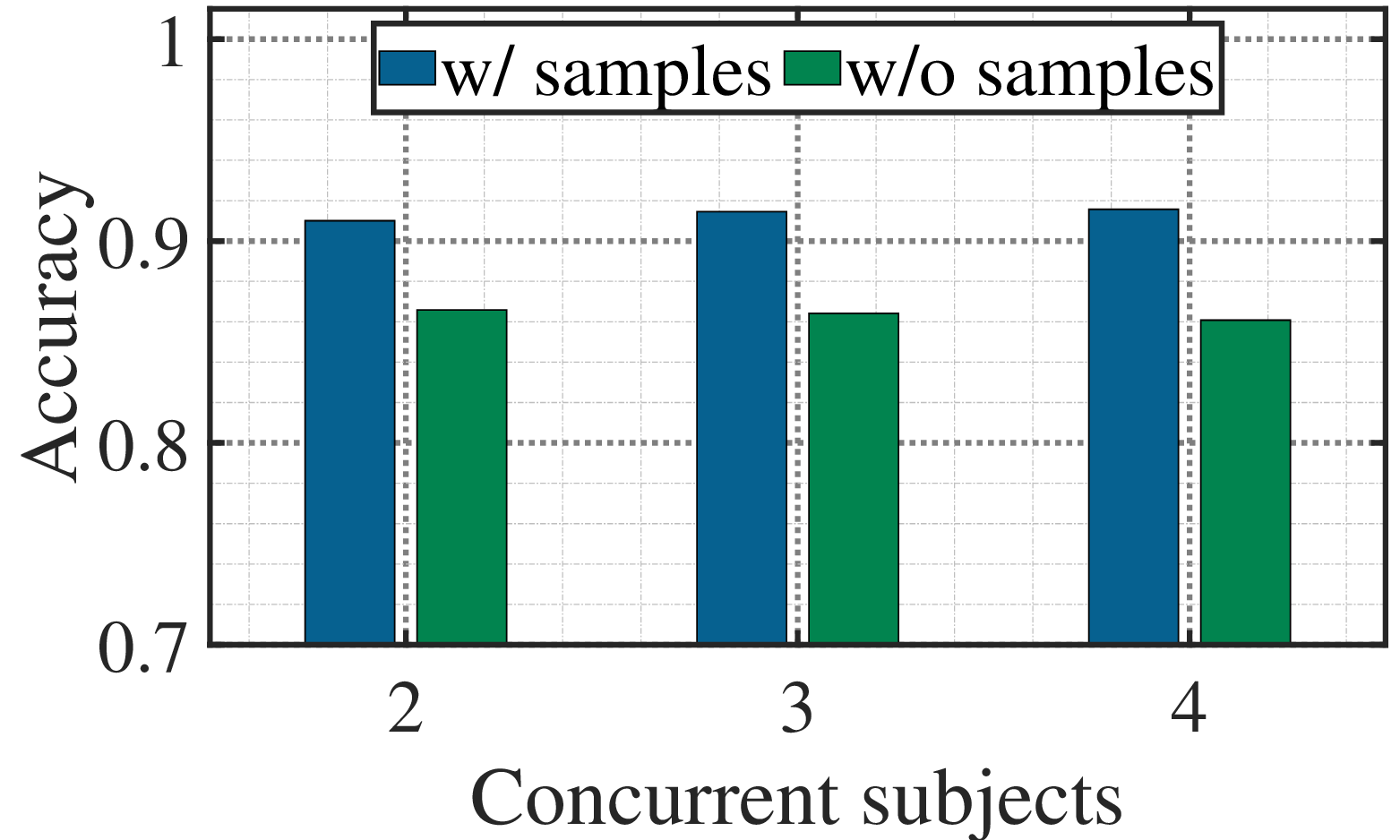}
		 \label{sfig:imp_subj_num}    
		}
        \\
	\subfigure[Impact of subject.]
		{
			\centering         
			\includegraphics[width=0.95\linewidth]{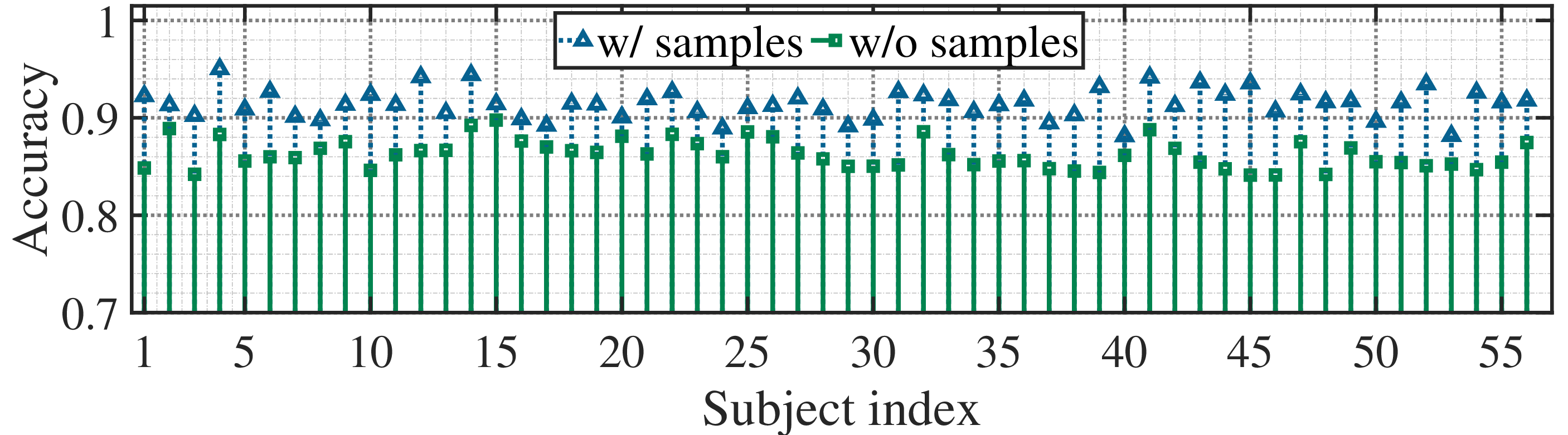}
			\label{sfig:imp_subj}  
		}
	\caption{Impacts of environment and subject.}
     \label{fig:imp_env_subj}
\end{figure}

To evaluate the impact of the environment on HAR performance, we analyze the average accuracy of all subjects across different environments, as shown in Fig.~\ref{sfig:imp_env}.
The results indicate that the accuracies of categories with and without FT samples vary only slightly across environments.
A closer examination shows that the accuracies in the DR and OR environments are relatively lower than those in other scenarios.
This is primarily due to the small and crowded nature of these rooms, which severely complicates multipath propagation and increases the likelihood of interference during activity execution, thereby negatively impacting recognition performance.

\rev{We next analyze the impact of the number of concurrent subjects, as shown in Fig.~\ref{sfig:imp_subj_num}.
The recognition accuracies across activity categories remain around 91\% and 86\% with and without FT samples, respectively, indicating no significant fluctuations.
This can be primarily attributed to the near-field domination effect in our multi-person sensing system, where each subject induces dominant channel variations on its corresponding link.}
Furthermore, the recognition accuracies across all 56 subjects are shown in Fig.~\ref{sfig:imp_subj}.
For activity categories with FT samples, the accuracy for all subjects remains around 90\%, with Subject 4 achieving the highest accuracy of 95.0\%, while Subjects 40 and 50 record relatively lower accuracies of 88.1\% and 89.6\%, respectively.
For activity categories without FT samples, most subjects achieve accuracies around 85\%, with Subject 48 having the lowest accuracy of 84.2\% and Subject 15 achieving the highest accuracy of 89.8\%.
Based on the experiment observations, this discrepancy may be attributed to differences in inter-class and intra-class similarity caused by variations in motion amplitudes.
Overall, satisfactory recognition results are achieved regardless of variations in environment or subjects, demonstrating the generalization capability of our \name framework.

\subsubsection{Activity Category} \label{sssec:factor_act}

%
\begin{figure}[b]
	\centering 
	\setlength{\abovecaptionskip}{2pt}  
	\subfigure[Absence activity category.] 
		{
		 \centering
		 \includegraphics[width=0.46\linewidth]{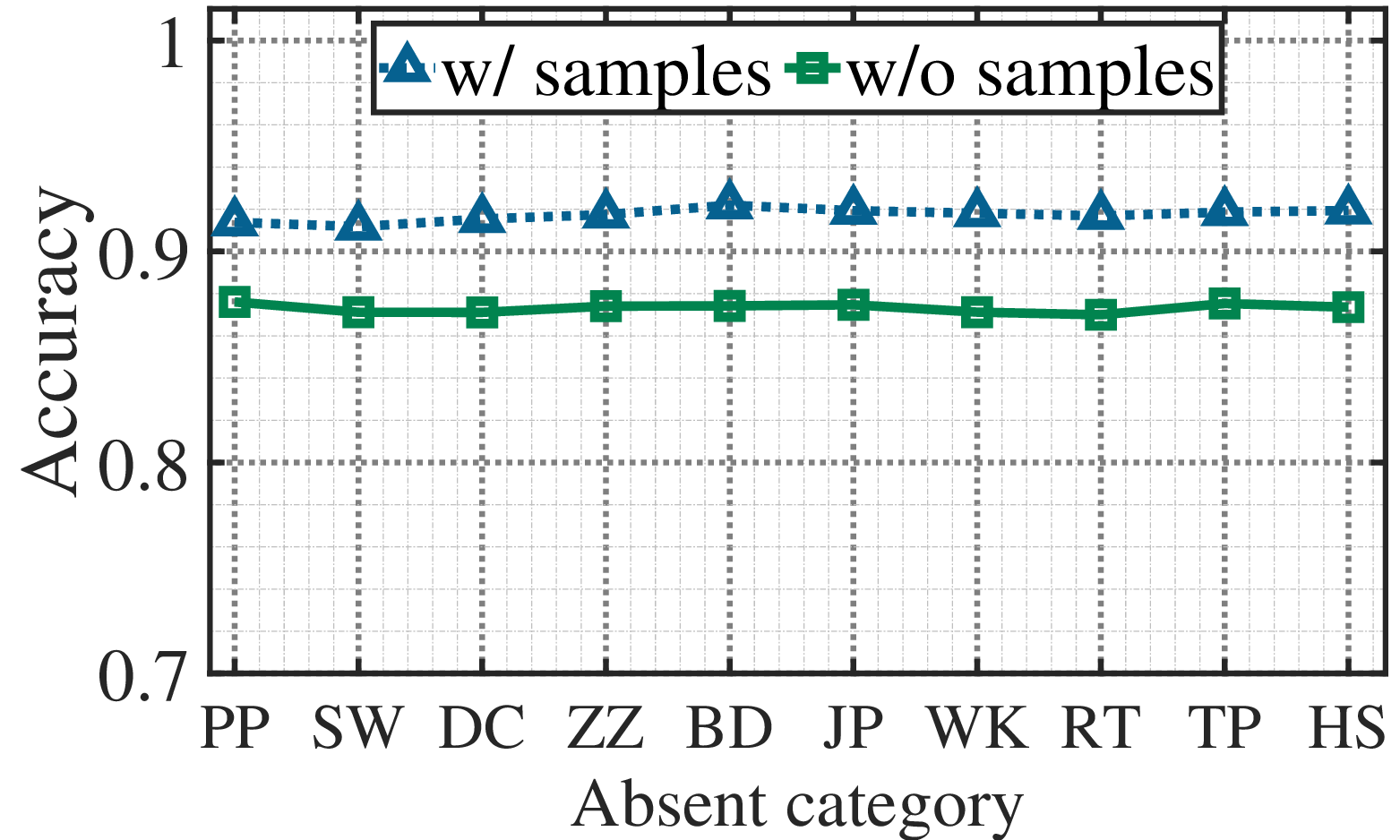}
		 \label{sfig:act_1}    
		}
        \hfill
	\subfigure[Absent category number.]
		{
			\centering         
			\includegraphics[width=0.46\linewidth]{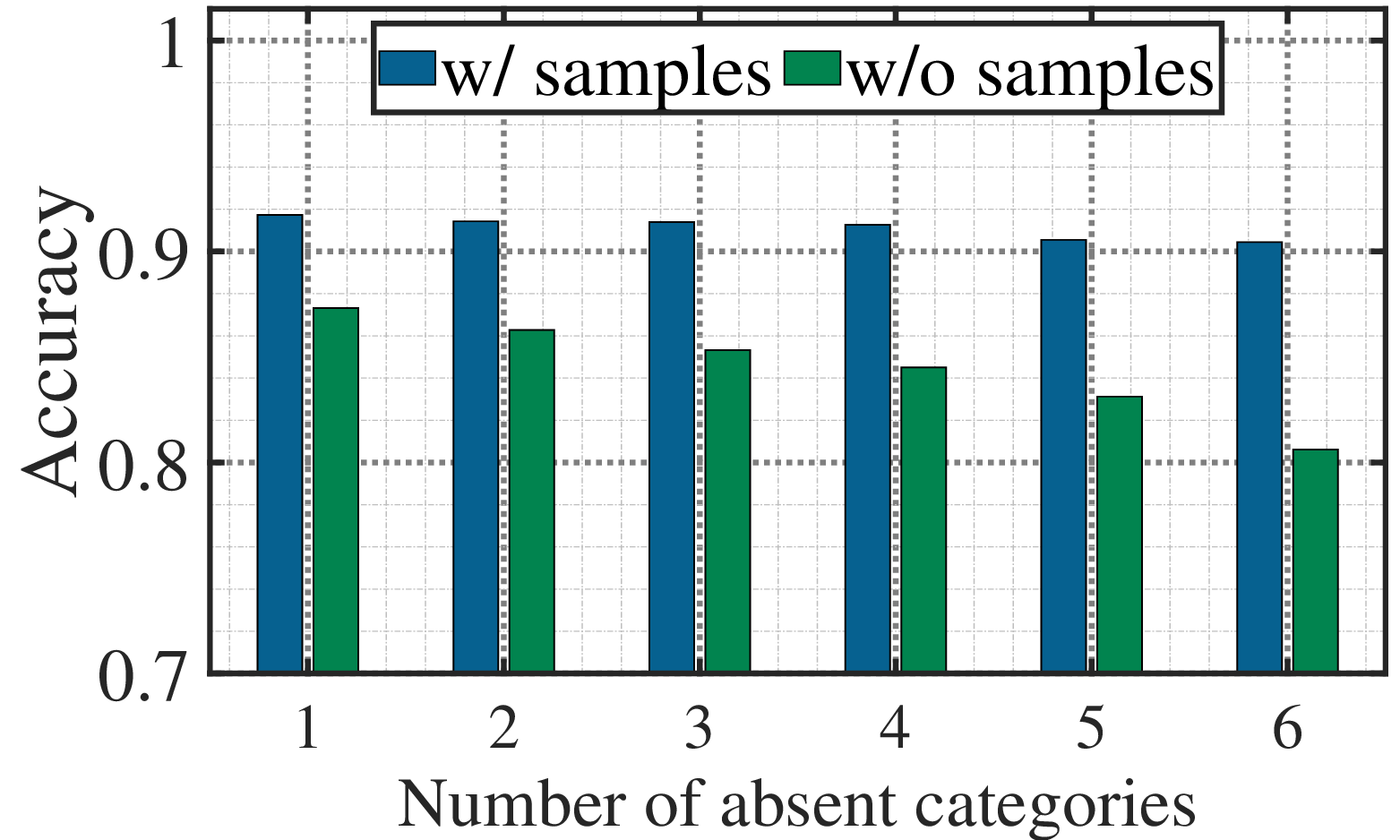}
			\label{sfig:act_number}  
		}
	\caption{Impact of absent activity.}
     \label{fig:impact_act}
\end{figure}

To evaluate the impact of activity category without FT samples, we analyze each activity individually, as shown in Fig.~\ref{sfig:act_1}.
The results indicate that the accuracy of categories with FT samples remains above 90\%, while the accuracy of the absent activity improves to approximately 87\%, demonstrating that our \name framework can handle scenarios with various absent categories.
\rev{Moreover, the average recognition accuracies for gesture, body, and interactive activities in the absence of FT samples are 87.3\%, 87.2\%, and 87.5\%, respectively.
The results indicate that \name framework effectively reduces the performance disparity across different activity types, compared with Fig.~\ref{sfig:FT_miss}.}

We further analyze the effect of the number of activity categories without FT samples, as shown in Fig.~\ref{sfig:act_number}.
The results reveal that the recognition accuracy of categories with FT samples fluctuates only slightly.
However, as the number of absent categories increases, the recognition accuracy of these activities gradually decreases, dropping to approximately 80\% when six categories are absent.
This decline is primarily due to the limited available samples, which do not provide sufficient information for the anchor matching algorithm introduced in Section~\ref{ssec:sys_ft} to learn subject-specific filtering characteristics.
Nevertheless, our \name framework consistently demonstrates significant performance in categories without FT samples, in comparison with Fig.~\ref{sfig:FT_miss}, while ensuring high recognition accuracy for categories with FT samples.

\subsubsection{Training Data Size} \label{sssec:factor_datasize}


\begin{figure}[b]
	\centering 
	\setlength{\abovecaptionskip}{2pt}  
        \subfigure[Impact of PT data size.] 
		{
		 \centering
		 \includegraphics[width=0.95\linewidth]{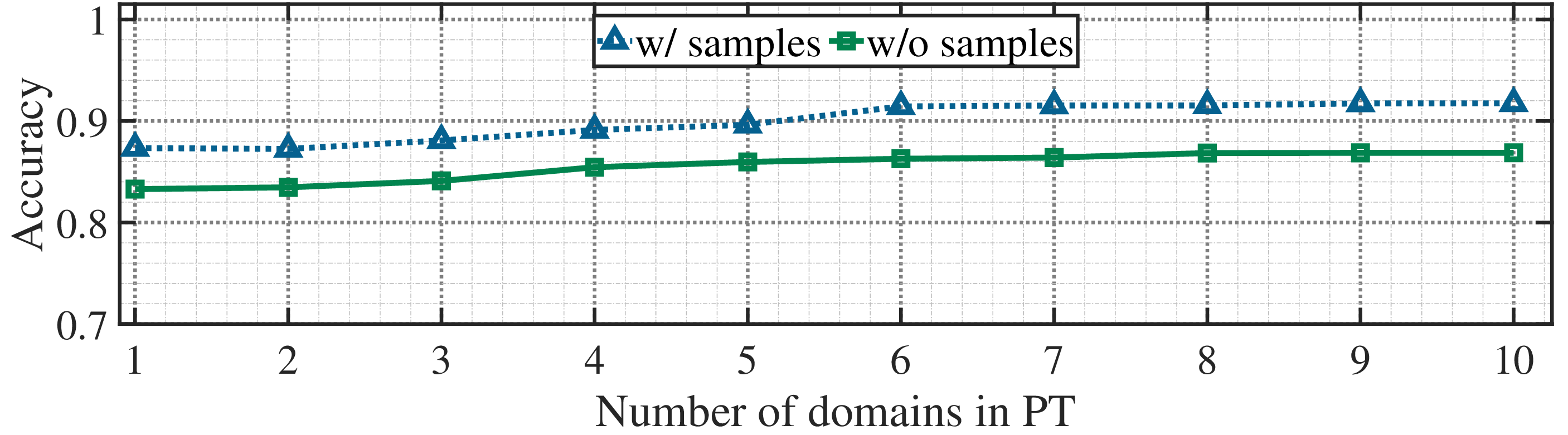}
		 \label{sfig:impact_PTsize}    
		}
        \\
	\subfigure[\rev{Impact of FT data size.}]
		{
			\centering         
			\includegraphics[width=0.46\linewidth]{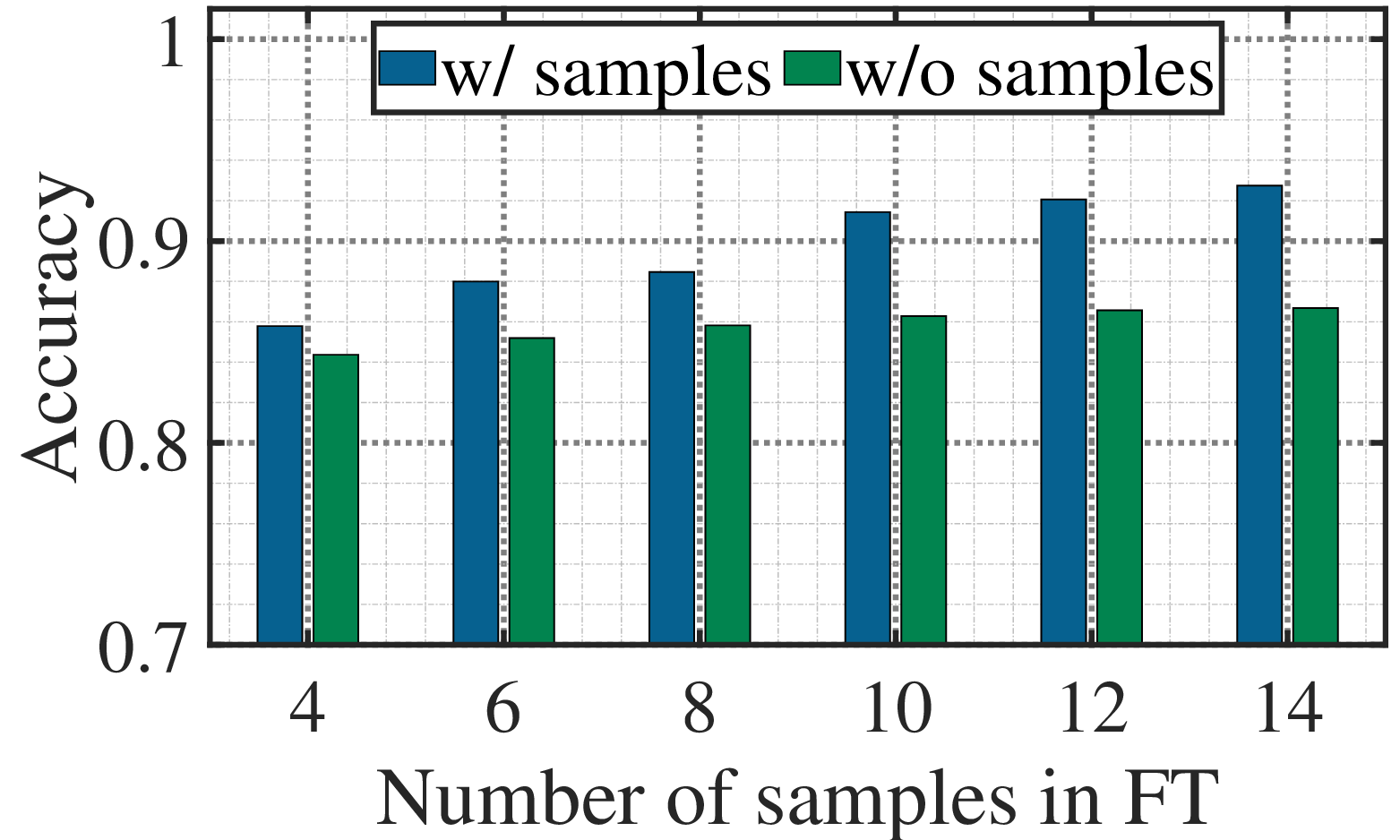}
			\label{sfig:impact_FTsize}  
		}
        \hfill
	\subfigure[\rev{Impact of anchor data size.}]
		{
			\centering         
			\includegraphics[width=0.46\linewidth]{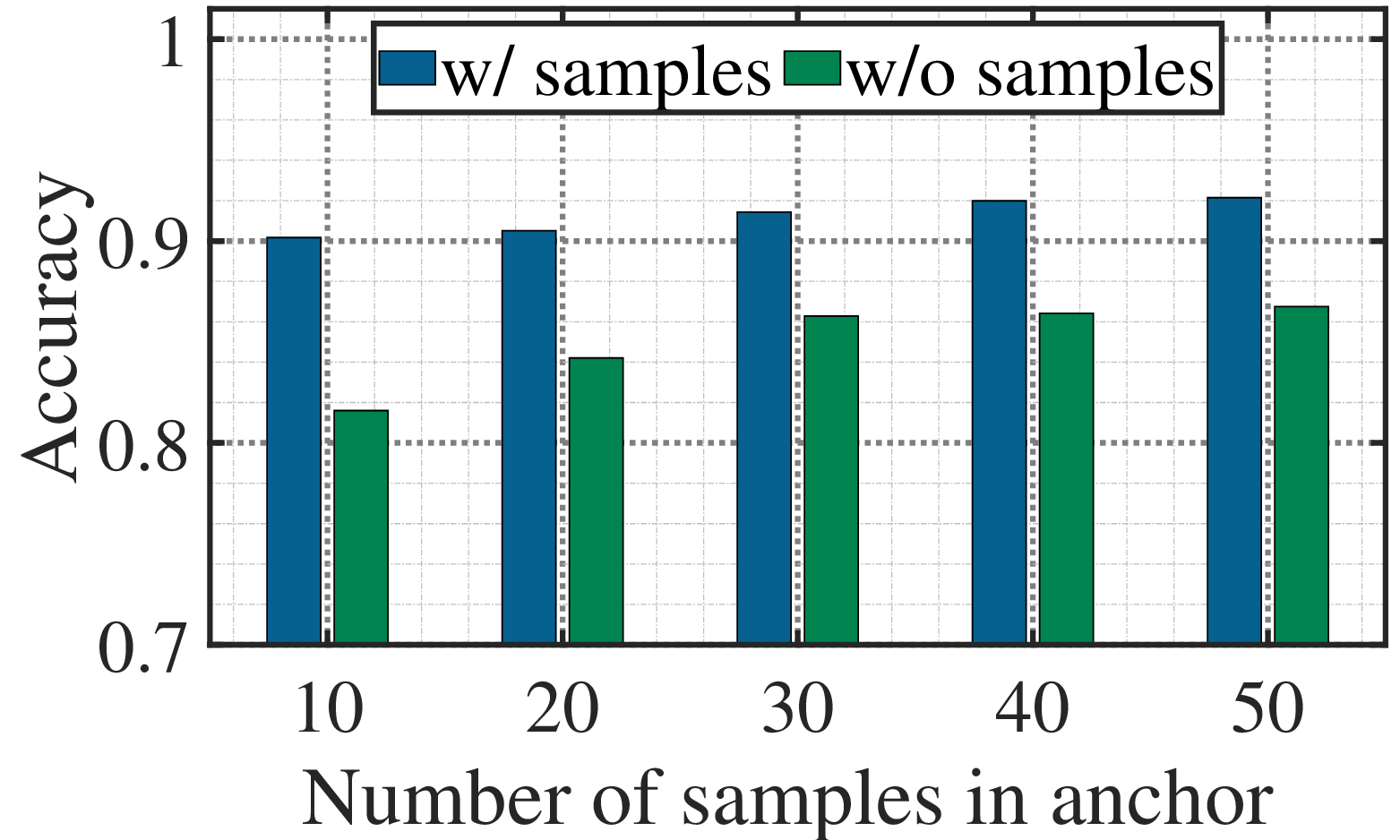}
			\label{sfig:impact_ANsize}  
		}
	\caption{Impact of training data size.}
     \label{fig:impact_datasize}
\end{figure}

To evaluate the impact of the data size used in the PT stage, we analyze the recognition results under different numbers of source domains (subjects).
As shown in Fig.~\ref{sfig:impact_PTsize}, the recognition accuracy of all activities increases with more PT data; however, once the number of source domains reaches six, the improvement becomes negligible.
This indicates that the model requires sufficient information to capture the distribution of activity features for better handling of unseen subjects. Nevertheless, once the training data diversity reaches a certain scale, simply increasing the data volume no longer provides additional cross-domain recognition benefits.
Therefore, we select data from six subjects as source domains for training our model.

We further evaluate the impact of the number of available FT samples per category on model performance, as shown in Fig.~\ref{sfig:impact_FTsize}, where RT and HS remain absent categories.
The results demonstrate that the recognition accuracy of activities with FT samples is strongly affected by their data size, but the performance gains saturate beyond 10 samples per category, indicating that the HAR model has already learned stable activity feature distributions.
In contrast, the number of available FT samples does not significantly affect the accuracy of activities without samples, since even a small and diverse set drawn from multiple categories provides sufficient information for the HAR model to filter subject-specific interference.
Overall, selecting 10 samples per available category for FT is adequate to achieve satisfactory performance.
\rev{Given that each activity takes 2 seconds, providing the FT dataset requires only 160 seconds, which is a reasonable effort.
Moreover, this is not mandatory, as the overall recognition rate of 85.1\% can still be achieved with just 4 samples per activity.}
\norev{Besides,} compared with Fig.~\ref{sfig:FT_size}, where 30 FT samples per category are needed to reach saturation, our \name framework substantially reduces the
\norev{burden on users.}

\rev{The anchor data, pre-stored on Wi-Fi devices, is another influencing factor in our \name framework.
The variation in results with the number of samples per activity category in the anchor data is shown in Fig.~\ref{sfig:impact_ANsize}.
The accuracy of activities with FT samples does not show significant variation, remaining around 91\%.
The accuracy of activities without FT samples increases with the anchor data size, reaching saturation upon reaching 30 samples per activity.
The reason is that our \name framework requires reliable cluster centers to learn the filtering characteristics necessary for recognizing activities with absent samples.
Furthermore, the anchor data is approximately 1~\!MB after specialized compression, which does not impose a heavy storage burden on Wi-Fi devices.}

\subsubsection{\rev{Hyper-parameter and Model Architecture}} \label{sssec:factor_arc}

\begin{figure}[b]
	\centering 
	\setlength{\abovecaptionskip}{2pt}  
	\subfigure[GRU model structure.] 
		{
		 \centering
		 \includegraphics[width=0.46\linewidth]{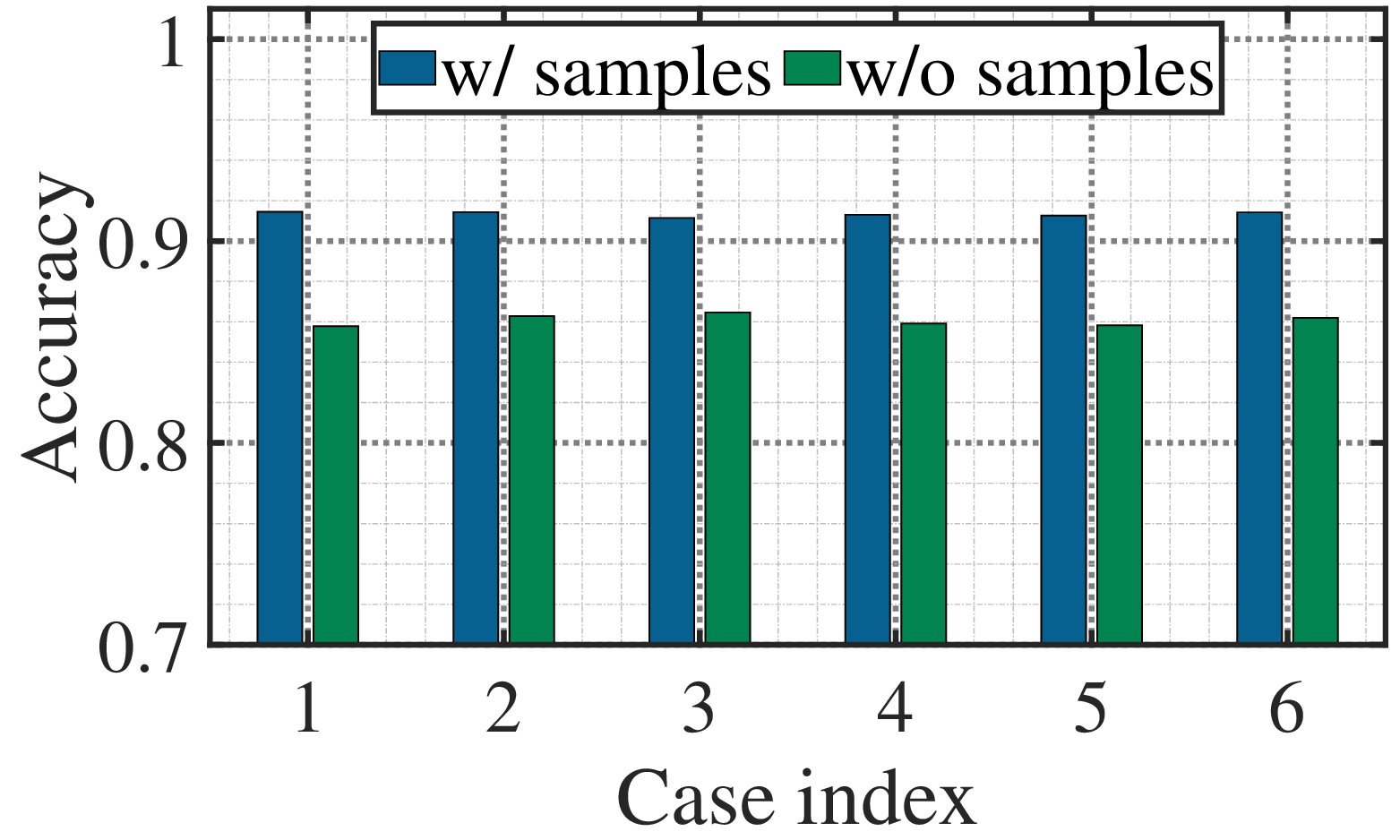}
		 \label{sfig:gru}    
		}
        \hfill
	\subfigure[\rev{Inference strategy weight.}]
		{
			\centering         
			\includegraphics[width=0.46\linewidth]{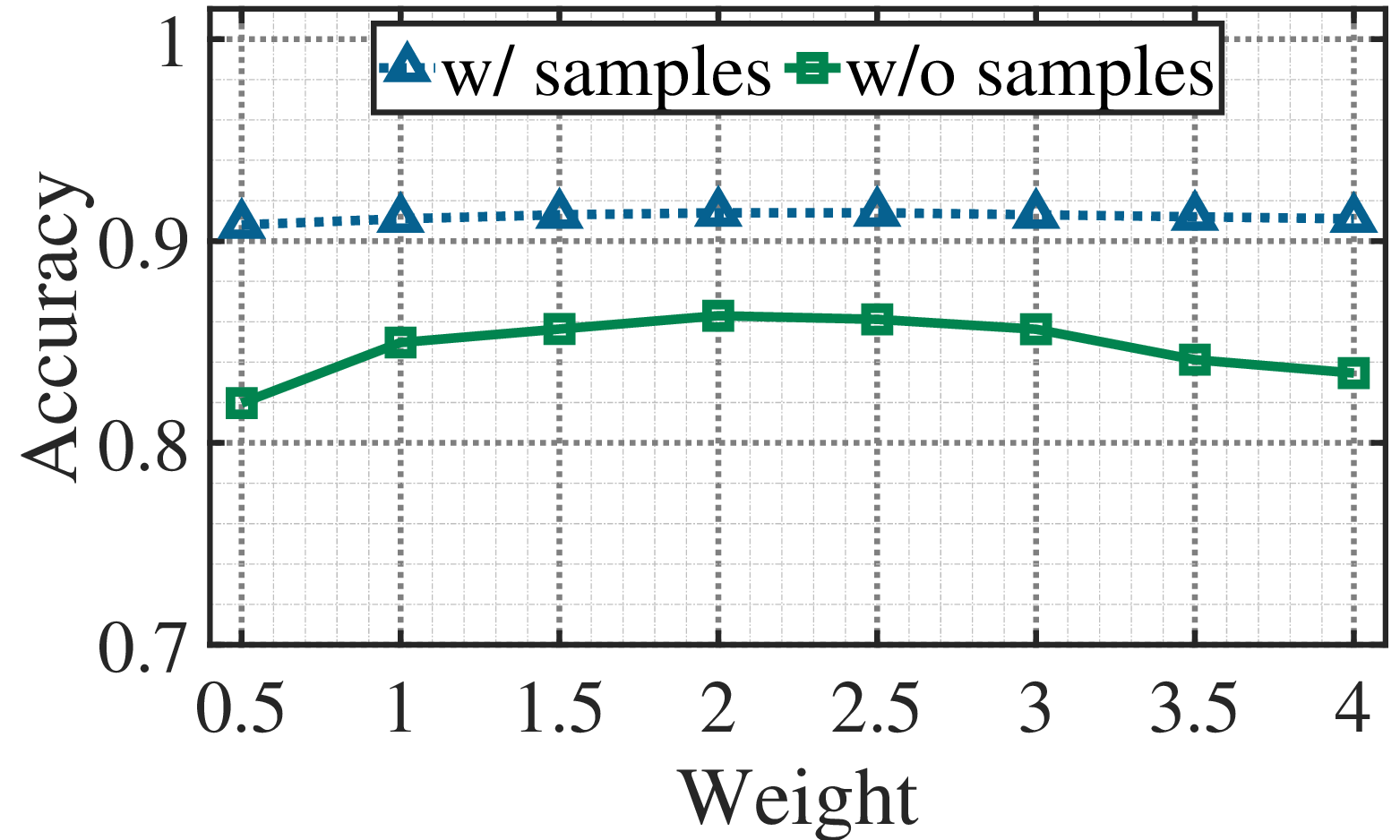}
			\label{sfig:lambda3}  
		}
	\caption{Impact of hyper-parameters.}
     \label{fig:impact_hyper}
\end{figure}

To evaluate the impact of the model structure parameters, we first configure the GRU with one layer and vary the hidden size as 32, 64, and 128, denoted as Cases 1–3, and then increase the number of layers to two for these hidden sizes, denoted as Cases 4–6.
As shown in Fig.~\ref{sfig:gru}, the accuracy differences are marginal, indicating the generalizability of our \name framework to different neural network structures.
In addition, the recognition accuracy of activities without FT samples is slightly lower in the two cases with a hidden size of 32.
Considering both performance and model simplicity, Case 2, i.e., one layer with a hidden size of 64, is selected as our configuration.
\rev{This model has only about $84,000$ parameters, with an inference time of less than 50~\!$\mu s$, making it highly promising for deployment on edge devices.}
\rev{The inference strategy weight, i.e., $\lambda_3$ in Eqn.~\eqref{eqn:decision}, is then analyzed, as shown in Fig.~\ref{sfig:lambda3}.
This weight primarily affects the recognition accuracy of activities without FT samples, with the best results obtained at $\lambda_3=2$.
Moreover, satisfactory results are achieved for $\lambda_3$ values between 1.5 and 3.5, demonstrating the framework's tolerance for hyper-parameter selection.}

\begin{figure}[t]
	\centering 
	\setlength{\abovecaptionskip}{2pt}  
	\subfigure[CNN model.]
		{
			\centering         
			\includegraphics[width=0.46\linewidth]{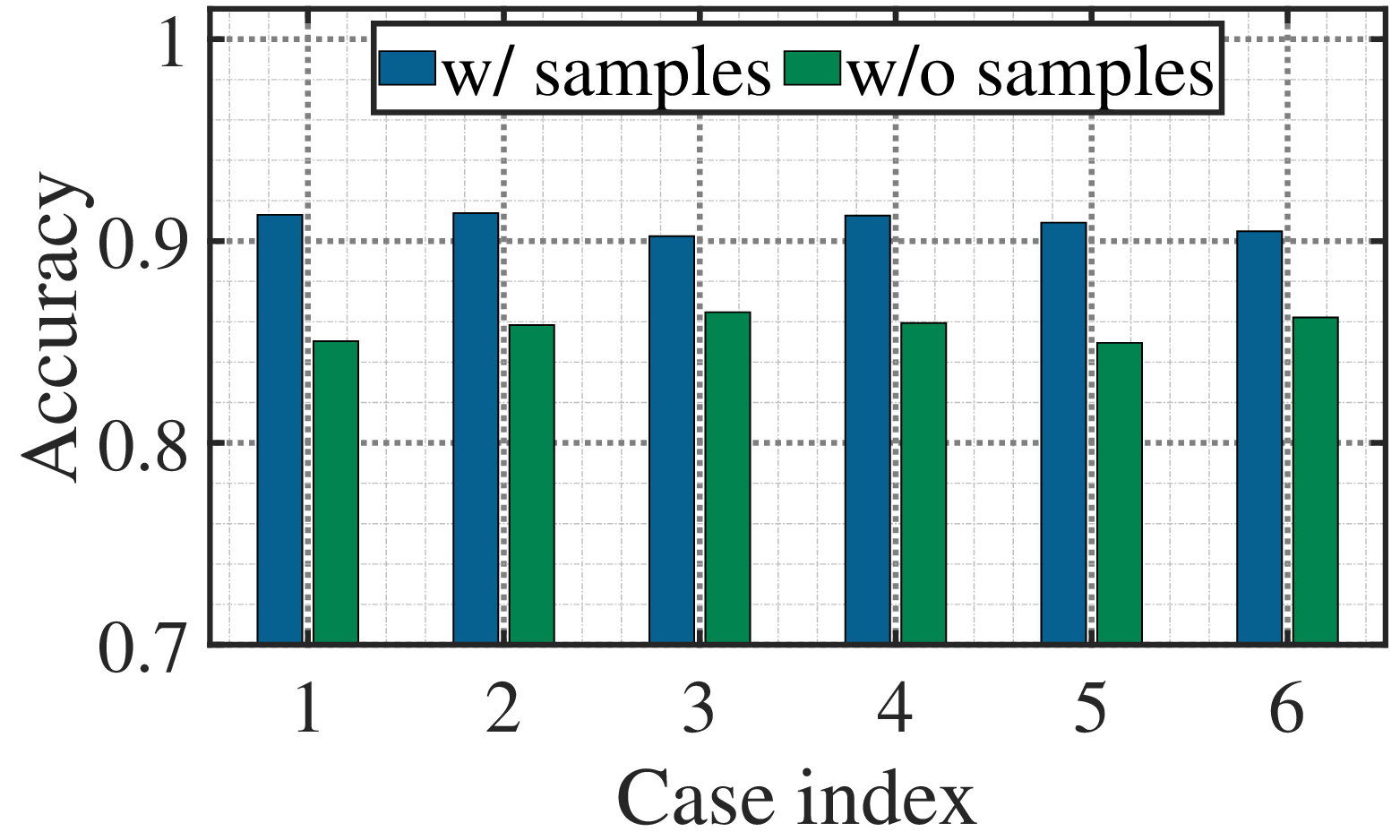}
			\label{sfig:cnn}  
		}
        \hfill
        \subfigure[\rev{Transformer model.}] 
		{
		 \centering
		 \includegraphics[width=0.46\linewidth]{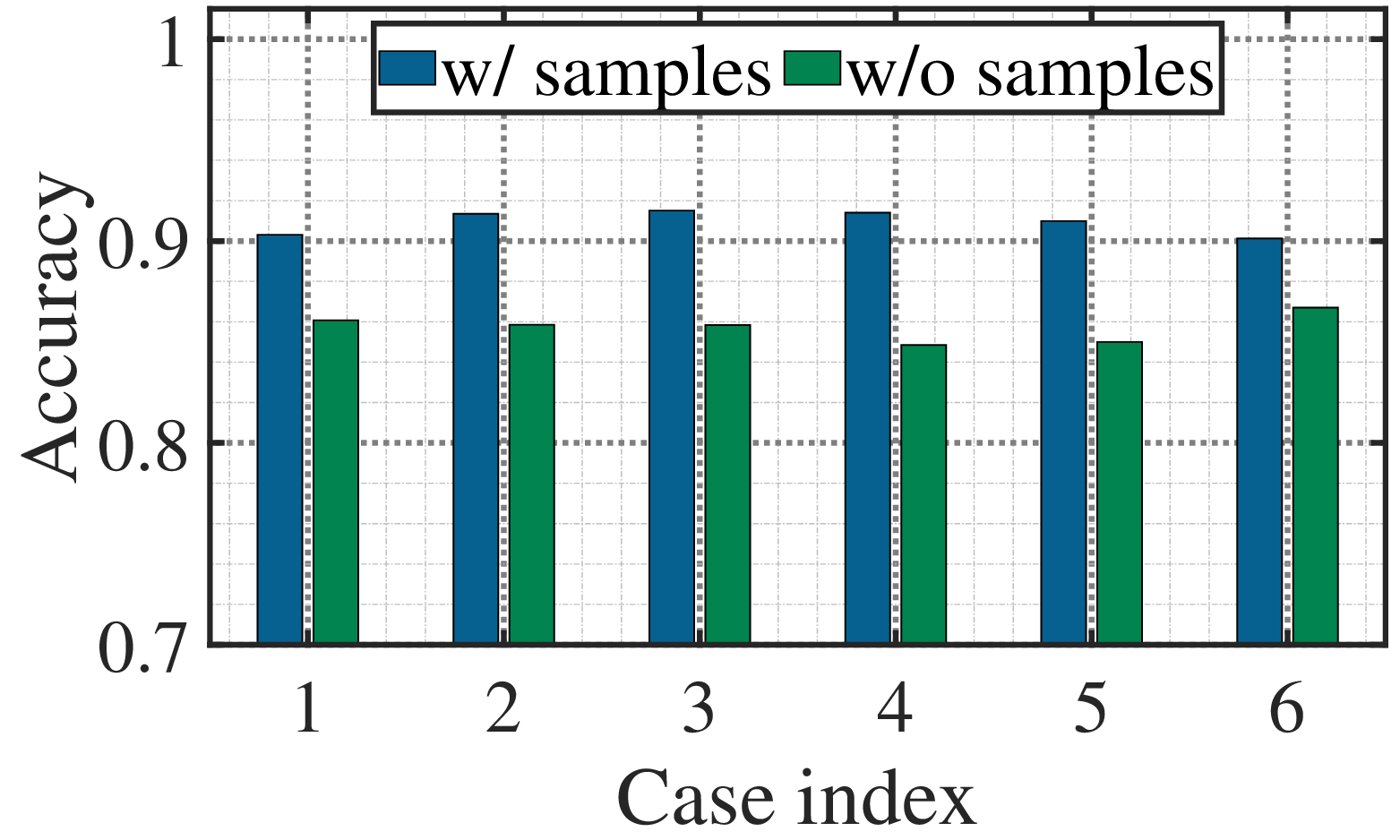}
		 \label{sfig:transf}    
		}
	\caption{Impact of model architectures.}
     \label{fig:impact_arc}
\end{figure}

To further assess the influence of the model architecture, we replace the GRU units with \rev{alternative modules.}
\rev{We first use 1D CNNs, setting} the number of convolutional layers to one with kernel sizes of 3 and 5, denoted as Cases 1 and 2, and then increase the number of layers to two and three, denoted as Cases 3–6.
\rev{The results are shown in Fig.~\ref{sfig:cnn}.}
\rev{We next use Transformers, setting the embedding dimension to 32 with both the encoder and decoder having 2, 3, and 4 layers, denoted as Cases 1–3, and then increase the embedding dimension to 64, denoted as Cases 4–6.
The results are shown in Fig.~\ref{sfig:transf}.
These recognition accuracies are insensitive to variations in the module and structure,}
and the overall performance does not exhibit a significant difference compared with the GRU-based model.
These results demonstrate that our \name framework can effectively adapt to diverse network architectures.
It is worth emphasizing that the core of this work focuses on the design of the training framework rather than the neural network architecture, while the development of high-performance models remains an open avenue.


\subsubsection{Ablation Study} \label{sssec:factor_ablation}

\begin{figure}[b]
	\centering 
	\setlength{\abovecaptionskip}{2pt}  
		 \includegraphics[width=0.95\linewidth]{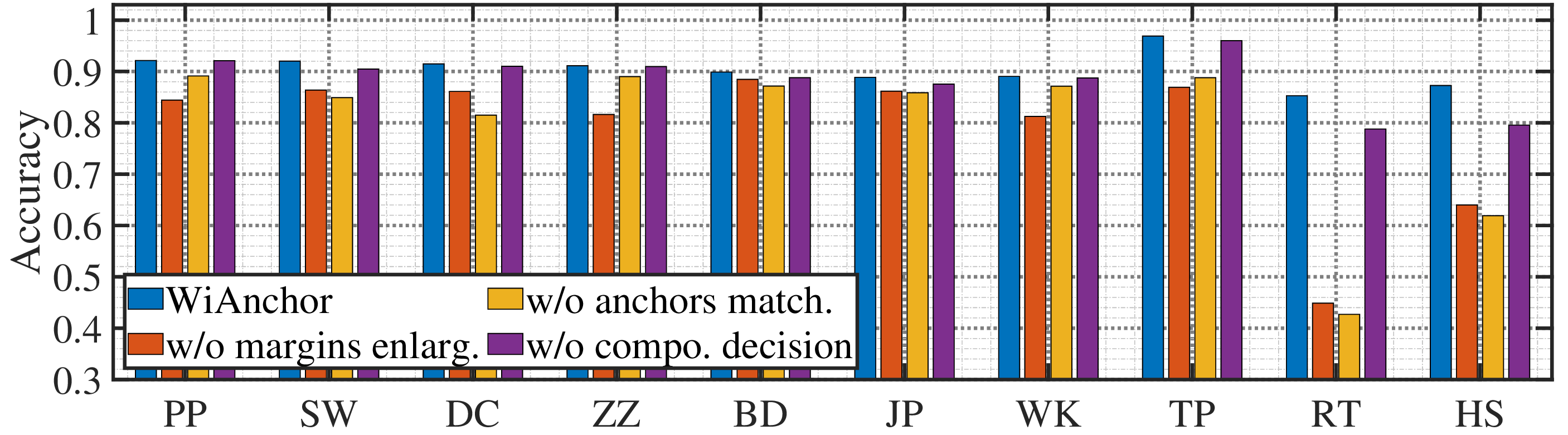}
	\caption{Impact of algorithms in \name.}
     \label{fig:ablation}
\end{figure}

To evaluate the importance of each algorithmic component in our \name framework, we perform HAR analysis by removing them individually, with the results shown in Fig.~\ref{fig:ablation}, where RT and HS remain as categories without FT samples.

First, we remove the inter-class margin enlarging applied throughout both PT and FT stages, with the results indicated in red.
The average recognition accuracy of activities with FT samples drops to 85.2\%, while that of activities without FT samples decreases significantly to 54.4\%.
This can be intuitively explained by the t-SNE visualization of features in Fig.~\ref{fig:vis_tsne}:
the decision boundaries between categories are relatively blurred, which inevitably leads to performance degradation when FT samples are scarce, and makes activities with no FT samples even harder to distinguish.

Second, we remove the anchor matching algorithm used in the PT stage, with the results shown in yellow.
The average recognition accuracy of activities with FT samples is 86.7\%, slightly higher than the previous case but still lower than the full \name framework;
for activities without FT samples, the average accuracy drops to 52.3\%, even lower than the previous case.
This suggests that, despite clear decision boundaries between categories, using only feature extraction without feature-matching filtering hinders accurate recognition of RT and HS, which have no FT samples.

Finally, we remove the composite inference strategy, with results indicated in purple.
The average accuracy of activities with FT samples is 90.7\%, while that of activities without FT samples decreases to 79.2\%.
Although the performance drop is less pronounced than in the first two cases, the accuracy of RT and HS falls below 80\%.
This highlights the benefit of the composite decision, which effectively leverages the advantages of both the PT and FT stages, yielding superior generalization compared with relying solely on the softmax function.

\section{Conclusions and Discussions} \label{sec:conclusion}

\revhu{We have developed a practical multi-person Wi-Fi HAR system that leverages the near-field domination effect to establish a dedicated link for each subject and capture diverse information from the physical layer.}
\revhu{This system effectively advances Wi-Fi sensing to realistic, real-world scenarios, with the ability to recognize previously unseen subjects during deployment through on-site FT.}
To address the challenge posed by the absence of FT samples in certain categories, we develop the \name framework.
Our \name first captures temporal irregularity patterns in CSI data through time information embedding.
HAR model training is then divided into two stages: during the PT stage, \name enlarges inter-class margins to improve category separability, while in the FT stage, it learns subject-specific filtering characteristics through an anchor matching mechanism.
In the inference stage, a composite decision strategy is employed to further enhance recognition performance.
Due to the lack of publicly available datasets, we construct a unique dataset comprising approximately 65,000 multi-person near-field sensing samples to evaluate our \name framework.
Extensive evaluation shows that \name achieves 91.4\% accuracy for categories with FT samples and 86.3\% for those without, while also demonstrating robust generalization to various impact factors.

\rev{The near-field sensing system relies on users carrying their dedicated devices, which remains a practical solution, as individuals often keep their personal Wi-Fi devices close and it does not disrupt communication.
Moreover, this paradigm, combined with the general \name training framework, has the potential to facilitate the large-scale application of Wi-Fi sensing in scenarios such as VR~\cite{wang2025vr}.}
Although improving recognition performance, \name framework introduces potential privacy risks, \rev{as manufacturers may deliberately conceal the full capabilities of Wi-Fi APs from users and infer unauthorized activities.
Therefore, protecting user rights to information and strengthening permission constraints are critical issues.
Given that manufacturers may not fully cooperate,} we have proposed a poisoning-based approach~\cite{hu2025poison} to safeguard user privacy and intend to investigate additional efficient strategies in future work.
\rev{Different applications generate varying traffic patterns, manifested in irregularity and packet arrival rate~\cite{muzi, favor, li2016art}, which in turn affect sensing performance.
The time information embedding method proposed in Section~\ref{ssec:sys_time} mitigates the former, while the latter remains a primary focus of our future work.}

\bibliographystyle{IEEEtran}


\end{document}